\documentclass{article}

\usepackage{arxiv}
\usepackage{amsmath}
\usepackage[utf8]{inputenc} 
\usepackage[T1]{fontenc}    
\usepackage{url}            
\usepackage{booktabs}       
\usepackage{amsfonts}       
\usepackage{nicefrac}       
\usepackage{microtype}      
\usepackage{graphicx}
\usepackage{natbib}
\usepackage{doi}
\usepackage{hyperref}       
\usepackage{fancyhdr}

\usepackage{enumitem}
\usepackage[mathlines]{lineno} 
\usepackage{subcaption}
\usepackage{caption}
\usepackage{amsmath, amsthm, amssymb, amsfonts}

\usepackage{tikz}
\usetikzlibrary{shapes.geometric, arrows.meta, positioning} 

\title{Bayesian Inversion Beyond the Generator's Range: Reducing Non-physical Constraints from Generative Neural Networks}

\date{} 					

\author{\hspace{1mm}Pratyay Roy \\
	School of GeoSciences\\
	University of Edinburgh\\
	Edinburgh, UK \\
	\texttt{pratyay.roy@ed.ac.uk} \\
	\And
	\hspace{1mm}Xuebin Zhao \\
	School of GeoSciences\\
	University of Edinburgh\\
	Edinburgh, UK \\
	\texttt{Xuebin.Zhao@ed.ac.uk} \\
    \And
	\hspace{1mm}Andrew Curtis \\
	School of GeoSciences\\
	University of Edinburgh\\
	Edinburgh, UK \\
	\texttt{Andrew.Curtis@ed.ac.uk} \\
}

\renewcommand{\headeright}{}

\renewcommand{\undertitle}{A preprint}
\renewcommand{\shorttitle}{Bayesian Inversion Beyond the Generator's Range: Reducing Non-physical Constraints from Generative Neural Networks}
\hypersetup{
pdftitle={Bayesian Inversion Beyond the Generator's Range: Reducing Non-physical Constraints from Generative Neural Networks},
pdfsubject={physics.geo-ph, cs.LG},
pdfauthor={Pratyay Roy, Xuebin Zhao, Andrew Curtis},
pdfkeywords={Bayesian inference, Inverse theory,  Machine learning, Monte Carlo Methods, Probability distributions,  Seismic tomography},
}

\begin{document}
\label{firstpage}

\maketitle

\begin{abstract}
    To address non-uniqueness in geophysical inversions, Generative Neural Networks (GNNs) are often used to parameterise geological or other prior information in a lower-dimensional space (the so-called latent space), thereby reducing the number of inversion parameters. However, the training samples that represent this information are then always incorrect, because Earth structure is not intrinsically low-dimensional, and all GNNs are subject to training errors. Consequently, even when solutions are mapped back to the original higher-dimensional space, they inherit this imperfection and remain restricted to a low-dimensional manifold. From a mathematical point of view, latent space inversion thus fails fundamentally: generally, it can not even reproduce the best, higher-dimensional approximation to the true Earth. To overcome this limitation, we introduce an `escaping prior' distribution, which provides the necessary freedom to find solutions beyond the manifold of GNN-generated solutions, while still using the GNN as a guide. We use a hierarchical Bayesian approach to adapt the strength of prior information to the observed data and synthetic forward function \textit{a posteriori}. We illustrate our method on low-dimensional problems where the escaping prior is defined over a 1D manifold in 2D space. We then compare our approach with standard latent space and model space inversions using a uniform prior distribution in synthetic tomography problems, ranging from simple to more geologically realistic cases. Furthermore, we tested scenarios where the true solution is either far from or close to the manifold. In each case, our escaping prior shows significantly better results than the alternatives. It results in solutions beyond each GNN's generative capabilities, solutions which include the best possible model, and clearly demonstrates the limitations of latent space inversion.
\end{abstract}

\vspace{1cm}

\section{Introduction}

Solutions to physical scientific inverse problems are inherently non-unique because the observed data are sparse, noisy, or insufficient to constrain the solution. Additional \textit{a priori} knowledge is therefore essential to further constrain the solution \citep{jackson_use_1979}. Here we examine the nature and role of prior information in subsurface geophysical tomography. Generally, pre-existing or prior \textit{geological} information must be introduced to guide the inversion toward solutions that are not only mathematically plausible but also geologically realistic \citep{curtis_geological_2004}. While some of our examples are therefore geologically focused, our conclusions apply to other disciplines and domains of application.

In a Bayesian approach to inverse problems with at least some continuous parameters, the prior information about model parameters is represented by a probability density function (pdf). Here, the word model is used to mean a physical representation of the subsurface, describing, for example, the distribution of seismic velocities or rock densities. This prior pdf is combined with the likelihood function, which describes information provided by the observed data, to calculate a pdf representing the resultant, \textit{a posteriori} information \citep{tarantola_inverse_2005}. 

Typically, nonlinear inverse problems are either solved by linearising the physics relating data and parameter values \citep{tarantola_generalized_1982} or by solving the fully nonlinear problem using Monte Carlo methods to explore the set of feasible parameter values (e.g., \cite{mosegaard_monte_1995}). Unfortunately, as the dimensionality (number of unknown model parameters) of an inverse problem increases, the volume of the parameter space grows exponentially so that it becomes impossible to explore, a property referred to as the curse of dimensionality \citep{curtis_prior_2001}. 

Generative Neural Networks (GNNs) provide a way to address the curse in certain classes of problems. GNNs are machine learning architectures composed of artificial neural networks, trained to approximate complex, high-dimensional probability distributions based on a training set of examples (samples) of the desired distribution. For example, particular network architectures such as Variational Autoencoders (VAEs, \cite{kingma_autoencoding_2014}) and Generative Adversarial Networks (GANs, \cite{goodfellow_generative_2014}) can be trained to generate new examples of complex, geologically plausible structures similar to those represented in the training set. In geological problems the training examples might come from geodynamic simulations or existing geological interpretations, for example. The generative networks parameterise this information in a lower-dimensional space, called latent space. If the prior information is genuinely intrinsically lower dimensional than the number of inverse problem parameters, then representing the solution in latent space can produce more tractable, well posed inverse problems.

Previous Geophysical examples include \cite{laloy_inversion_2017, laloy_trainingimage_2018} who reparametrised binary images of channelised alluvial and categorical fold aquifers in learned latent spaces for inverse problems in ground-penetrating radar (GPR) travel-time tomography and hydraulic tomography. \cite{mosser_stochastic_2020} reparametrised cross-sections of fluvial-dominated geological systems for seismic waveform inversion, \cite{meles_bayesian_2023} and \citep{levy_variational_2023} river channel structures for GPR travel-time tomography, \cite{bloem_introducing_2023} braided river systems and marine para-sequences for travel-time tomography,  \cite{santos_refining_2024} 2D synthetic models of the Earth's upper mantle featuring marble-cake structures \citep{alder_quantifying_2017} for elastic homogenization tomography, and \cite{holden_bayesian_2022, liu_uncertainty_2022, bao_variational_2022, chen_fracture_2023, liu_geostatistical_2024, hu_bayesian_2025, taufik_latent_2025} applied similar methods to a variety of other problems. After training in each case, a GNN embodies an approximate mapping from lower-dimensional samples $\mathbf{z}$ in the latent space to higher-dimensional models $\mathbf{m}$ in the geological (or geophysical) model space, using a generator network. All of the above studies performed inversion in this latent space (so-called \textit{latent space inversion}): this involves finding the family of latent parameter values that generate corresponding high-dimensional models which are consistent with the observed data. This reduced the number of parameters to be solved (the dimensionality) from the number of model parameters to the number of latent parameters. To draw samples from the posterior pdf over the latent parameters \cite{laloy_inversion_2017} used a differential evolution adaptive Metropolis algorithm \citep{vrugt_accelerating_2009, laloy_highdimensional_2012}, \cite{mosser_stochastic_2020} implemented approximate Metropolis-adjusted Langevin sampling (MALA-approx) \citep{roberts_exponential_1996, roberts_optimal_1998, nguyen_plug_2017}, \cite{meles_bayesian_2023} and \citep{santos_refining_2024} applied the Metropolis-Hastings Markov chain Monte Carlo (MH-MCMC) algorithm \citep{metropolis_equation_1953, hastings_monte_1970}, and \cite{bloem_introducing_2023} trained mixture density networks \citep{bishop_pattern_2006} to infer the posterior distributions over latent parameters. Regardless of the method used to characterise solutions, the final inversion results in high-dimensional model parameter space were obtained by passing latent samples $\mathbf{z}$ through the GNN's generator network, $\mathcal{G}(\mathbf{z}) = \mathbf{m}$, which implies that each posterior model sample $\mathbf{m}$ is a direct output of the GNN. Therefore, latent space inversion is limited to finding solutions within the range of models that the GNN can generate for a given latent distribution. 

Latent space inversion thus requires that the best model space estimate of the true Earth lies within the family of models that GNNs can generate. Since the true Earth inevitably lies outside of the range of the GNN-generated models (even if the generator embodies some of its principal characteristics), then the entire distribution of those models must fail to include the true solution. \cite{bloem_introducing_2023} demonstrated this by showing that using incorrect geological prior information parametrised by GNNs leads to considerably less accurate results: after training a GNN with images of marine stratigraphic para-sequences, they performed the inversion using data from a true Earth model featuring a river channel structure. Since the true Earth model lay outside of the range of models that the GNN could generate (marine para-sequences), inversion results were inaccurate and did not reflect the true solution.

Most experiments \citep{ruderman_statistics_1994, scholkopf_nonlinear_1998, roweis_nonlinear_2000, tenenbaum_global_2000, fefferman_testing_2016} support the idea that a lower-dimensional manifold exists within a high-dimensional parameter space across various natural images. High-dimensional images of natural geological structures (e.g., river channels, sediment deposits, and marine para-sequences) may lie on a lower-dimensional subspace or manifold within the higher-dimensional parameter space in which these high-dimensional images exist. However, the complex nature of geological processes leads to a degree of apparent randomness in true Earth structures \citep{burgess_sensitive_2004}. It is therefore much more likely that such manifolds describe only general patterns, those recognisable by the human eye, which may be interpretable and classifiable by geologists; true Earth structures in all their glory occupy a far higher-dimensional space around, perhaps in some sense close to, those manifolds. In this paper, we refer to the manifold that represents genuine, natural geological variability as the \textit{geological manifold}. After training, GNN-generated examples that contain geologically plausible structures in a high-dimensional model space also lie on a lower-dimensional manifold, which we refer to as the \textit{generative manifold}.

During training, a GNN learns only a finite-dimensional and approximate subset of Earth's true complex heterogeneity, due to training errors, limited latent parameter dimensionality, and imperfect or unrepresentative geological information embodied within the set of training examples. From here on, we denote the true geological manifold by $\mathcal{M}$, and the GNN-learned generative manifold by $\mathcal{M}_{gen}$.

To produce examples that capture a range of geological features such as river channels, facies, stratigraphic para-sequences, and folded or faulted structures, in this work we simulate geological processes including river flow, sedimentary dispersion, deposition, erosion, and sea level variations using geological process modelling software, \textit{SedSimple} \citep{tetzlaff_stratigraphic_2023}. This software is designed for rapid simulation, and while it contains a reasonably broad set of sedimentary processes that create models with recognisable and interpretable geological characteristics, it is not expected to generate structures that would occur exactly in the real Earth. We would nevertheless like to use the output examples as prior information, to guide inverse problem solutions but without constraining the solution to lie amongst them. We train three GNNs: a Variational Autoencoder (VAE, \cite{kingma_autoencoding_2014}), an Adversarial Autoencoder (AAE, \cite{makhzani_adversarial_2016}), and a Denoising Adversarial Autoencoder (DAAE, \cite{creswell_denoising_2018}), using many geological image examples from SedSimple. Once trained, these GNNs produce examples that contain geological features; they lie on the lower-dimensional generative manifold $\mathcal{M}_{gen}$, which approximates the geological manifold $\mathcal{M}$. Latent space inversion is therefore always restricted to find solutions within the range of this unrealistic generative manifold $\mathcal{M}_{gen}$.

This work investigates a technique that enables solutions to be found outside of $\mathcal{M}_{gen}$, while still using the generative manifold to guide the inversion process. To maintain $\mathcal{M}_{gen}$ as a source of information yet broaden the solution space beyond it, we introduce a so-called \textit{escaping prior} pdf that provides the necessary freedom to account for both on-manifold geological features and off-manifold variations in Bayesian model space inversion. The appropriate extent of freedom to offer is usually unknown \textit{a priori}, since we do not know how far from manifold $\mathcal{M}_{gen}$ in model parameter space a true model (the best representation of nature) might be. Selecting a fixed level of freedom would therefore be likely to lead to biased posterior uncertainty estimates \citep{malinverno_expanded_2004}. In this work, we assign uncertainty both to model parameters and to a parameter that defines the escaping prior's variance, to identify the posterior combination that best explains the observed data given the forward function that predicts data from model parameters. Since this necessarily involves the forward function, these parameter estimates (including the freedom to escape the prior manifold) are all made a posteriori \citep{mosegaard_inconsistency_2024}. This extended uncertainty is quantified using a hierarchical Bayesian formulation \citep{malinverno_expanded_2004, malinverno_two_2006,bodin_transdimensional_2012, galetti_transdimensional_2017}. The escaping prior, combined with a likelihood function, is used to formulate a posterior pdf, from which samples are drawn using Metropolis-Hastings Markov chain Monte Carlo (MH-MCMC, \cite{metropolis_equation_1953, hastings_monte_1970}). We test our approach using simple 2D problems, then non-linear travel-time tomography experiments, and compare the results with those of standard latent space inversion, and inversions with uniform prior information. These posterior distributions illustrate the value of allowing solutions to escape from imperfect prior manifolds.

This study serves three main purposes: first, to identify the likely limitations of latent space inversion in real-world geoscientific problems; second, to show that escaping priors provide a better alternative by integrating GNNs into a model-space inversion framework that goes beyond the generative manifold; and third, to show how to estimate appropriate uncertainty in the escaping prior \textit{a posteriori} through a hierarchical Bayesian algorithm.

Section \ref{subsection:Generative Neural Networks} briefly describes the three types of GNN used in our study, and in section \ref{subsec:Latent Space Inversion} we discuss latent space inversion. In section \ref{subsec:Generative Manifolds as Bayesian Priors} we explain the learning process to construct generative manifolds within a high-dimensional model spaces using GNNs. We then present the full formulation of our escaping prior and apply it to derive the posterior pdf in section \ref{subsec:Manifold-Guided MCMC Sampling in Model Space}. The hierarchical Bayesian algorithm is detailed in section \ref{subsec:Hierarchical Bayesian Inversion}. In section \ref{sec:Illustrative Low-Dimensional Problems} we demonstrate the methodology through simple problems, in section \ref{sec:Travel-Time Tomography} we present results when it is applied to non-linear travel-time tomography experiments using a standard machine-learning test data set, and in section \ref{subsec:Geological Velocity Models} we apply the method to genuinely geological models. Finally, we discuss the implications of these results in section \ref{sec:Discussion} before concluding. 
In Appendix \ref{app:Training of Generative Neural Networks}, we provide more detail about the training process for VAE, AAE and DAAE networks. Appendix \ref{app:Escaping prior samples} offers additional information about escaping prior samples and how they relate to the GNN latent spaces. Appendix \ref{app:Hyperprior distributions} shows the implementation of different hyperprior distributions, and Appendix \ref{app:prior_correlations} provides supplementary information about the prior correlation matrices used in our study.

\section{Methodology}

In a Bayesian framework, the solution to an inverse problem is described by the \textit{posterior} pdf, \(p(\mathbf{m}| \mathbf{d}_{obs})\), the probability that model $\mathbf{m}$ is true given that we have observed data $\mathbf{d}_{obs}$. This is obtained by combining a \textit{prior} distribution \(p(\mathbf{m})\), which embodies information known about $\mathbf{m}$ before the newly observed data were considered, with a data \textit{likelihood} \(p(\mathbf{d}_{\rm obs}| \mathbf{m})\). Thus, \textit{a priori} beliefs about model parameters are revised based on the additional observed data. This is formally achieved by Bayes' theorem \citep{bayes_essay_1763}:
\begin{equation}
p(\mathbf{m}| \mathbf{d}_{obs}) = \frac{p(\mathbf{d}_{obs}| \mathbf{m})\,p(\mathbf{m})}{p(\mathbf{d}_{obs})}
\label{eq:bayesian}
\end{equation}
The likelihood term \(p(\mathbf{d}_{obs}| \mathbf{m})\) represents the probability of observing the data if a given set of model parameter values $\mathbf{m}$ were true, evaluated by computing the consistency of the observed data with the data simulated from the given model $\mathbf{m}$. The term $p(\mathbf{d}_{obs})$, known as the \textit{evidence}, acts as a normalisation constant for the posterior distribution. It is equal to the integral of the numerator over the entire parameter space $\mathbb{R}^k$ of values of $\mathbf{m}$, which has dimensionality $k$, written explicitly as:
\begin{equation}p(\mathbf{d}_{obs}) = \int_{\mathbf{m} \in \mathbb{R}^k}  p(\mathbf{d}_{obs}| \mathbf{m})p(\mathbf{m})d\mathbf{m}
\label{eq:evidence}
\end{equation}
In practice, calculating the evidence is often computationally intractable due to the high dimensionality of parameter space $\mathbb{R}^k$. In such cases, we can sometimes use Monte Carlo sampling methods to draw samples from the posterior distribution, which avoids computing the evidence integral (equation \ref{eq:evidence}). This approach allows us to characterise the posterior pdf by the sample density, providing a description of solution uncertainty and parameter trade-offs as long as a sufficiently large number of samples can be taken to reach statistical equilibrium.

To ensure solutions are geologically plausible, prior geological knowledge about model parameters is often incorporated in the prior pdf $p(\mathbf{m})$. However, effectively parameterising this geological information in a high-dimensional model space is difficult. To address this, Generative Neural Networks (GNNs) can be trained to parameterise realistic, geologically plausible structures, serving as a source of prior information.

\subsection{Generative neural networks}
\label{subsection:Generative Neural Networks}

Machine learning architectures, such as Generative Neural Networks (GNNs) can be used to parameterise our prior information over model parameters inferred from a set of training images, where we use the word ``images" to represent instantaneous snapshots of geological characteristics in 1, 2 or 3 dimensional space, with the additional dimension of time if appropriate. The probability distribution of real geological training images is visibly complex and almost certainly infinite-dimensional in nature, but any dataset of images used to train GNNs can be assumed to exist within a high-dimensional space $\mathbb{R}^k$ of dimensionality $k$. GNNs are trained to learn this complex, high-dimensional probability distribution by constructing a mapping through a generator network $\mathbf{\mathcal{G}: \mathbb{R}^r \rightarrow \mathbb{R}^k}$, which maps a low-dimensional ($r$-dimensional) latent sample $\mathbf{z}$, generally drawn from a standard normal distribution $\mathcal{N}(0, \mathbf{I}_r)$, to a high-dimensional training model $\mathbf{m}$ that is a new sample from the learned distribution. This mapping  $\mathbf{\mathcal{G}: z \rightarrow m}$ can be trained, for example through an adversarial training method (as in GANs, \cite{goodfellow_generative_2014}) or by minimising a reconstruction loss with an Encoder-Generator network \citep{kingma_autoencoding_2014, makhzani_adversarial_2016, creswell_denoising_2018}. In this work, we parameterise our prior information using three different GNNs, each based on an Encoder-Generator architecture: a Variational Autoencoder (VAE), an Adversarial Autoencoder (AAE), and a Denoising Adversarial Autoencoder (DAAE).

\begin{figure}
    \centering
    \includegraphics[width=1\columnwidth]{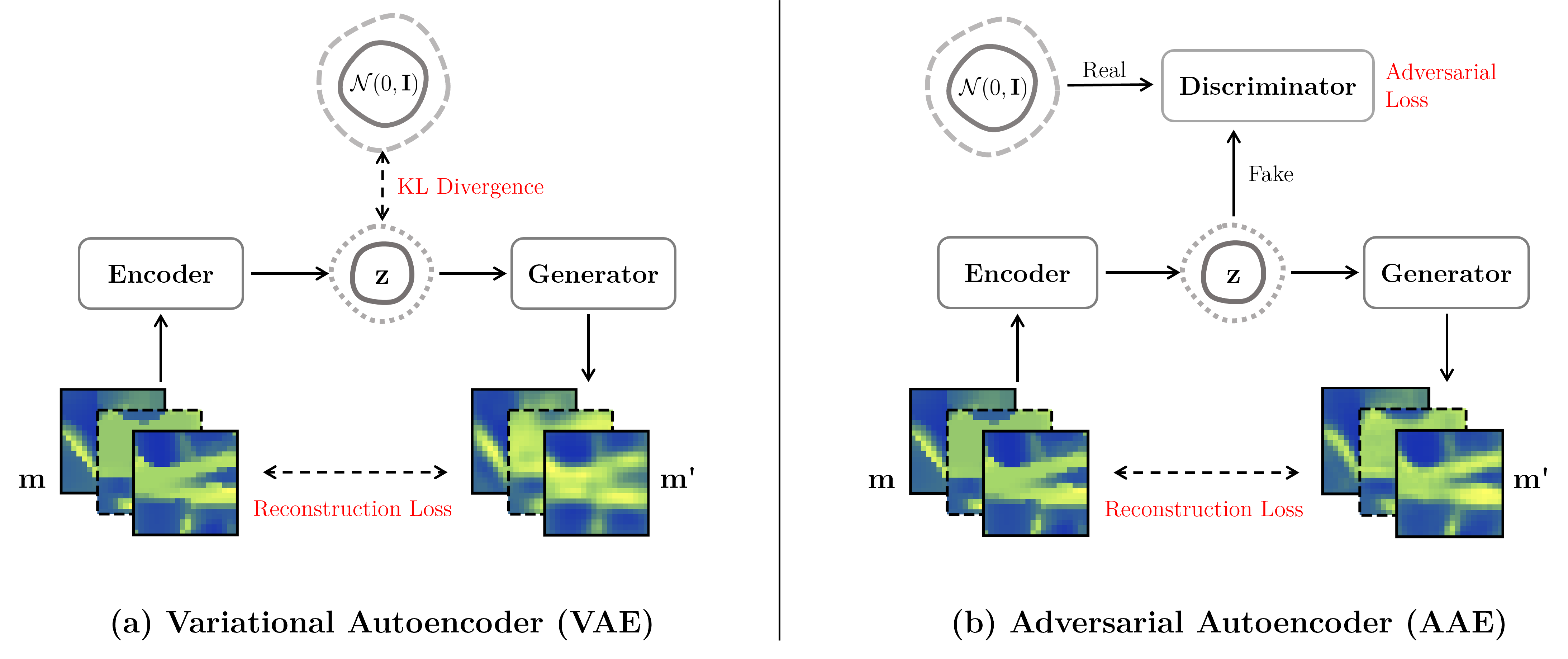}

    \vspace{1em}
    
    \includegraphics[width=0.95\columnwidth]{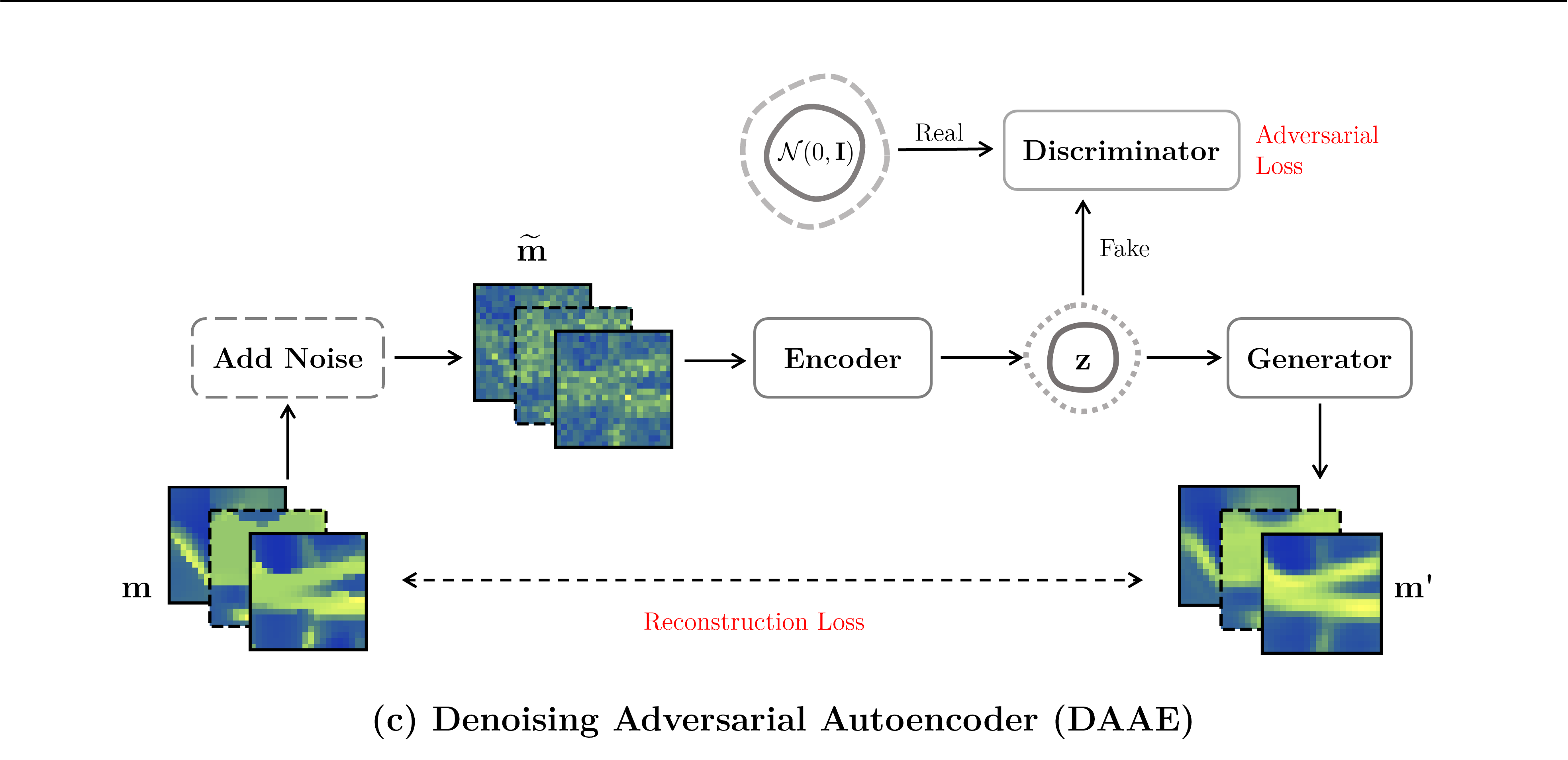}
    \caption{Comparison of neural network architectures, with components of the objective function minimised or maximised during training annotated in red. (a) Standard VAE, (b) AAE, (c) DAAE incorporating additional noise.}
    \label{fig:GNN_architectures}
\end{figure}

The Variational Autoencoder \citep{kingma_autoencoding_2014} consists of two major neural networks: a probabilistic encoder $\mathbf{E}$ and a generator $\mathcal{G}$ (Figure \ref{fig:GNN_architectures}a), which are trained to learn a high-dimensional probability distribution over training images $p_{model}(\mathbf{m})$ and to parameterise our prior information over model parameters. In our case, we will use 2D images of geology in which each pixel value of the training image represents seismic velocity of the geological medium, and these pixelated values comprise our vector of model parameters $\mathbf{m}$. In this learning process, first, the high-dimensional models $\mathbf{m}$ (with dimensionality $k$) from the training samples are mapped to a low-dimensional ($r$-dimensional) space, called the latent space, represented by the latent parameters $\mathbf{z}$. This mapping $\mathbf{E: \mathbb{R}^k \rightarrow \mathbb{R}^r}$ is learned by the encoder network $\mathbf{E}$ and is composed of a series of nested simple mappings, each defined by a sequence of intermediate neural network layers, which progressively capture more abstract features from the training images \citep{zeiler_visualizing_2013}. Thus, the high-dimensional models $\mathbf{m}$ from the training images are mapped to the arbitrarily lower-dimensional latent samples $\mathbf{z} \in \mathbb{R}^r$, which contain the most abstract features of the training images. The latent space is defined to be the space of all values of the latent parameter $\mathbf{z}$, which is regularised to approximate a standard normal distribution $\mathcal{N}(0, \mathbf{I}_r)$ by minimising the Kullback-Leibler divergence (KL, \cite{kullback_information_1951}) between the encoder output distribution and $\mathcal{N}(0, \mathbf{I}_r)$. After encoding the training images into the latent space through the encoder $\mathbf{E}$, a VAE reconstructs the training images from the latent samples by a mapping $\mathbf{\mathcal{G}: \mathbb{R}^r \rightarrow \mathbb{R}^k}$, where the $r$-dimensional latent parameters are mapped back to the higher-dimensional model parameters through the generator network $\mathcal{G}$.

An Adversarial Autoencoder \citep{makhzani_adversarial_2016} uses the same encoder-generator architecture as the VAE, but employs a different regularisation strategy. Instead of an explicit KL-divergence, it uses an additional discriminator network to regularise its latent space (Figure \ref{fig:GNN_architectures}b).

A Denoising Adversarial Autoencoder \citep{creswell_denoising_2018} extends the AAE by integrating a denoising criterion. The network is trained to reconstruct the original, clean model $\mathbf{m}$ from a \textit{corrupted} or noisy input version $\widetilde{\mathbf{m}}$ (Figure \ref{fig:GNN_architectures}c). This corruption process (e.g., adding uncorrelated Gaussian noise) forces the DAAE to learn more robust (and hopefully meaningful) features of the training set - rather than details, the distribution of which may not be captured adequately by a finite training set.

In Appendix \ref{app:Training of Generative Neural Networks}, we describe the training process for each of these GNNs. Each training session was conducted on an NVIDIA L4 GPU. After suitable training, GNNs can produce examples from a similar distribution to that of the training samples by passing latent samples $\mathbf{z}$, through the generator:
\begin{subequations} 
\label{eq:generator_block}
\begin{align}
    \mathbf{z} \sim \mathcal{N}(0, \mathbf{I}_r),\quad \mathbf{z} \in \mathbb{R}^r \label{eq:latent_z} \\
    \mathcal{G}: \mathbf{z} \to \mathbf{m}, \quad \mathbf{m} \in \mathbb{R}^k \label{eq:generator}
\end{align}
\end{subequations}

\subsection{Latent space inversion}
\label{subsec:Latent Space Inversion}
The latent space consists of all possible values of the latent parameters, which are generally predefined to follow a standard normal distribution $\mathcal{N}(0, \mathbf{I}_r)$. Latent space inversion involves identifying values of latent parameters $\mathbf{z} \sim \mathcal{N}(0, \mathbf{I}_r)$ which, when passed through the generator $\mathcal{G}$, produce a model that is consistent with the observed data. By seeking values of latent parameters $\mathbf{z}$ rather than model parameters $\mathbf{m}$, this process reduces the total number of parameters that need to be solved for, thereby significantly reducing the dimensionality of the inverse problem.

Within a Bayesian framework, the posterior pdf for inversion in the latent space can be expressed as
\begin{equation}
    p(\mathbf{z} | \mathbf{d}_{obs}) \propto p(\mathbf{d}_{obs}|\mathbf{z}) \: p(\mathbf{z})
    \label{eq:latent_space_posterior_pdf_short}
\end{equation}
where the latent space posterior pdf $p(\mathbf{z}|\mathbf{d}_{obs})$ represents the posterior probability density of a latent variable $\mathbf{z}$, given the observed data $\mathbf{d}_{obs}$. Here, the prior distribution of the latent parameters is specified by the pdf $p(\mathbf{z})$, which is assumed to be a standard normal distribution $\mathcal{N}(0, \mathbf{I})$ if training has been performed adequately:
\begin{equation}
    p(\mathbf{z}) \propto \exp\left( - \frac{\|\mathbf{z}\|^2_2}{2} \right)
    \label{eq:latent_prior_pdf}
\end{equation}
The likelihood function $p(\mathbf{d}_{obs}|\mathbf{z})$ measures how well a given latent parameter $\mathbf{z}$ explains the observed data $\mathbf{d}_{obs}$. The evaluation of this term involves the following three steps: first, a latent sample $\mathbf{z} \sim \mathcal{N}(0, \mathbf{I}_r)$ is passed through the GNN's generator $\mathcal{G}$, to produce a high-dimensional model: $\mathbf{m} = \mathcal{G}(\mathbf{z})$. This model $\mathbf{m}$ is then used in the forward function ($f$) to create synthetic data $f(\mathbf{m}) = f(\mathcal{G}(\mathbf{z}))$. The consistency between the synthetic data ($\mathbf{d}_{\text{cal}}$) and observed data ($\mathbf{d}_{obs}$) is calculated by the likelihood
\begin{equation}
    p(\mathbf{d}_{obs}|\mathbf{z}) \propto \exp\left( - \frac{\|\mathbf{d}_{obs} - f(\mathcal{G}(\mathbf{z}))\|^2_2}{2\sigma_d^2} \right)
    \label{eq:latent_likelihood_pdf}
\end{equation}
where $\sigma_d^2$ represents data uncertainties, assumed here to follow a Gaussian distribution, and $\sigma^2_d$ quantifies uncertainty in observed data $\mathbf{d}_{obs}$. In this study (which involves only synthetic data) we assume an uncorrelated data covariance matrix in which each datum has the same uncertainty, but equation \ref{eq:latent_likelihood_pdf} can easily be changed to account for a more general Gaussian data covariance matrix.

We use a MH-MCMC to draw a set of $N$ samples in the low-dimensional latent space, $\widetilde{\mathbf{z}} = \{ \mathbf{z}^{(i)} \}_{i=1}^{N}$, from the posterior distribution $p(\mathbf{z} | \mathbf{d}_{obs})$ in equation \ref{eq:latent_space_posterior_pdf_short}. Samples of the final inversion result (in terms of high-dimensional model parameters) are then obtained by passing all latent samples in this set through the, geological generator network: $\mathbf{m}^{(i)} = \mathcal{G}(\mathbf{z}^{(i)})$, where $\mathbf{z}^{(i)} \in \widetilde{\mathbf{z}}$, or by storing these $\mathbf{m}^{(i)}$ when they were evaluated during the Monte Carlo process.  We can characterise or visualise our results by calculating local sample densities or other statistical estimates from these high-dimensional model samples. For example, the posterior mean model ($\overline{\mathbf{m}}$) is estimated by averaging the decoded samples,
\begin{equation}
    \overline{\mathbf{m}} = \frac{1}{N} \sum_{i=1}^{N} \mathcal{G}(\mathbf{z}^{(i)}) = \frac{1}{N} \sum_{i=1}^{N} \mathbf{m}^{(i)}
\end{equation}
and the standard deviation of the samples
$$
\boldsymbol{\sigma}_{\mathbf{m}}
= \sqrt{
\frac{1}{N}
\sum_{i=1}^{N}
\left(\mathcal{G}(\mathbf{z}^{(i)})-\overline{\mathbf{m}}\right)^2
}
$$
can be used to estimate the posterior uncertainty of our inversion results in model parameter space. In this way, the resulting distribution over high-dimensional model parameters is forced to satisfy the prior information represented by the generator. However, this reveals the fundamental limitation of latent space inversion: every final model sample $\mathbf{m}^{(i)}$ is a direct output of the generator $\mathcal{G}: \mathbf{z}^{(i)} \rightarrow \mathbf{m}^{(i)}$, so the entire posterior distribution of generated models is strictly confined to a lower-dimensional manifold, the generative manifold $\mathcal{M}_{gen}$, embedded within a higher-dimensional space $\mathbb{R}^k$. If the true Earth model is outside of $\mathcal{M}_{gen}$, the MCMC sampler will produce a family of Earth model samples that can never approach the true solution.

\subsection{Geological generative manifolds as Bayesian priors}
\label{subsec:Generative Manifolds as Bayesian Priors}

\begin{figure}
    \centering
    \includegraphics[width=1\linewidth]{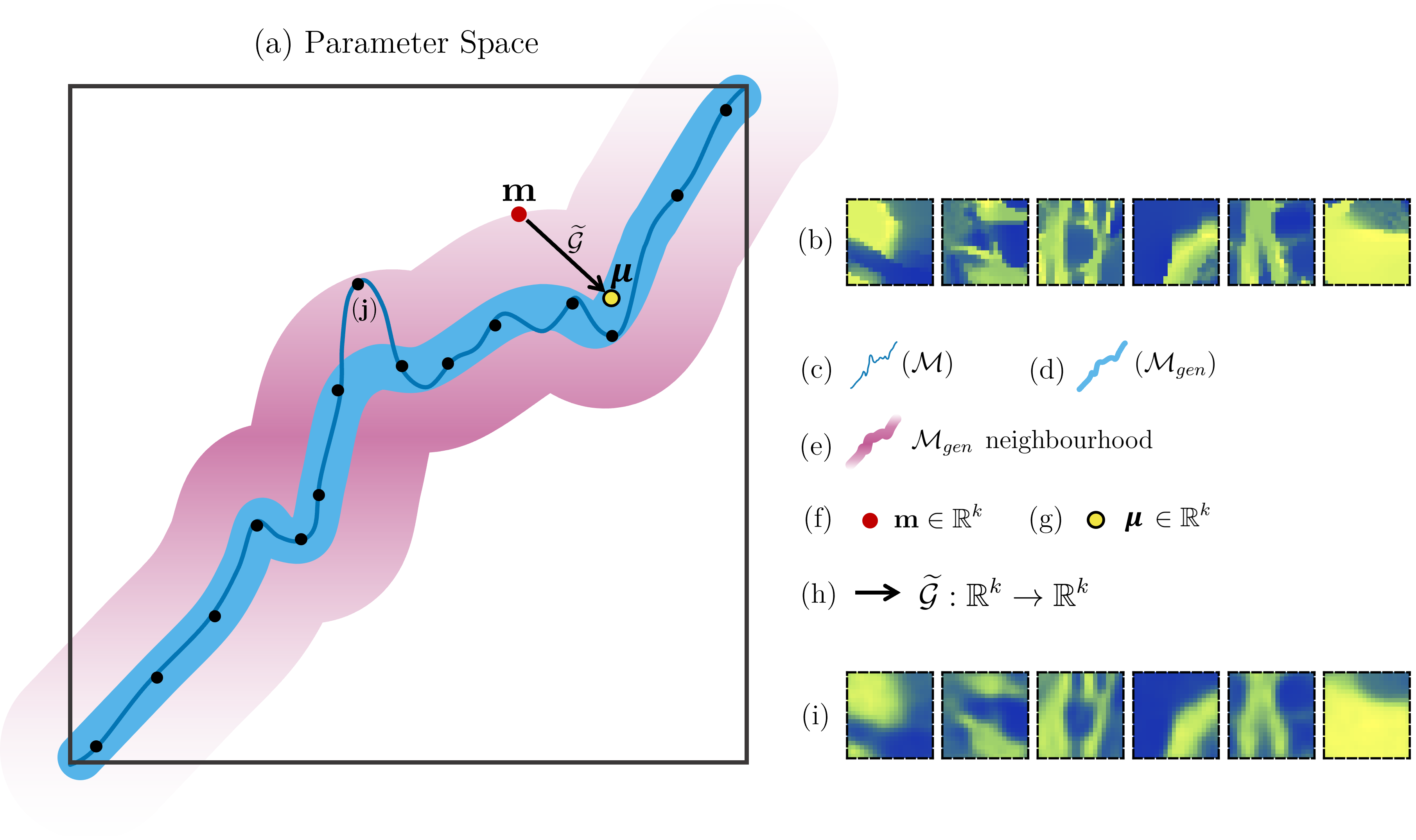}
    \caption{Schematic representation of manifolds in a (a) high-dimensional parameter space, $\mathbb{R}^k$. Black dots show (b) high-dimensional training images concentrated on a lower-dimensional manifold (c) $\mathcal{M}$ (thin dark blue curve). (d) The learned generative manifold $\mathcal{M}_{gen}$ (thick blue curve) is a close approximation of $\mathcal{M}$. (f) Any model $\mathbf{m}$ (red dot) is mapped onto $\mathcal{M}_{gen}$ by (h) GNN ($\widetilde{\mathcal{G}}: \mathbb{R}^k \to \mathbb{R}^k$) (black arrow) to (g) a GNN-generated model $\boldsymbol{\mu}$ (yellow dot). (e) The neighbourhood (pink hue) around $\mathcal{M}_{gen}$ represents the probability region of the prior pdf (equation~\ref{eq:prior_integral_iso}), and its width is controlled by the variance $\tau^2$ (equation \ref{eq:prior_integral_iso}). (i) Examples of GNN-generated images. (j) Part of the $\mathcal{M}$ not accurately captured by $\mathcal{M}_{gen}$.}
    \label{fig:manifolds_in_HD}
\end{figure}

The manifold hypothesis assumes that high-dimensional model parameters often lie on a much lower-dimensional subspace (i.e., the manifold) \citep{ruderman_statistics_1994, scholkopf_nonlinear_1998, roweis_nonlinear_2000, tenenbaum_global_2000, fefferman_testing_2016}. High-dimensional training sets, for example, composed of natural images, are often represented as points within a high-dimensional space, $\mathbb{R}^k$ (Figure \ref{fig:manifolds_in_HD}). These points (training images in $\mathbb{R}^k$) are not distributed randomly throughout this space; rather, they are concentrated on or around some lower-dimensional manifold $\mathcal{M}$ (dark blue curve in Figure \ref{fig:manifolds_in_HD}c), which is embedded within $\mathbb{R}^k$ (the full space in Figure \ref{fig:manifolds_in_HD}a). 

In practice, training images include noise and other natural variability, so they need not lie exactly on manifold $\mathcal{M}$ \citep{vincent_extracting_2008}. In other words, the distribution of natural images is expected to have non-zero variance \textit{around} manifold $\mathcal{M}$. For this reason, probabilistic GNNs such as Variational Autoencoders (VAEs, \cite{kingma_autoencoding_2014}), Adversarial Autoencoders (AAEs, \cite{makhzani_adversarial_2016}), Denoising Adversarial Autoencoders (DAAEs, \cite{creswell_denoising_2018}) and diffusion models \citep{sohl-dickstein_deep_2015, ho_denoising_2020} do not learn to produce samples confined only to this manifold $\mathcal{M}$ (i.e., samples that lie exactly on the thin blue curve in Figure \ref{fig:manifolds_in_HD}c). Instead, they define a non-zero probability density over the full parameter space $\mathbb{R}^k$ which in principle may then contain models that describe the true solution (conditioned on the real world being adequately representable within our parameter spaces). The objective of training a GNN is to learn this probability density so that most of the GNN-generated samples concentrate around $\mathcal{M}$ with some variance ($\mathcal{M}_{gen}$, thick blue curve in Figure \ref{fig:manifolds_in_HD}d).

The model that best represents the true Earth $\mathbf{m}_{\text{true}}$ (henceforth referred to as the ``true Earth model") is not expected to lie on $\mathcal{M}_{gen}$ (Figure \ref{fig:manifolds_in_HD}d). One reason could be that the generative manifold $\mathcal{M}_{gen}$, learned by the GNN, is an approximation of the true geological manifold $\mathcal{M}$ which includes all geologically possible Earth models including the true Earth model. Nevertheless, the true Earth exhibits complexity and heterogeneity at every scale, with large-scale geological structures and infinitely complex small-scale variability. The GNN-learned generative manifold $\mathcal{M}_{gen}$ could, in principle, represent structures over a certain range of scales (for example, the macro-scale patterns of folded sediments) if it were trained on a set of geologically realistic images, but will never accurately capture geological features at all scale lengths (Figure \ref{fig:manifolds_in_HD}j) due to training errors, imperfect training images, and the limited number of latent parameters.

Another reason for inaccuracy in $\mathcal{M}_{gen}$ could be the use of an inappropriate class of geological prior information. For instance, if the GNN is trained solely on geological images representing the structures of marine para-sequences, then any Earth model that is not structurally related to marine para-sequences, such as a river channel, will inevitably fall outside of $\mathcal{M}_{gen}$ \citep{bloem_introducing_2023}. In that case, attempting to recover a river channel structure will result in incorrect solutions during latent space inversion. 

Therefore, to recover the true Earth model, it is necessary to expand the search space beyond the strict boundaries of $\mathcal{M}_{gen}$. In medical imaging, \cite{obmann_sparse_2020} proposed a manifold penalty (regularisation) term,
\begin{equation} \label{eq:Obmann_penalty}
\|\mathbf{m} - \widetilde{\mathcal{G}}(\mathbf{m})\|^2
\end{equation}
to allow this freedom in linear inverse problems in X-ray tomography within an optimisation framework, where the mapping $\widetilde{\mathcal{G}}: \mathbb{R}^k \to \mathbb{R}^k$ is defined by an autoencoder network. Equation \ref{eq:Obmann_penalty} quantifies the distance from a randomly chosen model parameter $\mathbf{m}$ in the full parameter space to $\mathcal{M}_{gen}$ and encourages solutions to be close to the generative manifold but not necessarily on it. Unlike \cite{obmann_sparse_2020}, which employs the manifold penalty term within an optimisation framework to find a single, optimal solution, we reformulate a related manifold penalty in a fully Bayesian approach and use a hierarchical algorithm to estimate the magnitude of this penalty (details in Section \ref{subsec:Hierarchical Bayesian Inversion}).

In geophysical inversion methods, geological realism is often neglected as a criterion for solutions, due to challenges in quantifying the required geological information and the computational complexity involved. Consequently, the resulting solutions are generally geologically implausible.  \cite{bloem_introducing_2023} addressed the problem by implementing geological process models \citep{tetzlaff_stratigraphic_2023} and parameterising the geological prior information using GNNs for latent space inversion. However, process models always simplify natural geological processes so the true model always lies outside of the learned $\mathcal{M}_{gen}$; latent space inversion results, even if they appear geologically realistic, are still considered to be incorrect.

In the Bayesian approach, we can incorporate \textit{a priori} knowledge about model parameters through a prior pdf. In this work, we introduce an escaping prior pdf that broadens the solution space beyond $\mathcal{M}_{gen}$ while still maintaining geological realism by using $\mathcal{M}_{gen}$ as a guide to geologically realistic structures and properties. We express this by defining a Gaussian prior probability density function $p_{\mathcal{M}_{gen}}(\bf{m})$ over the entire model parameter space $\mathbb{R}^k$:
\begin{equation} \label{eq:prior_integral}
p_{\mathcal{M}_{gen}}(\mathbf{m}) \propto \int_{\boldsymbol{\mu} \in \mathcal{M}_{gen}} \mathcal{N}(\mathbf{m}; \boldsymbol{\mu}, \mathbf{\Sigma}) d\boldsymbol{\mu}
\end{equation}
\noindent
Here, $\boldsymbol{\mu}$ represents a model on the generative manifold $\mathcal{M}_{gen}$, and $\mathcal{N}(\mathbf{m}; \boldsymbol{\mu}, \mathbf{\Sigma})$ in equation~\eqref{eq:prior_integral} represents a multivariate Gaussian distribution centred on mean $\boldsymbol{\mu}$ with covariance matrix $\mathbf{\Sigma}$. If different model parameters have different numerical scales (e.g., porosity lies in the range 0 to 1, whereas distance is greater than 0), then an anisotropic Gaussian distribution represented by a general covariance matrix $\mathbf{\Sigma}$ may be appropriate. However, for the examples herein, we assume an isotropic, uncorrelated covariance $\mathbf{\Sigma} = \tau^2\mathbf{I}$, where $\mathbf{I}$ is the identity matrix and $\tau^2$ is a scalar variance. This simplifies the pdf to be proportional to the exponential term:
\begin{equation} 
\label{eq:gaussian_prior}
\mathcal{N}(\mathbf{m}; \boldsymbol{\mu}, \tau^2\mathbf{I}) \propto \exp\left(-\frac{\|\mathbf{m} - \boldsymbol{\mu}\|_2^2}{2\tau^2}\right)
\end{equation}
where $\tau^2$ defines the prior uncertainty for the off-manifold solutions, controlling the spread of the escaping prior pdf $p_{\mathcal{M}_{gen}}(\mathbf{m})$ around the manifold $\mathcal{M}_{gen}$. If we had almost no information about the solution, $\tau$ would assume a very large value resulting in an extremely broad and almost flat prior; on the other hand, a small value of $\tau$ leads to a prior distribution strongly peaked around $\mathcal{M}_{gen}$. Substituting the Gaussian distribution in equation \ref{eq:gaussian_prior} into equation \ref{eq:prior_integral}, the escaping prior probability density function becomes:
\begin{equation} \label{eq:prior_integral_iso}
p_{\mathcal{M}_{gen}}(\mathbf{m}) \propto \int_{\boldsymbol{\mu}\in\mathcal{M}_{gen}} \exp\left(-\frac{\|\mathbf{m}-\boldsymbol{\mu}\|_2^2}{2\tau^2}\right) \, d\boldsymbol{\mu}
\end{equation}

For any point $\mathbf{m} \in \mathbb{R}^k$, within a cylindrical neighbourhood of $\mathcal{M}_{gen}$, there exists a model $\boldsymbol{\mu}\in\mathcal{M}_{gen}$ obtained by mapping $\mathbf{m}$ onto $\mathcal{M}_{gen}$ (Figure \ref{fig:manifolds_in_HD}h) through a GNN ($\widetilde{\mathcal{G}}: \mathbb{R}^k \to \mathbb{R}^k$). For example, when using an autoencoder (encoder-generator) network,
\begin{equation} \label{eq:E_D_mapping}
\widetilde{\mathcal{G}}(\mathbf{m})=\mathcal{G}(\mathbf{E}(\mathbf{m})) = {\boldsymbol{\mu}} \:
\end{equation}
In equation \ref{eq:E_D_mapping}, ${\boldsymbol{\mu}}$ is the mapping of $\mathbf{m}$ onto $\mathcal{M}$ through the autoencoder, and this $\boldsymbol{\mu}$ is used in equations \ref{eq:prior_integral} to \ref{eq:prior_integral_iso}.

\subsection{Manifold-guided MCMC sampling in model space}
\label{subsec:Manifold-Guided MCMC Sampling in Model Space}

We incorporate prior information from the generative manifold $\mathcal{M}_{gen}$ within a MCMC sampling framework to sample the posterior distribution (equation \ref{eq:bayesian}) in model space using the Metropolis-Hastings algorithm \citep{metropolis_equation_1953, hastings_monte_1970}.

The likelihood of observing data $\mathbf{d}_{obs}$ given model $\mathbf{m}$ is an uncorrelated Gaussian distribution:
\begin{equation}
    p(\mathbf{d}_{obs}|\mathbf{m}) \propto \exp \left(-\frac{\|\mathbf{d}_{obs} - f(\mathbf{m})\|_2^2}{2\sigma_d^2}\right) \:
    \label{eq:likelihood}
\end{equation}
The forward problem $f(\mathbf{m})$ predicts the synthetic data that we would observe in the absence of measurement noise, if model $\mathbf{m}$ represented the true Earth. The measurement error, and hence the noise in the observed data, is assumed to be Gaussian with variance $\sigma_d^2$.

The prior pdf $p(\mathbf{m})$ must contain our prior beliefs. First, we define a uniform prior $p_0(\mathbf{m})$ to constrain parameters within valid physical bounds $[\mathbf{m}_{min}, \mathbf{m}_{max}]$:
\begin{equation}
p_0(\mathbf{m}) =
\begin{cases}
1/(\mathbf{m}_{max} - \mathbf{m}_{min})^k, & \text{if } \mathbf{m}_{min} \le \mathbf{m} \le \mathbf{m}_{max} \\[6pt]
0, & \text{otherwise} \:
\end{cases}
\label{eq:uniform_distribution}
\end{equation}
Next, we further constrain the solution space with the escaping prior $p_{\mathcal{M}_{gen}}(\mathbf{m})$, thereby defining the overall prior $p(\mathbf{m})$ as a product of $p_{\mathcal{M}_{gen}}(\mathbf{m})$ and the uniform prior $p_0(\mathbf{m})$:
\begin{equation}
    p(\mathbf{m}) \propto p_0(\mathbf{m}) \times p_{\mathcal{M}_{gen}}(\mathbf{m}) \:.
    \label{eq:total_prior_p(Mgen)Xuniform}
\end{equation}
Thus, with a simplified version of $p_{\mathcal{M}_{gen}}(\mathbf{m})$ from equation \ref{eq:prior_integral_iso}, our overall prior pdf will be:
\begin{equation}
    p(\mathbf{m}) \propto p_0(\mathbf{m}) \times \exp\left(-\alpha \|\mathbf{m}-\boldsymbol{\mu}\|_2^2\right) \, ,\quad \text{where } \boldsymbol{\mu} = \mathcal{G}(\mathbf{E}(\mathbf{m})) \:
    \label{eq:modelspace_prior_pdf}
\end{equation}
In equation \ref{eq:modelspace_prior_pdf}, $\alpha = 1/(2\tau^2)$ represents the precision or strength of our prior belief in the manifold $\mathcal{M}_{gen}$, and $\tau^2$ represents the variance around this manifold. A smaller $\alpha$ indicates higher uncertainty (a more weakly informative prior) surrounding $\mathcal{M}_{gen}$, and vice versa. Note that we simplify the escaping prior pdf $p_{\mathcal{M}_{gen}}(\mathbf{m})$ (Equation \ref{eq:prior_integral_iso}) into the form shown in equation \ref{eq:modelspace_prior_pdf} for cases where the model probability is computed one sample at a time. Therefore, by combining data-likelihood (equation \ref{eq:likelihood}) with the prior pdf (equation \ref{eq:modelspace_prior_pdf}) defined inside the bounds $[\mathbf{m}_{min} \le \mathbf{m} \le \mathbf{m}_{max}]$, we can draw samples from the posterior pdf (equation \ref{eq:bayesian}) in high-dimensional model space using MCMC methods.

Selecting the hyperparameter $\alpha$ for a particular inverse problem also requires prior knowledge. Since the appropriate value to use can not be known in advance (we never know how far the true model is from $\mathcal{M}_{gen}$), an inversion performed with a specific $\alpha$ value is likely to introduce bias into the posterior. One could consider choosing a small value of $\alpha$ to make the prior pdf extremely broad and flat; however, doing so reduces the information from the escaping prior to near zero because in equation \ref{eq:modelspace_prior_pdf}: $\alpha \to 0, \: p(\mathbf{m}) \to p_0(\mathbf{m})$. Our goal in this work is not to reduce the generative manifold information to zero, but to find $\alpha$ that represents a balance between our prior state of information about the true Earth model from $\mathcal{M}_{gen}$, and new information from the observed data and the forward model. Therefore, posterior uncertainties are determined not only for the model parameters but also by the hyperparameters.

\subsection{Hierarchical Bayesian inversion}
\label{subsec:Hierarchical Bayesian Inversion}
As suggested by \cite{malinverno_expanded_2004}, we treat the value of $\alpha$ as an unknown and estimate it during inversion using a hierarchical Bayesian approach \citep{malinverno_two_2006,bodin_transdimensional_2012, galetti_transdimensional_2017}. In this approach, we allow $\alpha$ to have its own degree of uncertainty rather than assigning a fixed value. Thus, we estimate the uncertainty in $p(\mathbf{m})$, which improves its consistency with data. As a result, the distributions of $\alpha$, or equivalently $\tau$, and hence the final distribution $p(\mathbf{m})$, become \textit{posterior} distributions \citep{mosegaard_inconsistency_2024}. To avoid the introduction of more complicated terms, we will still refer to $p(\mathbf{m})$ as the prior distribution, but its posterior nature should be understood implicitly.

\begin{figure}
    \centering
    \includegraphics[width=0.75\linewidth]{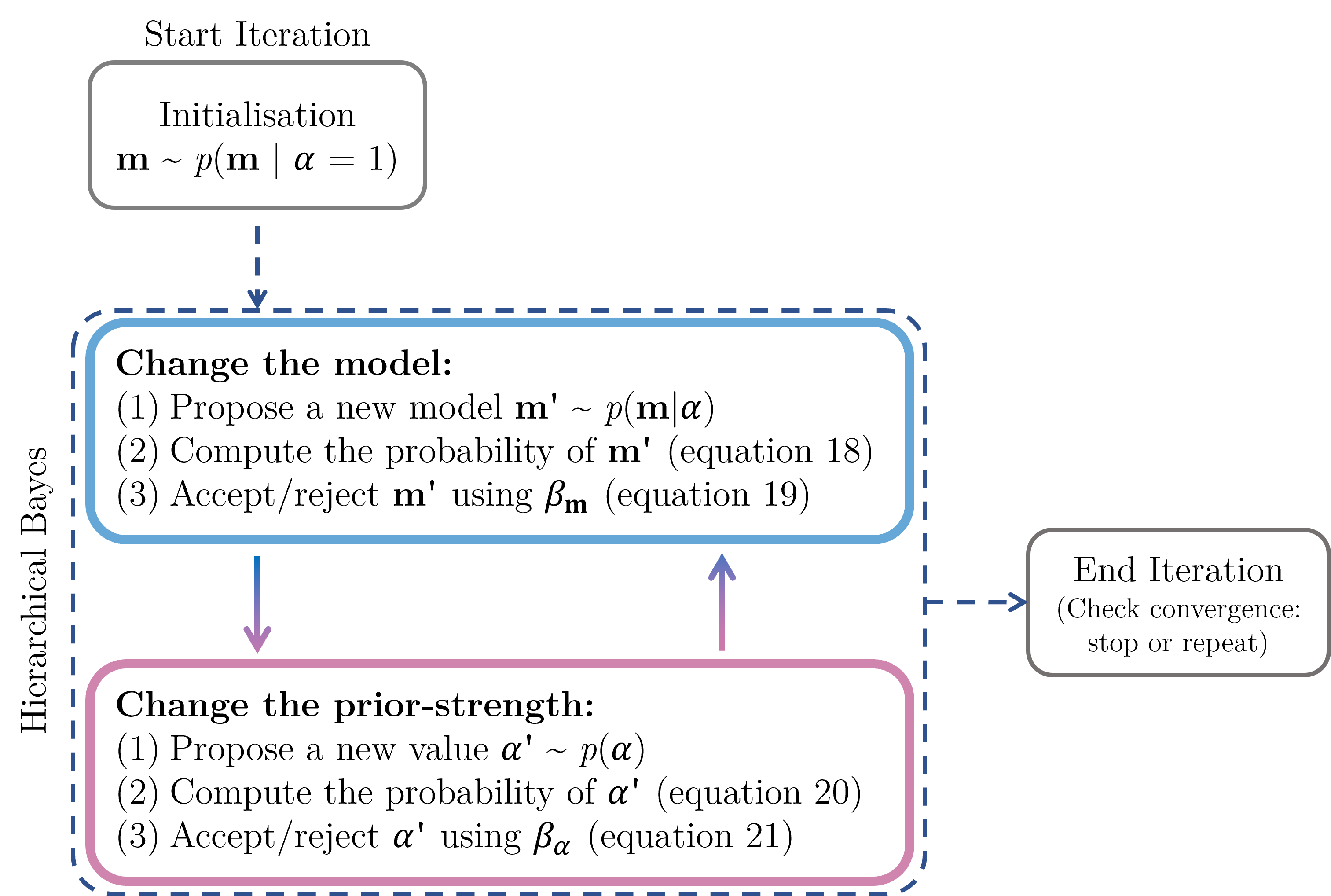}
    \vspace{5pt}
    \caption{The hierarchical Bayesian inversion algorithm involves two main steps after initialisation. During each iteration, the model $\mathbf{m}$ (in blue) or prior-strength $\alpha$ (in pink) can be revised based on the previously sampled values.}
    \label{fig:hier_bayes_diagram}
\end{figure}

To represent this extended uncertainty in both the model parameters and the hyperparameter, we estimate their joint posterior distribution,
\begin{equation}
    p(\mathbf{m}, \alpha | \mathbf{d}_{obs}) \propto  p(\mathbf{d}_{obs}|\mathbf{m}) \: p(\mathbf{m}|\alpha) \: p(\alpha)
    \label{eq:hier_posterior}
\end{equation}
To draw samples from this joint posterior $p(\mathbf{m}, \alpha | \mathbf{d}_{obs})$, we use an MCMC sampler with the Metropolis-within-Gibbs algorithm \citep{chib_understanding_1995}. For further information about the Gibbs sampling algorithm, see \cite{casella_explaining_1992}. In our implemented hierarchical Bayesian algorithm, the sampling process for the joint posterior (equation \ref{eq:hier_posterior}) follows a dependency structure: $\alpha \rightarrow \mathbf{m} \rightarrow \mathbf{d}_{obs}$ (i.e., the data distribution $p(\mathbf{d}_{obs}|\mathbf{m})$ depends solely on $\mathbf{m}$, and the hyperparameter $\alpha$ influences $\mathbf{d}_{obs}$ only through $\mathbf{m}$), and can therefore be divided into two stages. We initialise the MCMC sampling by drawing a random model $\mathbf{m}$ from a uniform distribution (equation \ref{eq:uniform_distribution}), and a prior-strength value $\alpha = 1$ which is always consistent with its prior distribution in the examples herein. Following the initialisation, each MCMC step is divided into two stages, performed sequentially, to change the model $\mathbf{m}$ and the prior-strength $\alpha$ (Figure \ref{fig:hier_bayes_diagram}):

\begin{enumerate}[leftmargin=*, label=(\roman*), itemsep=1ex]
\item  \textbf{Change the model}: at iteration $t$, with an accepted value of prior-strength hyperparameter $\alpha_{t-1}$ from the previous iteration, a new model $\mathbf{m}'$ is proposed by randomly perturbing the current model parameters according to an adaptive Gaussian multivariate normal proposal distribution $q(\mathbf{m})$.
\begin{enumerate}
    \item Compute the probability of the proposed model $\mathbf{m}'$, given observed data $\mathbf{d}_{obs}$ and hyperparameter $\alpha_{t-1}$ using:
\begin{equation}
p(\mathbf{m}'|\mathbf{d}_{obs}, \alpha_{t-1}) \propto p(\mathbf{d}_{obs}|\mathbf{m}') \: p(\mathbf{m}'|\alpha_{t-1})
    \label{eq:hier_post_stage_1}
\end{equation}

\item Accept or reject the proposed model $\mathbf{m}'$ by calculating the acceptance probability using $\beta_\mathbf{m}$ (equation \ref{eq:hier_post_stage_1_acctp}). Set $\mathbf{m}_t = \mathbf{m'}$ if $u \leq \beta_\mathbf{m}$; otherwise $\mathbf{m}_t = \mathbf{m}_{t-1}$. Here, $u$ is a random number drawn from a uniform distribution $\mathcal{U}(0, 1)$, and
\begin{equation}
   \beta_\mathbf{m} = \min\left(1, \frac{p(\mathbf{d}_{obs} | \mathbf{m}') \: p(\mathbf{m}' | \alpha_{t-1}) \: q(\mathbf{m}_{t-1} | \mathbf{m}')}{p(\mathbf{d}_{obs} | \mathbf{m}_{t-1}) \:  p(\mathbf{m}_{t-1} | \alpha_{t-1}) \: q(\mathbf{m}' | \mathbf{m}_{t-1})}\right)
    \label{eq:hier_post_stage_1_acctp}
\end{equation}
\end{enumerate}

\item  \textbf{Change the prior-strength}: centred at its current value $\alpha_{t-1}$, a prior-strength $\alpha'$ is proposed by randomly perturbing its value according to a Gaussian normal proposal distribution $q(\alpha)$.
\begin{enumerate}
    \item Calculate the probability of the proposed $\alpha'$ for the current model $\mathbf{m}_t$, using:
\begin{equation}
    p(\mathbf{m}_t, \alpha') = p(\mathbf{m}_t | \alpha') \: p(\alpha')
    \label{eq:hier_post_stage_2}
\end{equation}
    \item Accept ($\alpha_t = \alpha'$) if $u \leq \beta_\mathbf{\alpha}$ as $u \sim \mathcal{U}(0, 1)$; otherwise reject ($\alpha_t = \alpha_{t-1}$) the proposed $\alpha'$, based on acceptance probability

\begin{equation}
   \beta_\alpha = \min\left(1, \frac{p(\mathbf{m}_t | \alpha') \:  p(\alpha') \: q(\alpha_{t-1} | \alpha')}{p(\mathbf{m}_t | \alpha_{t-1}) \:  p(\alpha_{t-1}) \: q(\alpha' | \alpha_{t-1})}\right)
    \label{eq:hier_post_stage_2_acctp}
\end{equation}
\end{enumerate}
\end{enumerate}
\noindent
In equations \ref{eq:hier_post_stage_1} and \ref{eq:hier_post_stage_2}, $p(\mathbf{m} | \alpha)$ can be derived from equation \ref{eq:modelspace_prior_pdf} as:
\begin{equation}
    p(\mathbf{m} | \alpha) = p_0(\mathbf{m}) \times \left(\frac{\alpha}{\pi}\right)^{k/2} \exp\left(-\alpha \|\mathbf{m}-\boldsymbol{\mu}\|_2^2\right) \,
    \label{eq:hier_prior_pdf_p(m|a)}
\end{equation}
where $\left(\frac{\alpha}{\pi}\right)^{k/2}$ acts as a normalisation constant with dimensionality $k$ and $\boldsymbol{\mu} = \mathcal{G}(\mathbf{E}(\mathbf{m}))$. 


Hierarchical Bayesian inversion also requires a prior distribution for the hyperparameter $\alpha$ (a so-called hyperprior, $p(\alpha)$ in equation \ref{eq:hier_post_stage_2}). If there is very little information, a uniform hyperprior distribution is sometimes the most sensible choice within fixed bounds \citep{gelman_bayesian_2013,malinverno_expanded_2004}. For simplicity, \cite{malinverno_two_2006} and \cite{bodin_transdimensional_2012} use a non-informative uniform hyperprior to account for uncertainties in data noise.

In our case, the hyperprior is defined over the prior-strength hyperparameter (or prior precision; $\alpha = 1/(2\tau^2)$). The higher-dimensional parameter space far from $\mathcal{M}_{gen}$ (more likely to be selected using low $\alpha$ values) occupies a substantially larger hypervolume (volume of spaces with dimensionality greater than 3) than the space close to $\mathcal{M}_{gen}$ (more likely to be selected using high $\alpha$ values) because $\mathcal{M}_{gen}$ exists within a lower-dimensional subspace of $\mathbb{R}^k$. This implies a higher probability that the true Earth lies outside the manifold; however, we do not know how far from the manifold it lies. For this reason, we use an exponential distribution as our hyperprior, which assigns higher probability to low $\alpha$ values (giving higher probability to the parameter space further from $\mathcal{M}_{gen}$); the prior probability of any model then decreases monotonically with higher $\alpha$ values (focusing the prior pdf closer to $\mathcal{M}_{gen}$):
\begin{equation}
   p(\alpha) = \exp \left(- \gamma \: (\alpha - \alpha_{lb})\right)
    \label{eq:hyperprior(alpha)}
\end{equation}
In equation \ref{eq:hyperprior(alpha)}, $\gamma$ denotes the decay rate with an acceptable lower bound $\alpha_{lb} > 0$,  which can be set as in Figure \ref{fig:hyperprior_test_near}d, though it could be lower. In Appendix \ref{app:Hyperprior distributions}, we present examples demonstrating other hyperpriors applied to a low-dimensional problem.

\section{Illustrative Low-Dimensional Problems}
\label{sec:Illustrative Low-Dimensional Problems}
\subsection{Problem setup}
\label{subsec:Problem setup}

\begin{figure}
    \centering
    \includegraphics[width=0.8\linewidth]{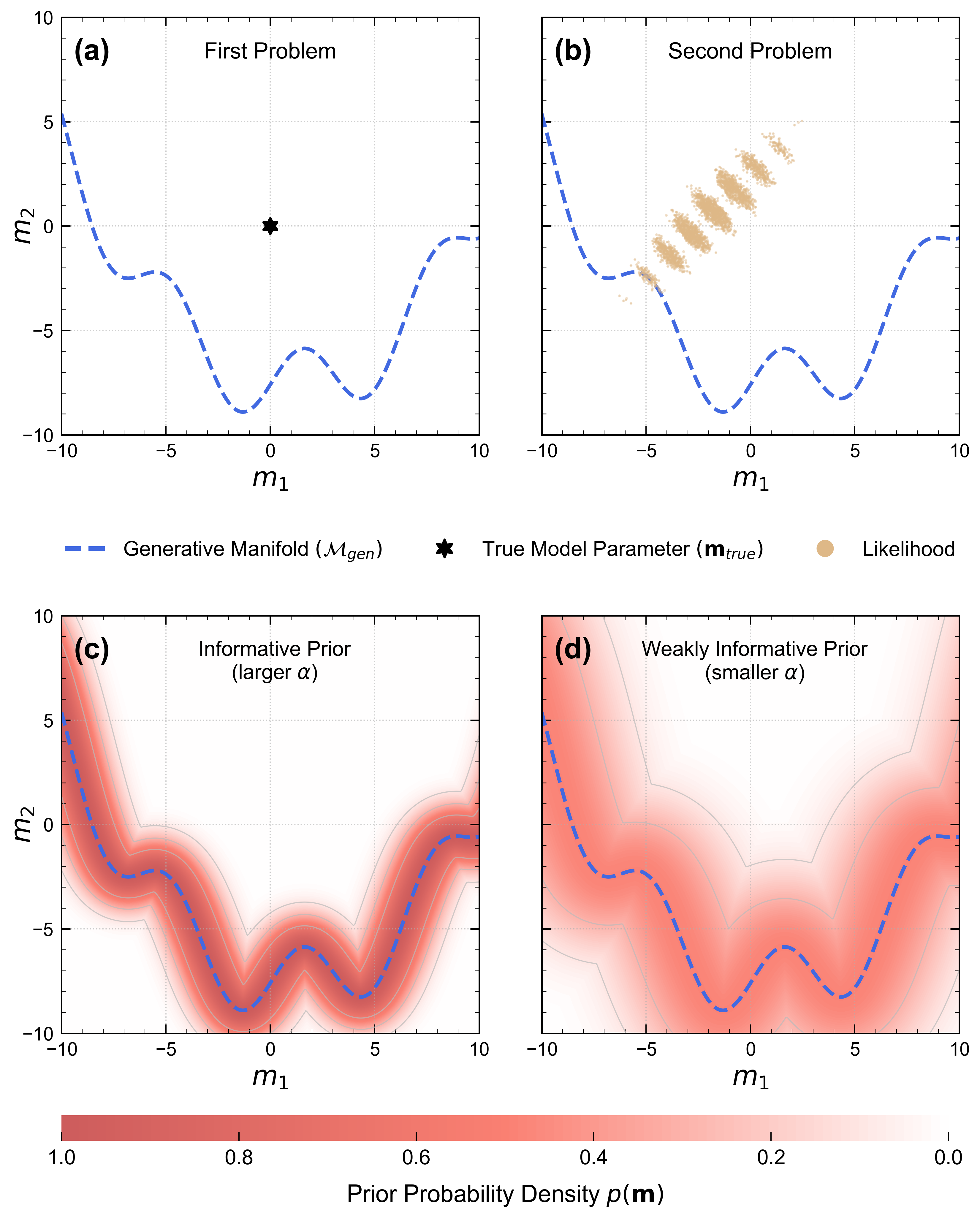}
    \caption{Two illustrative problems (panels a and b). The blue dashed curve represents a 1D generative manifold $\mathcal{M}_{gen}$ learned from $\mathcal{M}$ (equation \ref{eq:analytical_manifold}), embedded in a 2D parameter space. (a) The true model parameter $\mathbf{m}_{true}$ is shown as a black star. (b) A multimodal likelihood is visible from its brown samples. (c) A strongly informative and (d) weakly informative prior pdf defined by $p(\mathbf{m})$ (equation \ref{eq:modelspace_prior_pdf}). }
    \label{fig:toy_problem_prior}
\end{figure}

Consider a 1D non-linear function that combines polynomial (linear and quadratic) and trigonometric components shown in Figure \ref{fig:toy_problem_prior} (blue dashed curve):
\begin{equation}
    f(m_1, m_2) = A + B \cdot m_1 - C \cdot \sin(m_1) - Q \cdot m_1^2 + m_2= 0
    \label{eq:analytical_manifold}
\end{equation}
where $m_1$ and $m_2$ are the model parameters bounded in the range $[-10, 10]$, with constants $A=7.6$, $B=0.2$, $C=1.8$, and $Q=0.1$. Assume that this analytical function is our lower-dimensional (1D) manifold $\mathcal{M}$, embedded in a 2D (full) space. We use three different GNNs (VAE, AAE and DAAE) to learn the structure of this manifold $\mathcal{M}$. The learned structure, which is an approximation of $\mathcal{M}$, is the generative manifold $\mathcal{M}_{gen}$ (see section \ref{subsec:Generative Manifolds as Bayesian Priors}).

In Figure \ref{fig:toy_problem_prior}, we present two problems: the first (Figure \ref{fig:toy_problem_prior}a) has a unique true solution $\mathbf{m}_{true}$ at $(0, 0)$ (marked by a black star), which is far from the generative manifold $\mathcal{M}_{gen}$ (blue dashed curve); the second (Figure \ref{fig:toy_problem_prior}b) has a solution defined only by a multimodal likelihood, and $\mathcal{M}_{gen}$ passes through one of its modes. We define our prior pdf $p(\mathbf{m})$ by equation \ref{eq:modelspace_prior_pdf} (Figure \ref{fig:toy_problem_prior}), where $p_0(\mathbf{m})$ is bounded within the range $[-10,10]$ and the strength of the prior belief in $p_{\mathcal{M}_{gen}}(\mathbf{m})$ is controlled by $\alpha$. A larger value of $\alpha$ produces a more informative prior (Figure \ref{fig:toy_problem_prior}c) and a smaller value defines a more weakly informative prior (Figure \ref{fig:toy_problem_prior}d). 

\begin{figure}
    \centering
    \includegraphics[width=1\linewidth]{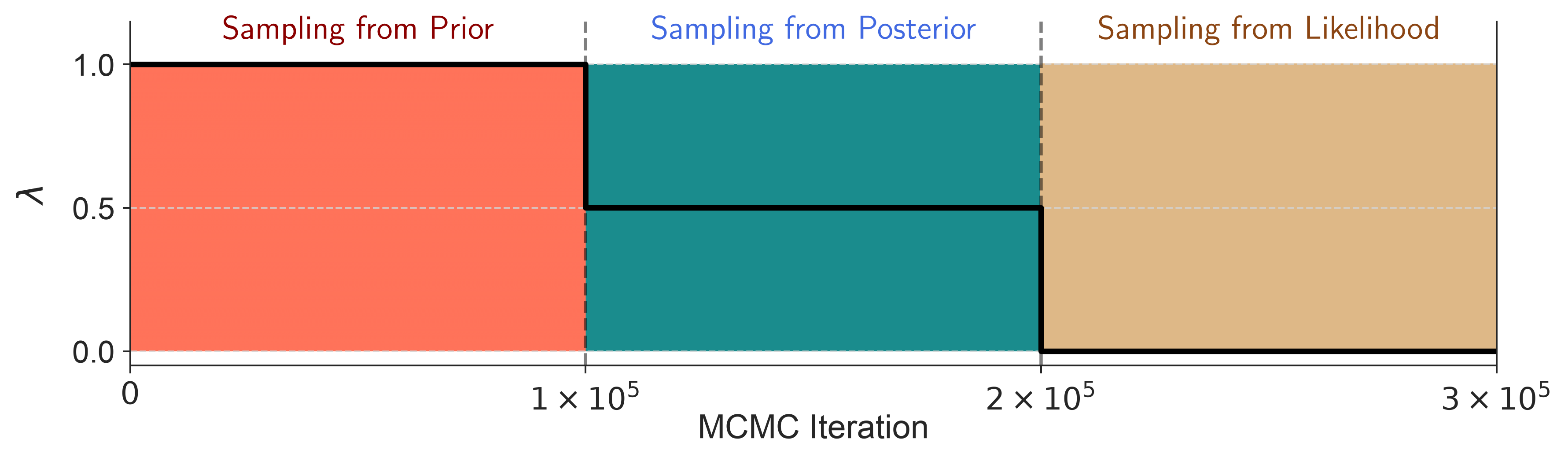}
    \caption{Variation of $\lambda$ with MCMC iterations. At $\lambda = 1$ (red coloured zone) the sampler only draw samples from the prior pdf; at $\lambda = 0.5$ (teal coloured zone) it draws samples from the posterior pdf; and at $\lambda = 0$ (brown coloured zone) it samples from the likelihood.  Colour scheme is maintained in Figures \ref{fig:toy_problem_1} to \ref{fig:toy_problem_3_hier}.}
    \label{fig:lambda_decay}
\end{figure}

The most appropriate value for the hyperparameter $\alpha$ is not known \textit{a priori}. We initially test three different $\alpha$ values, $\{1.0, 0.5, 0.1\}$ in the first problem and three others $\{1, 10, 50\}$ in the second, to evaluate the effect of hyperparameter $\alpha$ on the inversion results. For these examples, we define our target distribution $ \pi_\lambda(\mathbf{m})$:
\begin{equation}
\pi_\lambda(\mathbf{m}) \propto \exp \left(- (1 - \lambda)\frac{\|\mathbf{d}_{obs} - f(\mathbf{m})\|_2^2}{2\sigma_d^2}\right) \times  p_0(\mathbf{m}) \times \exp\left(- \lambda \:\alpha \|\mathbf{m}-\boldsymbol{\mu}\|_2^2\right)
\label{eq:target_distribution_with_lambda}
\end{equation}
where $\lambda$ can take values such as 1 (to sample prior pdf $\pi_1(\mathbf{m})$), 0.5 (to sample the posterior pdf $\pi_{0.5}(\mathbf{m})$), and 0 (to sample the likelihood $\pi_0(\mathbf{m})$) (see details in Figure \ref{fig:lambda_decay}), where each element of $\mathbf{m}$ is bounded within the range $[-10,10]$.

In the first problem, the forward operation is simply an identity function $f$, where $f(\mathbf{m}) = \mathbf{m}$, with $\mathbf{d}_{obs} = \mathbf{m}_{true}$ and the data likelihood is defined by the first term in equation \ref{eq:target_distribution_with_lambda}. The second problem has a complex target distribution within the 2D model parameter space. The forward operator $f$ in this case is defined to be non-linear, producing a non-unique inverse solution represented by a multimodal likelihood (Figure \ref{fig:toy_problem_prior}b):
\begin{equation}
    f(\mathbf{m}) = 1.5 \Big( 1 - \cos\big(3(m_1 + m_2 + 1.1)\big) \Big)+0.6 \left( (m_1 - m_2 + 2.5)^2 + \frac{(m_1 + m_2 + 1.1)^2}{16} + 1 \right)
    \label{eq:forward_opr_toy2}
\end{equation}
Therefore, if the model parameter vector is $\mathbf{m} = [m_1, m_2]^T \in \mathbb{R}^2$, then the forward function $f(\mathbf{m}): \mathbb{R}^2 \rightarrow \mathbb{R}$ maps the model parameters to the 1D data space.

Our goal is to sample the target distribution (equation \ref{eq:target_distribution_with_lambda}) using the prior information from $\mathcal{M}_{gen}$. We generate a total of $3 \times 10^5$ samples for both problems using the MCMC sampler in three stages (Figure \ref{fig:lambda_decay}):

\begin{itemize}
    \item Iterations: 1 to $10^5$ ($\lambda = 1$): Drawn from the prior pdf, $\pi_1(\mathbf{m})$.
    \item Iterations: $10^5$ to $2 \times 10^5$ ($\lambda = 0.5$): Drawn from the posterior pdf, $\pi_{0.5}(\mathbf{m})$.
    \item Iterations: $2 \times 10^5$ to $3 \times 10^5$ ($\lambda = 0$): Drawn from the likelihood, $\pi_0(\mathbf{m})$.
\end{itemize}

We run 10 MH-MCMC chains to draw samples from the target distributions, whose samples are thinned by a factor of 250, then amalgamated. All chains use random initialisations to ensure the reliability of the emerging patterns.

\subsection{Results}
\label{subsec:results_toyproblem}

The set of all possible outputs or generations from the generators of our trained GNNs defines the generative manifold $\mathcal{M}_{gen}$ discussed in section \ref{subsec:Generative Manifolds as Bayesian Priors}. We draw $10^5$ random samples from the analytical manifold defined by equation \ref{eq:analytical_manifold} to train the GNNs (VAE, AAE and DAAE). After training, models generated by the GNNs define the generative manifold $\mathcal{M}_{gen}$, which is used as prior information with some uncertainty described by equation \ref{eq:modelspace_prior_pdf} to constrain the 2D model space inversion defined by two problems.

\begin{figure}
    \centering
    \includegraphics[width=1\linewidth]{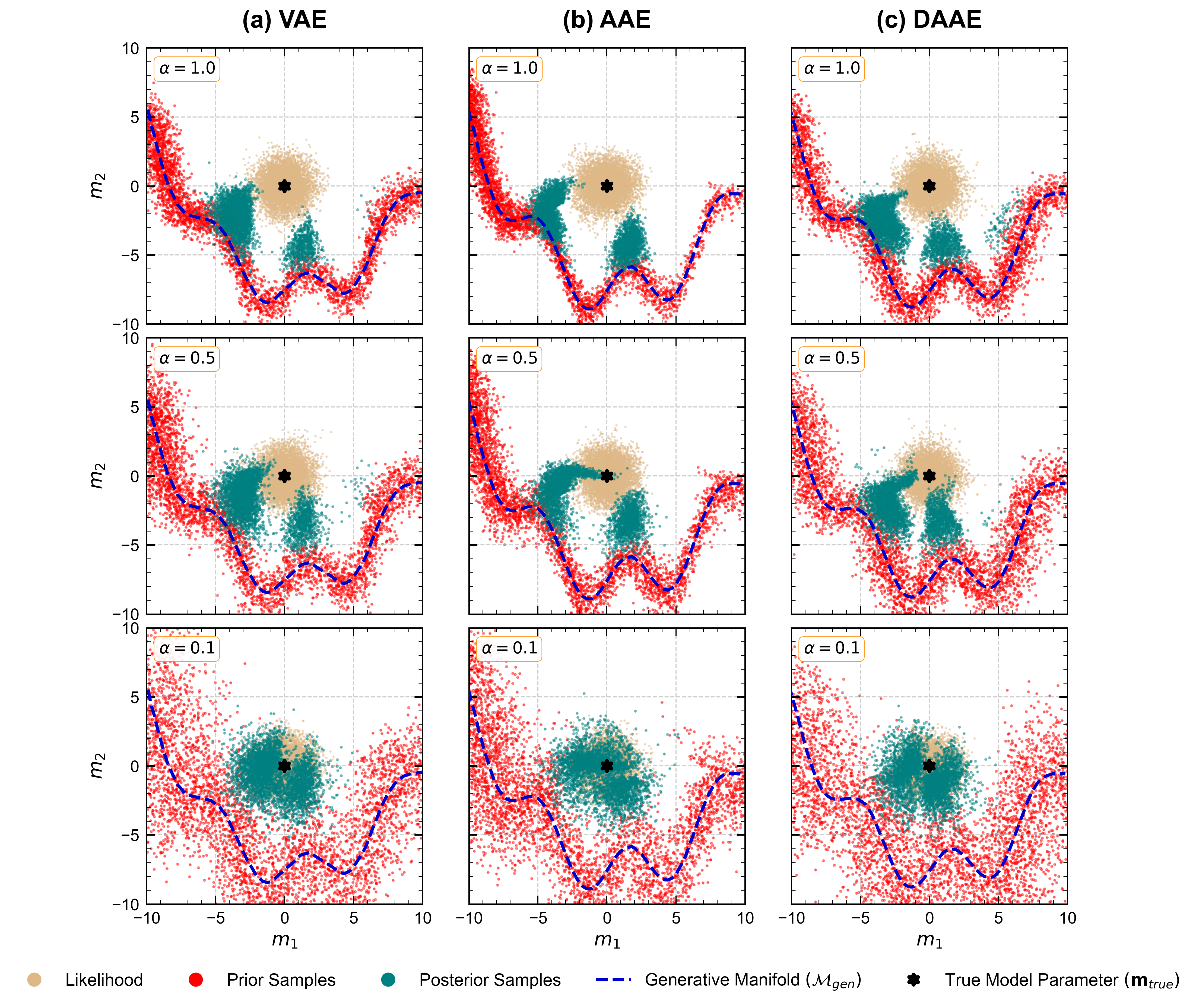}
    \caption{Results from the problem in Figure \ref{fig:toy_problem_prior}a. Every 250th sample from the prior pdf, posterior pdf, and likelihood are plotted in red, teal and brown, respectively. The generative manifold $\mathcal{M}_{gen}$ (blue dashed curve) is produced by (a) VAE, (b) AAE and (c) DAAE in each of the columns. Every row corresponds to a different value of $\alpha$, $\{1.0, 0.5, 0.1\}$ respectively.}
    \label{fig:toy_problem_1}
\end{figure}

In the first problem (Figure \ref{fig:toy_problem_prior}a), the true model parameter $\mathbf{m}_{true}$ is located at $(0, 0)$ and Figure \ref{fig:toy_problem_1} shows the corresponding inversion results. Samples are colour-coded based on the distribution from which they were drawn (Figure \ref{fig:lambda_decay}). Red samples represent samples drawn from the prior pdf (at $\lambda=1$), brown samples are drawn from the likelihood ($\lambda=0$), and teal coloured samples correspond to the posterior pdf ($\lambda=0.5$).

The results for $\alpha = 1$ (first row, Figure \ref{fig:toy_problem_1}) represent an informative prior as the distribution of the red samples shows lower variance. For $\alpha = 1$, the prior samples are more concentrated along $\mathcal{M}_{gen}$ than in the cases with $\alpha = 0.5$ and $\alpha = 0.1$. The posterior samples (teal) are more aligned with the prior than with the likelihood (brown). For $\alpha = 0.5$ and $\alpha = 0.1$ the prior is progressively less informative; the prior samples span a wider region, and the posterior samples are shifted towards the $\mathbf{m}_{true}$. The results illustrate that when the true solution is far from $\mathcal{M}_{gen}$, incorporating a less informative prior is beneficial.

\begin{figure}
    \centering
    \includegraphics[width=1\linewidth]{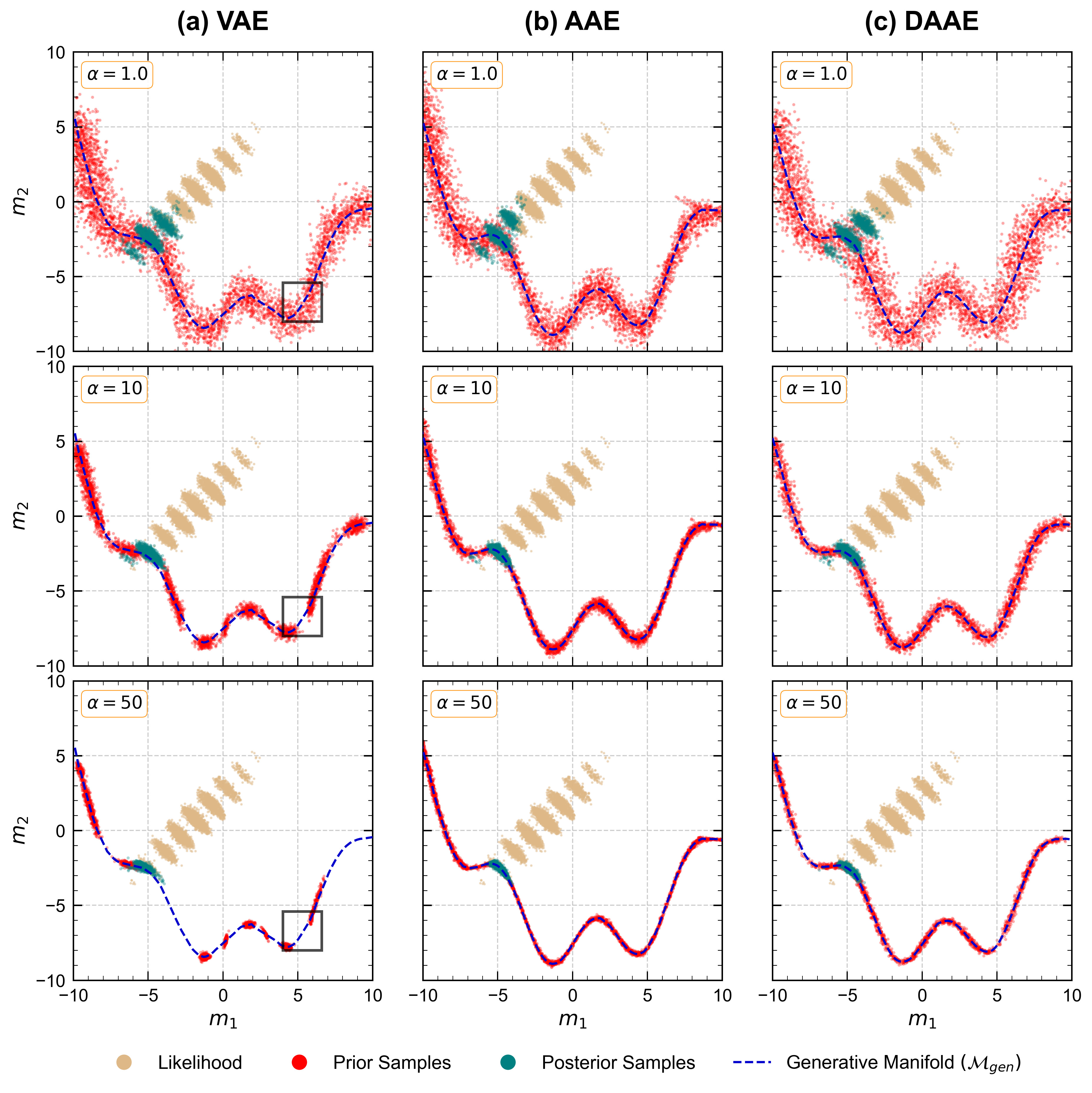}
    \caption{MCMC results from the problem in Figure \ref{fig:toy_problem_prior}b, comparing three escaping priors (a) VAE, (b) AAE and (c) DAAE (columns) at three different prior-strengths $\alpha = \{1, 10, 50\}$ (rows). Every 250th sample is plotted and coloured by their target distributions: prior (red), posterior (teal), and likelihood (brown). The learned generative manifold $\mathcal{M}_{gen}$ is shown in a blue dashed curve. Black squares mark regions with gaps in prior samples.}
    \label{fig:toy_problem_2}
\end{figure}

Figure~\ref{fig:toy_problem_2} shows results from the problem with a multimodal likelihood function in Figure \ref{fig:toy_problem_prior}b. We compare three prior-strengths, $\alpha = 1, 10,$ and $50$, shown in the top, middle and bottom rows, respectively. With a weakly informative prior $(\alpha = 1)$, the posterior samples (teal) are dispersed across several well-separated modes near $\mathcal{M}_{gen}$; more informative priors ($\alpha = 10$ and $50$) constrain the posterior samples to be closer to $\mathcal{M}_{gen}$. We also observe that the prior samples (at $\lambda=1$) are separated into clusters at higher prior-strengths for the VAE  (Figure \ref{fig:toy_problem_2}a). This is more visible at $\alpha = 50$, because the prior probability density is more concentrated around the manifold $\mathcal{M}_{gen}$ (Figure \ref{fig:toy_problem_prior}c). Such clustering occurs because the KL divergence in the VAE does not effectively ensure that the encoder's output follows a continuous Gaussian distribution during training. In Appendix \ref{app:Escaping prior samples}, comparing the escaping prior samples with high $\alpha$ values (implying samples on or very close to $\mathcal{M}_{gen}$) from VAE, AAE and DAAE within the region marked by the black square in Figure \ref{fig:toy_problem_2}a, we observe that the VAE's encoder output distribution in the latent space displays distinct high-density regions or modes and the VAE's generator maps latent samples from those modes to reconstructed model samples in parameter space that are spatially close on $\mathcal{M}_{gen}$. Unfortunately those samples are far from their actual locations on $\mathcal{M}$, thereby increasing the $L_2$ norm (implying a lower probability, equation \ref{eq:modelspace_prior_pdf}); this leads the MCMC sampler to reject these samples and results in those clusters.


\begin{figure}
    \centering
    \includegraphics[width=1\linewidth]{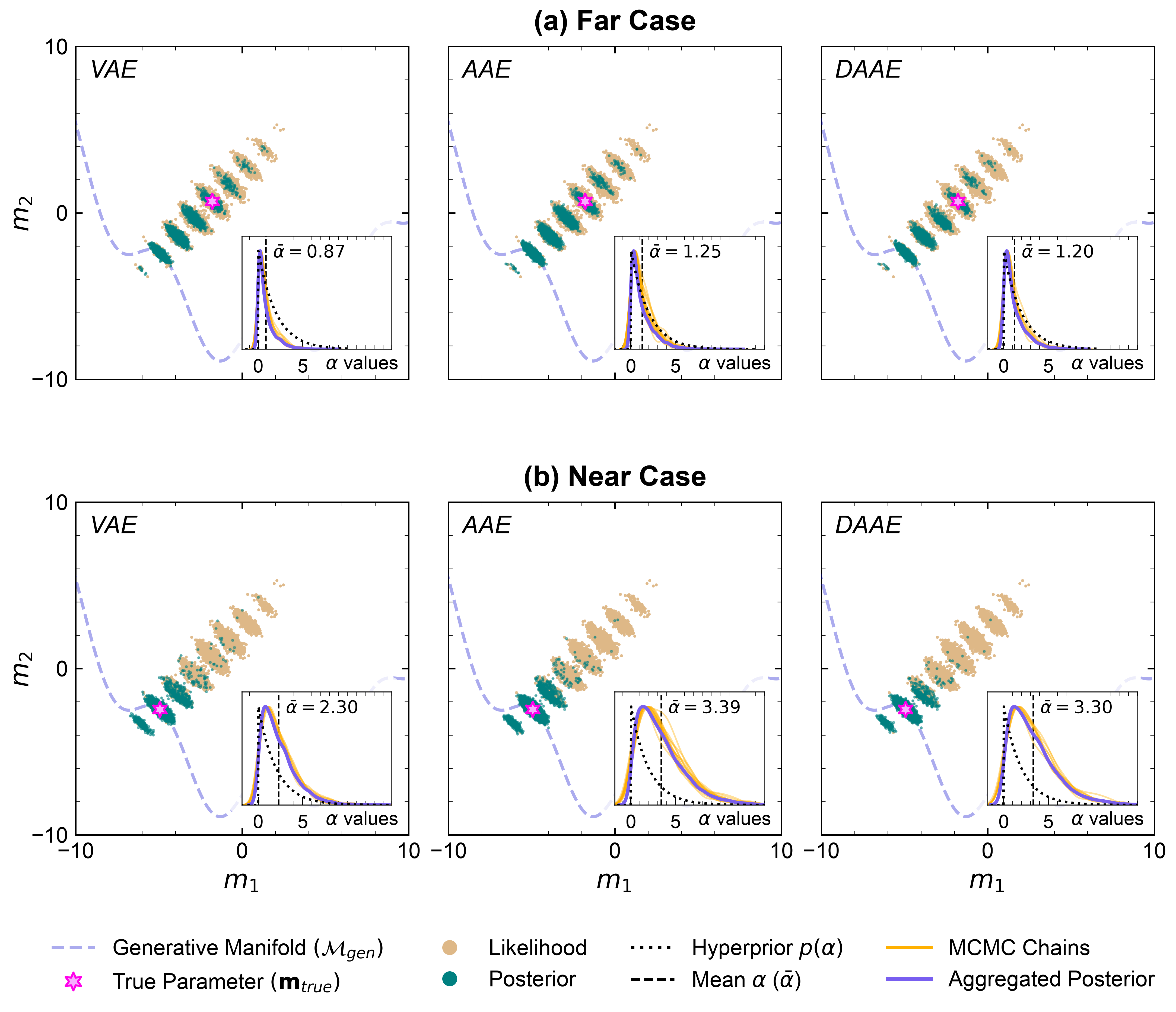}
    \caption{MCMC results from the hierarchical Bayesian inversion of the problem in Figure \ref{fig:toy_problem_2}, comparing two scenarios: (a) $\mathbf{m}_{true}$ (pink star) is far from $\mathcal{M}_{gen}$, and (b) $\mathbf{m}_{true}$ (pink star) is near to $\mathcal{M}_{gen}$ (the blue dashed curve), in top and bottom panels, respectively. VAE, AAE and DAAE-informed escaping priors are implemented to estimate the posterior $\pi_{0.5}(\mathbf{m})$ as defined in equation \ref{eq:target_distribution_with_lambda}. Posterior samples (teal) are plotted on top of likelihood samples (brown) in top and bottom panels. Inset plots compare the sampled $\alpha$ posteriors (orange) from 10 MCMC chains with their corresponding hyperpriors (black dotted curve), alongside the mean values ($\bar{{\alpha}}$, vertical black dashed line) of the aggregated $\alpha$ posteriors (violet).}
    \label{fig:toy_problem_3_hier}
\end{figure}

From examples shown in Figure  \ref{fig:toy_problem_1} and \ref{fig:toy_problem_2}, we see that the optimal value of $\alpha$ varies (if it exists at all) and clearly the posterior uncertainty changes with $\alpha$. We therefore treat the prior-strength $\alpha$ as an unknown, along with the model parameters $m_1$ and $m_2$. As explained in section \ref{subsec:Hierarchical Bayesian Inversion}, we implement the hierarchical Bayesian algorithm to draw samples from the joint distribution in equation \ref{eq:hier_posterior}, conditional on 1D data observations $\mathbf{d}_{obs}$ generated using equation \ref{eq:forward_opr_toy2} for the true solution $\mathbf{m}_{true}$, which is either far from $\mathcal{M}_{gen}$ (Figure \ref{fig:toy_problem_3_hier}a) or near $\mathcal{M}_{gen}$ (Figure \ref{fig:toy_problem_3_hier}b). For each case, we use a decay rate $\gamma = 0.5$ and set the lower bound $\alpha_{lb}$ at 0.01, according to the hyperprior specified in equation \ref{eq:hyperprior(alpha)}. We run 10 MCMC chains each with $3 \times 10^5$ iterations to draw samples from the joint distribution (equation \ref{eq:hier_posterior}) and after burn-in every 250th sample is displayed in the Figure \ref{fig:toy_problem_3_hier}.

In the case where $\mathbf{m}_{true}$ is far from $\mathcal{M}_{gen}$, the top panels in Figure \ref{fig:toy_problem_3_hier} show that the posterior samples are well distributed along $\mathbf{m}_{true}$, irrespective of the choice of GNNs (VAE, AAE and DAAE). The posterior density plots for the prior-strength $\alpha$ have samples concentrated near zero and mean values near $1$. For the case in which $\mathbf{m}_{true}$ is near $\mathcal{M}_{gen}$, the posterior density plot shows samples of $\alpha$ with higher values and mean greater than $2$. This indicates that the algorithm assigns posterior-strength values $\alpha$ that result in model parameters which explain the observed data while remaining consistent with the prior distribution. Thus, it focuses posterior sampling on the correct mode and effectively ignores the most distant modes of the likelihood, demonstrating the effectiveness of hierarchical Bayesian inversion.

\section{Travel-Time Tomography}
\label{sec:Travel-Time Tomography}
\subsection{Simple velocity models}
\label{subsec:Simple Velocity Models}

We now test the intuition gained in low-dimensional problems above by applying it to nonlinear travel-time tomography experiments. Figures \ref{fig:diff_alphas_7_Z_vels}a and b show two different true velocity models, both consisting of simple square-cellular structures of $10 \times 10$ model parameters with a similar low-velocity background. Figure \ref{fig:diff_alphas_7_Z_vels}a defines a sparse survey geometry with 4 sources and receivers (indicated by red triangles), resulting in 6 independent travel-time measurements. We do not add noise to the observed travel-time data to allow us to focus solely on the impact of escaping prior distributions on inversion problems.

The true velocity model in Figure \ref{fig:diff_alphas_7_Z_vels}a has a shape similar to the digit ``7". To generate samples from our prior information we used the MNIST images, which contain examples of handwritten digits \citep{lecun_mnist_1998}. We train the generative neural networks (GNNs) (each with 16 latent parameters) only on samples of the digit ``7", reshaped to $10\times10$ pixels/parameters giving greyscale values, re-normalised to a velocity range of [1, 3] km/s. Based on these training data, the GNNs ideally learn distinctive structural features of ``7" and can only generate structures similar to those. Models generated by the trained GNN will therefore theoretically lie on a lower-dimensional generative manifold $\mathcal{M}_{gen}$.

Figure \ref{fig:diff_alphas_7_Z_vels}b shows another simple tomography experiment. In this case, our aim is to recover a velocity model that resembles a ``$\mathcal{Z}$"-like shape, which presumably does not lie on $\mathcal{M}_{gen}$. We increase the number of sources and receivers to eight, to provide more information to the inversion from travel-time observations, and as in the first experiment no noise is added to the observed travel-time data. Our main objectives in this example are to demonstrate that even with perfect data, latent space inversion cannot produce correct inversion results when the true model lies outside the generative manifold $\mathcal{M}_{gen}$, and to assess whether our proposed method using an escaping prior could recover the off-manifold features - specifically the middle and bottom horizontal strokes of the true model ``$\mathcal{Z}$".

Our escaping overall prior pdf $p(\mathbf{m})$ (defined in equation \ref{eq:modelspace_prior_pdf}) consists of two components. The first is the uniform prior $p_0(\mathbf{m})$, which in this case, is bounded between 0.5 and 3.5 km/s in the tomography experiments. The second component is the escaping prior $p_{\mathcal{M}_{gen}}(\mathbf{m})$ (equation \ref{eq:prior_integral_iso}), which introduces the prior information from the generative manifold $\mathcal{M}_{gen}$. The posterior pdf was sampled using the MH-MCMC sampler. For inference, 40 independent MCMC chains were run in parallel, each drawing 2 million samples, to ensure that each chain had reasonably converged and that the results accurately reflect the different performances of each applied method. Each chain was initialised from a different, randomly sampled velocity model drawn from the uniform distribution $p_0(\mathbf{m})$.

\subsection{Testing variations of $\alpha$}
\label{subsec:Testing variations of alpha}

\begin{figure}
    \centering
    \includegraphics[width=1\linewidth]{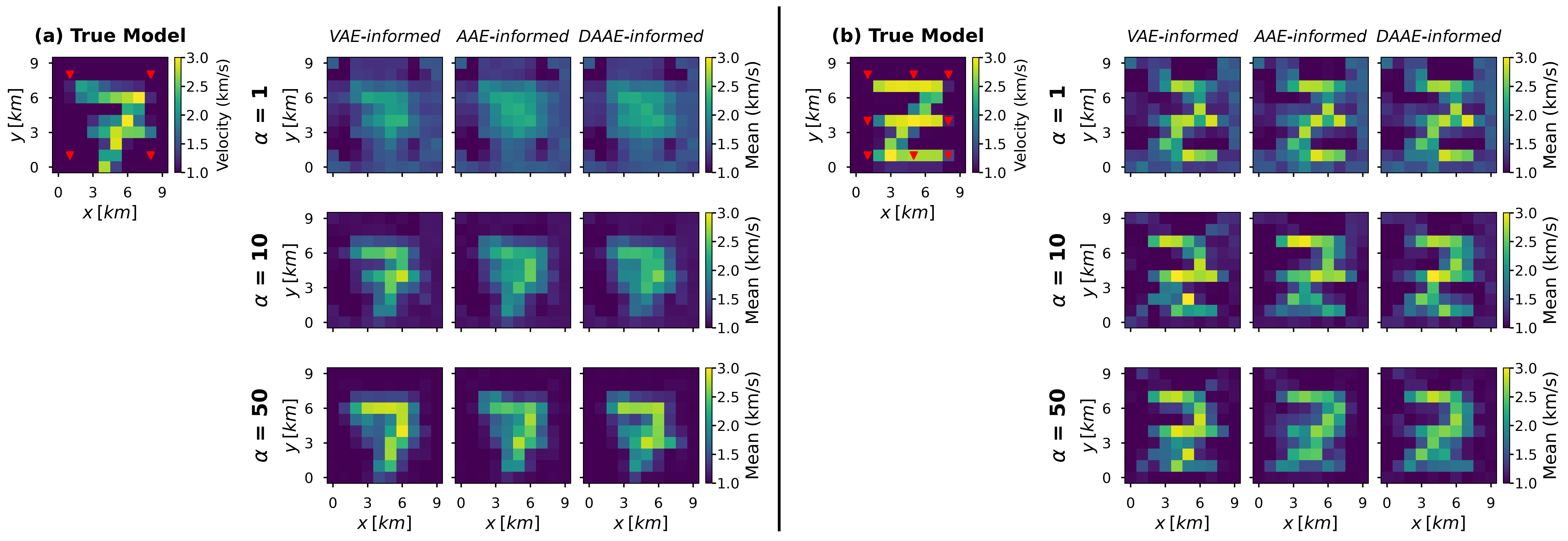}
    \caption{Posterior mean plots (rows) obtained from the escaping prior with three constant prior-strengths $\alpha = \{1, 10, 50\}$. In the left and right panels, the columns represent the true model (left) and the mean model parameter inversion results from three different GNNs (VAE, AAE and DAAE), respectively. Red stars represent points at which we assume both sources and receivers of wave energy are located, and between which we model travel-time data $\mathbf{d}_{obs}$, which are used to inform the inversion.}
    \label{fig:diff_alphas_7_Z_vels}
\end{figure}

Since there is no prior knowledge about the hyperparameter for prior-strength, it is uncertain which $\alpha$ is appropriate for the current inversion problems. We therefore initially test three different prior-strength values (equation \ref{eq:modelspace_prior_pdf}, with $\alpha = \{1, 10, 50\}$) to assess how the escaping prior influences the inversion results (Figure \ref{fig:diff_alphas_7_Z_vels}).
Comparing each result obtained from three different escaping priors to the true velocity model in Figure \ref{fig:diff_alphas_7_Z_vels}a, we observe that with $\alpha=1$ the posterior mean models are blurry, with no visible structural features. However, Figure \ref{fig:diff_alphas_7_Z_vels}b shows some visible structural features in the posterior mean model, although the resolution remains quite low. 

Figure \ref{fig:diff_alphas_7_Z_vels}a shows that increasing $\alpha$ (stronger prior information from $\mathcal{M}_{gen}$) improves results because the prior information is consistent with the ``7"-like true velocity model. However, for the ``$\mathcal{Z}$"-like true model, Figure \ref{fig:diff_alphas_7_Z_vels}b clearly shows that increasing $\alpha$ degrades the inversion outcomes, even given the additional travel-time observations. This occurs because the stronger prior overpowers any ``$\mathcal{Z}$"-like information in the likelihood. Hence, a proper balance between the likelihood and the escaping prior is necessary and selecting a fixed $\alpha$ value \textit{a priori} is likely to introduce systematic bias into the inference, as in sections \ref{subsec:Manifold-Guided MCMC Sampling in Model Space} and \ref{subsec:Hierarchical Bayesian Inversion}.

\newpage
\subsection{Hierarchical inversion}
\label{subsec:Hierarchical inversion}

We assign uncertainty to the prior-strength hyperparameter $\alpha$ by defining a hyperprior $p(\alpha)$ (equation \ref{eq:hyperprior(alpha)}) within a hierarchical Bayesian algorithm (see section \ref{subsec:Hierarchical Bayesian Inversion}). To compare the effects of the different priors, we performed three sets of inversions: using only the uniform prior pdf (i.e., equation \ref{eq:total_prior_p(Mgen)Xuniform} with  $p(\mathbf{m}) = p_0(\mathbf{m})$), latent space inversion with three different GNNs (VAE, AAE and DAAE), which only allows solutions that lie exactly on $\mathcal{M}_{gen}$ (equation \ref{eq:latent_space_posterior_pdf_short}), and the hierarchical Bayesian approach, which incorporates the escaping priors (equation \ref{eq:hier_posterior}).

\begin{figure}
    \centering
    \includegraphics[width=1\linewidth]{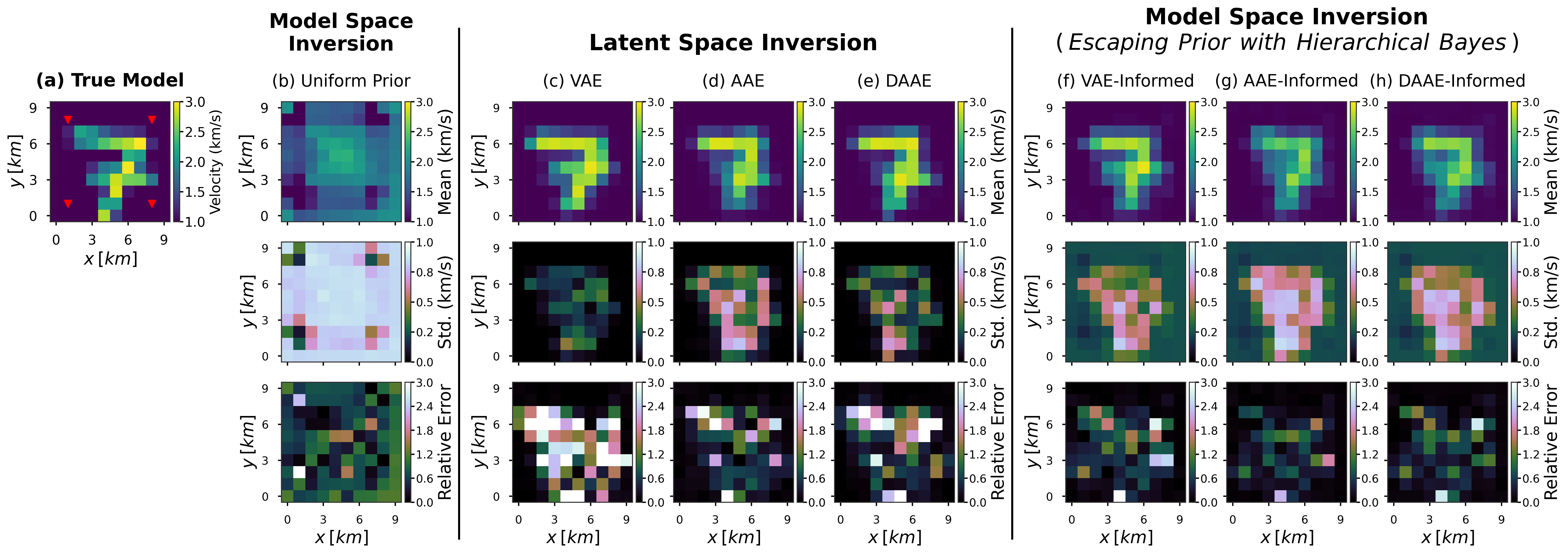}
    \caption{(a) A simple, ``7"-like true velocity model. The remaining panels show inversion results from different methods using travel-time observations between all non-identical pairs of 4 sources/receivers (red triangles). Subsequent columns compare (b) model space inversion using a uniform prior, (c-e) latent space inversion, and (f-h) model space hierarchical Bayesian inversion using the escaping prior. From top to bottom, each row represents the posterior mean, standard deviation, and relative error defined in the main text.}
    \label{fig:vel_7_lat_vs_hier}
\end{figure}

Figure \ref{fig:vel_7_lat_vs_hier} shows the inversion results. The posterior mean plots show that using a uniform prior results in a blurred posterior mean that fails to capture any structural features of the true model (Figure \ref{fig:vel_7_lat_vs_hier}b). The corresponding high posterior standard deviation across the entire model (middle row of panels) indicates significant uncertainty. We also show relative error models in the bottom row of Figure \ref{fig:vel_7_lat_vs_hier}b, defined at each point by the absolute difference between the true model and the posterior mean vectors, divided by the posterior standard deviation at each point.  This shows the number of standard deviations away from the mean is the true structure. The high standard deviations therefore lead to low relative errors.

The latent space inversion results in Figures \ref{fig:vel_7_lat_vs_hier}c,d,e have very low uncertainty due to the strong prior information parametrised by the GNNs. The VAE-based latent space inversion shows higher relative error due to its lower uncertainty than those from AAE and DAAE. Using hierarchical inversion (Figures \ref{fig:vel_7_lat_vs_hier}f,g,h) also recovers the ``7"-shaped mean structure present in the true model. So, using either latent space or hierarchical inversion provides posterior means that are more structurally coherent and correct than those obtained with a uniform prior. The posterior standard deviation is higher using hierarchical inversion than that from latent space inversion, resulting in lower relative errors. As expected, the escaping prior effectively constrains the solution and reduces uncertainty more than the relatively uninformative uniform prior.

\begin{figure}
    \centering
    \includegraphics[width=1\linewidth]{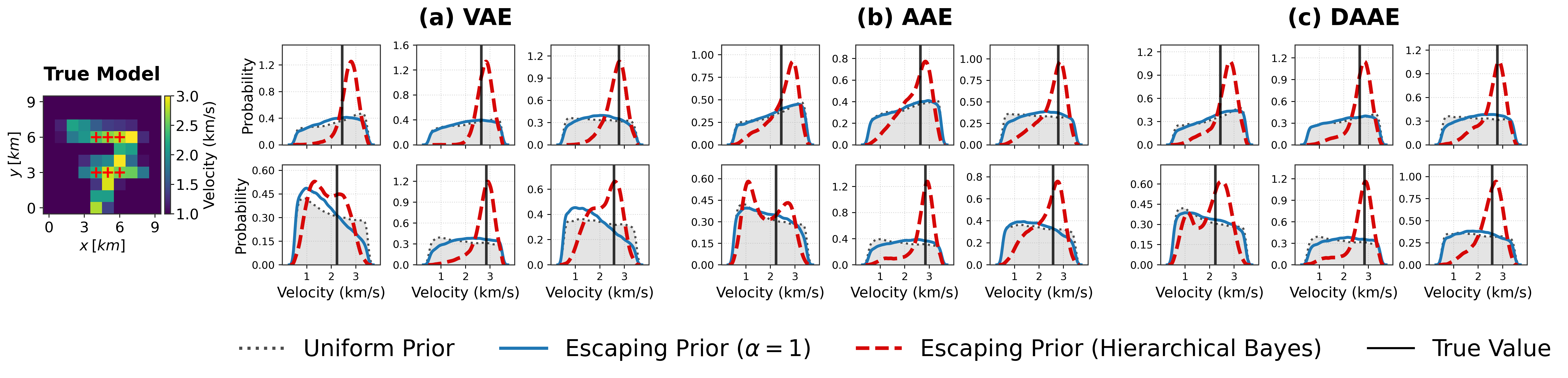}
    \caption{Marginal pdfs of the posterior MCMC samples at six points indicated by red crosses in the left panel. Results are obtained using different priors: a uniform prior (grey dotted curve), an escaping prior with $\alpha=1$ (blue curve), and an escaping prior with the hierarchical Bayesian algorithm (red dashed curve). Results are shown for three distinct GNNs: (a) VAE, (b) AAE, and (c) DAAE. Black vertical lines represent the true values of the selected model parameters.}
    \label{fig:pdf_7_2x2}
\end{figure}

Figure \ref{fig:pdf_7_2x2} shows the posterior marginal distributions of solutions at 6 different locations using a uniform prior, escaping priors with $\alpha =1$ from Figure \ref{fig:diff_alphas_7_Z_vels}a and escaping prior with hierarchical Bayes from Figure \ref{fig:vel_7_lat_vs_hier}. The hierarchical Bayesian inversion with escaping prior results show narrower marginal pdfs than those obtained with a uniform prior. They also show high posterior probability near the true value with the hierarchical Bayesian algorithm, but with $\alpha = 1$, the method introduces little or no additional prior information compared to a uniform prior, resulting in nearly identical posterior marginal pdfs.

\begin{figure}
    \centering
    \includegraphics[width=1\linewidth]{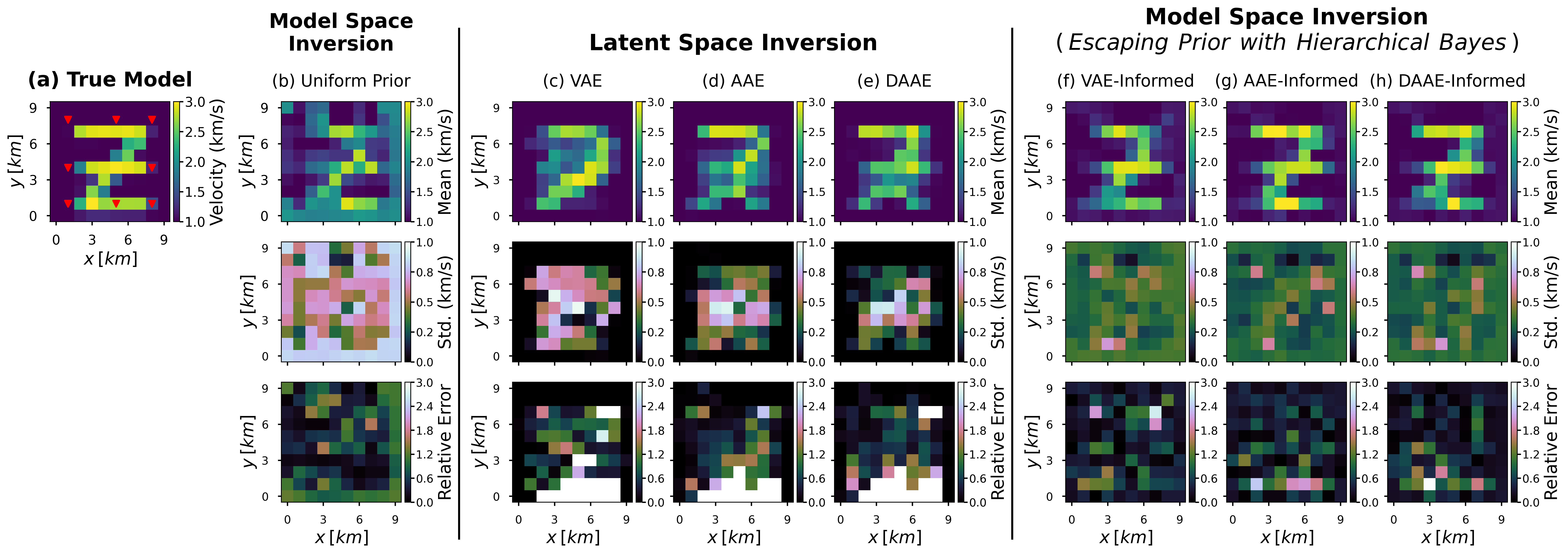}
    \caption{(a) A simple velocity model with ``$\mathcal{Z}$"-like shape. Other panels display inversion results from various methods, based on travel-time observations between all non-identical pairs of 8 sources/receivers (red triangles). Following columns compare (b) model space inversion with a uniform prior, (c-e) latent space inversion, and (f-h) model space hierarchical Bayesian inversion using the escaping prior. The rows, from top to bottom, show the posterior mean, standard deviation, and relative error defined in the main text.}
    \label{fig:vel_Z_lat_vs_hier}
\end{figure}

\begin{figure}
    \centering
    \includegraphics[width=1\linewidth]{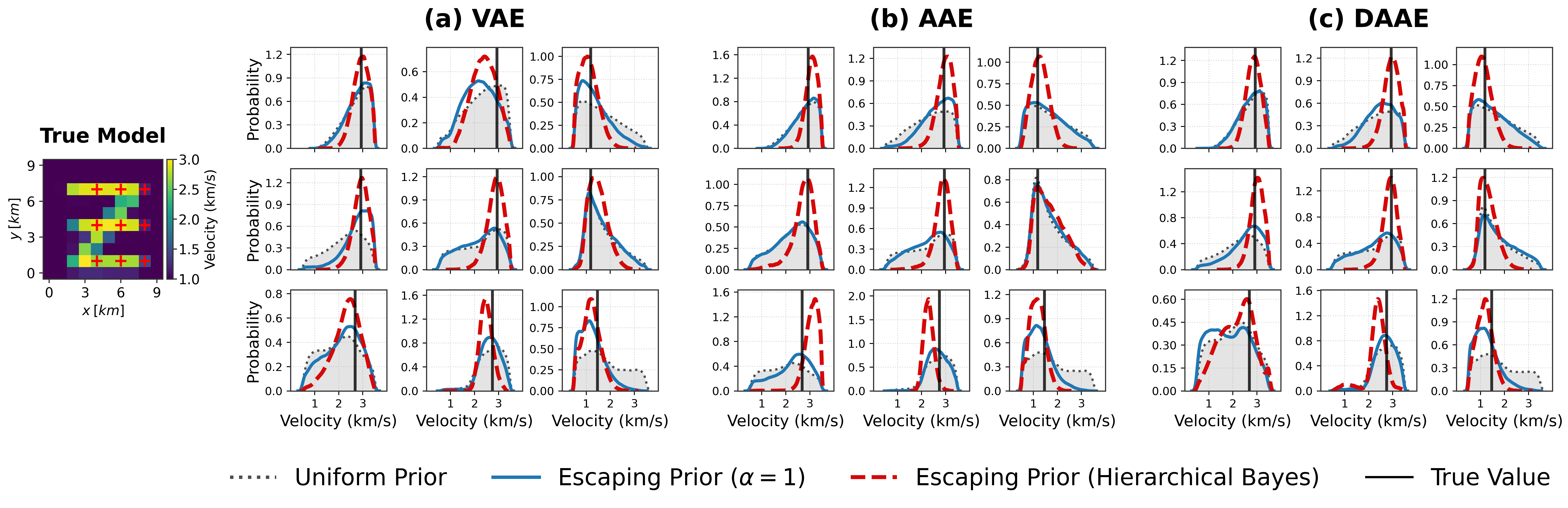}
    \caption{Marginal pdfs of the MCMC samples at six locations marked by the red crosses in the left panel. The results are obtained using a uniform prior (grey dotted curve), an escaping prior with $\alpha =1$ (blue curve), and an escaping prior with the hierarchical Bayesian algorithm (red dashed curve) for three different GNNs: (a) VAE, (b) AAE, and (c) DAAE. The true parameter values at the selected locations are indicated by the black vertical lines.}
    \label{fig:pdf_Z_3x3}
\end{figure}

In Figure \ref{fig:vel_Z_lat_vs_hier}, we show the inversion results for the case when the true model is ``$\mathcal{Z}$"-shaped, which shows that the latent space inversion fails to recover clear images of the off-manifold structural features of the true model (middle and bottom horizontal stroke), because of the incorrect prior information parametrised by GNNs trained using ``7"-shaped samples. Latent space inversion is unable to find the true solution because it lies outside of the generative manifold $\mathcal{M}_{gen}$. By contrast, the escaping prior with hierarchical Bayesian inversion successfully recovered the true model’s off-manifold structural features in the mean, with substantially lower uncertainty and relative error. The relative error of latent space inversions are higher along the bottom horizontal stroke of the true model, because this structural feature lies outside of $\mathcal{M}_{gen}$, illustrating the fundamental flaw in latent space inversion. The model space inversion with a uniform prior was unable to recover an interpretable ``$\mathcal{Z}$"-shaped structure of the true model and exhibited higher uncertainty than with an escaping prior. Therefore, a more adaptable escaping prior for model space inversion is again shown to be advantageous.

The posterior marginal pdfs for 9 locations, derived from the inversion results of the true model with ``$\mathcal{Z}$"-like structure (Figure \ref{fig:pdf_Z_3x3}), show that fixing the prior-strength at $\alpha=1$ (Figure \ref{fig:diff_alphas_7_Z_vels}) produced similar probability distributions to those obtained with a uniform prior (Figure \ref{fig:vel_Z_lat_vs_hier}b) in most locations. By contrast, the marginals from hierarchical Bayesian inversion with escaping prior (Figures \ref{fig:vel_Z_lat_vs_hier} f,g,h) showed narrower distributions and a higher probability of finding the true solution at most locations.

\subsection{Posterior distribution of $\alpha$}
\label{subsec:Posterior distribution of alpha}

\begin{figure}
    \centering
    \includegraphics[width=\columnwidth]{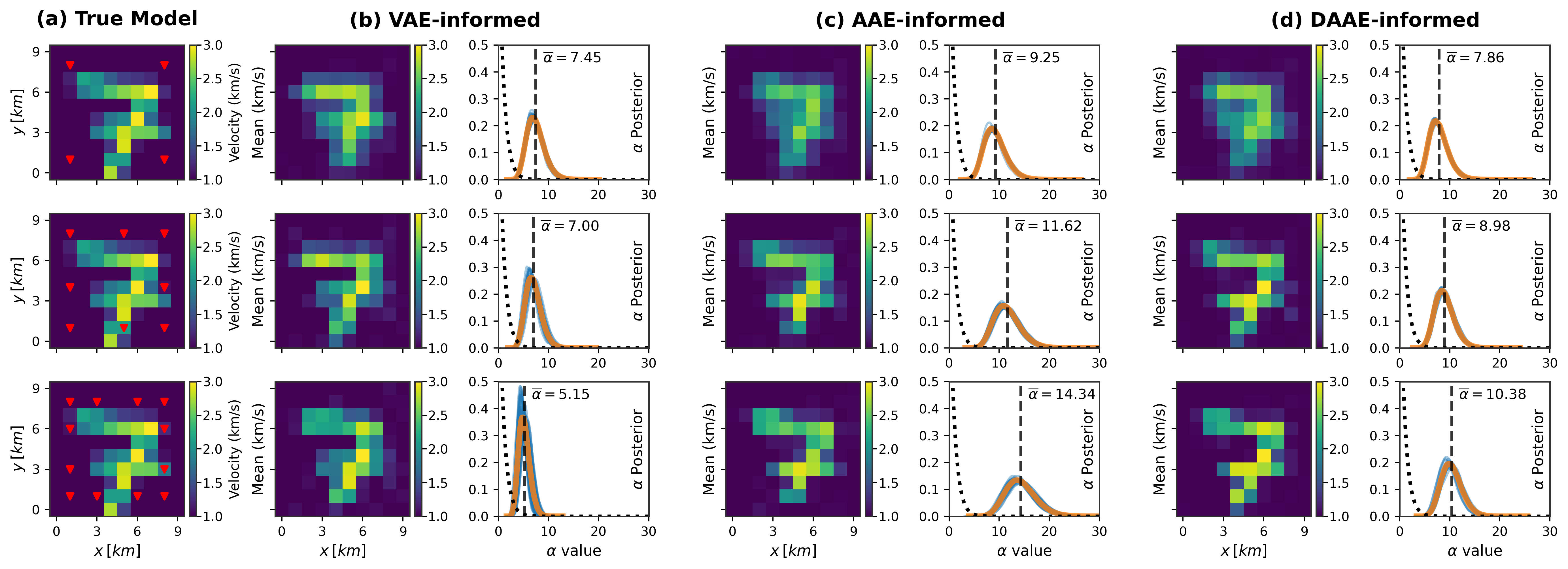}
    \caption{(a) ``7"-like true velocity model with data observations increasing from top to bottom row, where source/receivers are indicated by red triangles. Following columns compare (b) VAE, (c) AAE, and (d) DAAE-informed escaping priors showing posterior mean models (left panels) and $\alpha$ posterior distributions (right panels) from individual MCMC chains (blue curve), the aggregated posterior (orange curve), its mean $\overline{\alpha}$ (vertical black dashed line), and the $\alpha$ hyperprior (black dotted curve).}
    \label{fig:srcs_vs_alpha_7_K}

    \vspace{3em}
    
    \includegraphics[width=\columnwidth]{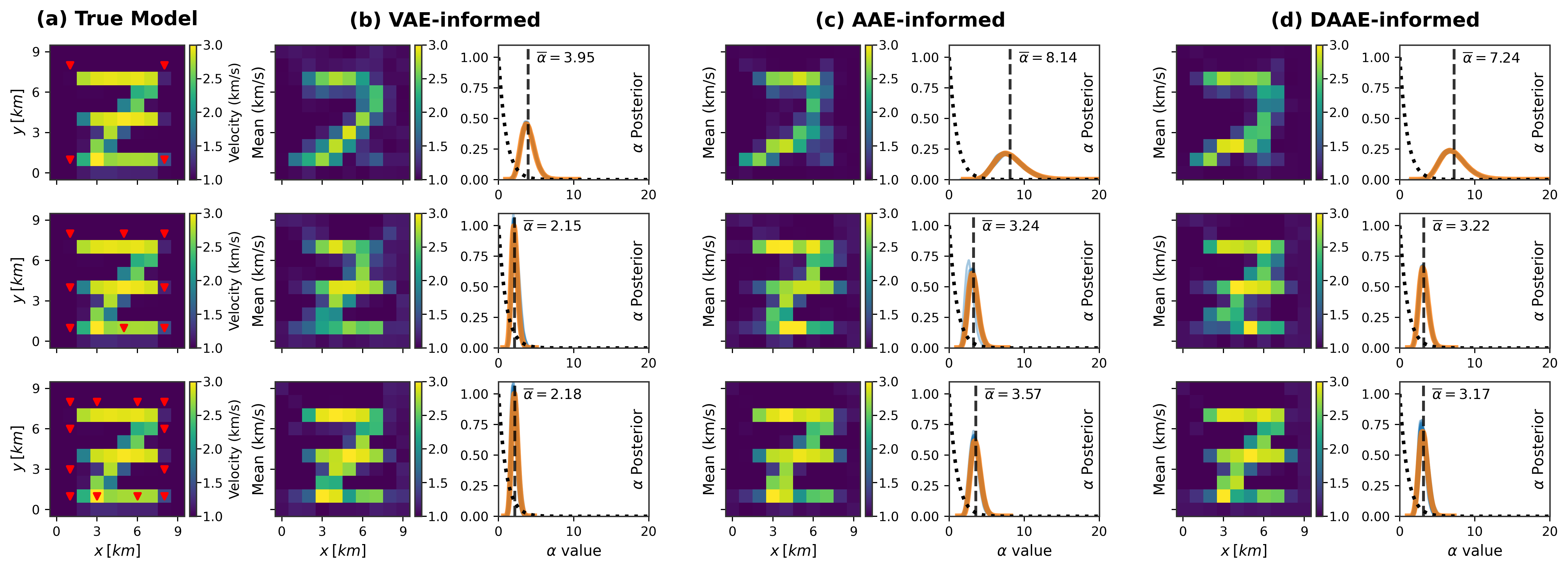}
    \caption{(a) ``$\mathcal{Z}$"-like velocity model, with the amount of data observations increasing from the top row to the bottom row (source/receivers marked in red). Subsequent columns compare escaping priors informed by (b) VAE, (c) AAE and (d) DAAE showing the posterior mean models (left panels) and the $ \alpha $ posterior distributions (right panels) with individual MCMC chains (blue curves) and the overall aggregated posterior (orange curve), the posterior mean $ \overline{\alpha} $ (vertical black dashed line) and the $ \alpha $ hyperprior (black dotted curve). Note the change in $\alpha$-axis scale compared to panels in Figure \ref{fig:srcs_vs_alpha_7_K}.}
    \label{fig:srcs_vs_alpha_Z_UN}
\end{figure}

In Figures \ref{fig:srcs_vs_alpha_7_K} and \ref{fig:srcs_vs_alpha_Z_UN} we also present the mean models and the posterior distribution of $\alpha$ values from hierarchical inversion results for different source/receiver configurations, using the escaping prior. In each case we set the decay rate $\gamma = 1$ and the lower bound to $\alpha_{lb} = 0.01$ for the hyperprior defined in equation \ref{eq:hyperprior(alpha)}. For the ``7"-shaped true velocity model in Figure \ref{fig:srcs_vs_alpha_7_K}, each escaping prior can recover the main structural features using only 6 independent travel-times from 4 source/receiver pairs. Adding more data observations through additional sources and receivers increases the quality of the mean models. Figures \ref{fig:srcs_vs_alpha_7_K}c,d show that adding more data to the inversion (by increasing the number of sources and receivers) also causes the $\alpha$ posterior to shift towards higher $\alpha$ values, clearly illustrated by the mean $\alpha$ values (black dashed vertical line) derived from the $\alpha$ posterior distributions (orange curves). This happens because the data show that the true solution is likely to lie on the $\mathcal{M}_{gen}$ embodied within the escaping priors using AAE and DAAE networks. However, the opposite is true when using the VAE-informed escaping prior. One possible explanation is that the true solution lies far from the manifold learned by a VAE. As shown in Figure \ref{fig:vel_7_lat_vs_hier}c, the result of latent space inversion based on the VAE shows a higher relative error, suggesting that the generative manifold learned by the VAE was less effective at capturing the structural variations (of ``7"-like samples) from the training set compared to the other GNNs. Adding more data from sources and receivers appears to contradict the poorly-learned prior, so the only way to find a consistent solution is to move further away from $\mathcal{M}_{gen}$ by relaxing the constraints of the VAE-informed escaping prior, achieved by sampling lower $\alpha$ values during inversion (Figure \ref{fig:srcs_vs_alpha_7_K}b).

Including additional observational data in the inversion for the ``$\mathcal{Z}$"-shaped true velocity model reveals distinct structural features and improved resolution in the mean models (Figure \ref{fig:srcs_vs_alpha_Z_UN}). Furthermore, the posterior distribution of $\alpha$ values tends to shift toward lower values as the number of sources and receivers grows. With more travel-time data, the hierarchical inversion prioritises minimising data misfit and relies less on the escaping prior. This occurs because the ``$\mathcal{Z}$"-shaped structure lies beyond the scope of $\mathcal{M}_{gen}$, and the only way to find it is by reducing the influence of the escaping prior. However, in Figures \ref{fig:srcs_vs_alpha_Z_UN}b-d, due to a sparse geometry with 4 sources/receivers, the seismic rays span only a limited region of the true model, completely ignoring the high-velocity middle horizontal stroke. Therefore, the limited information from the travel-time data causes the inversion process to incorporate more escaping prior information, restricting solutions to $\mathcal{M}_{gen}$ (consisting of ``7"-like shapes) by sampling higher $\alpha$ values and producing mean models that resemble a ``7"-like shape.

\section{Geological Velocity Models}
\label{subsec:Geological Velocity Models}

\begin{figure}
    \centering
    \includegraphics[width=\columnwidth]{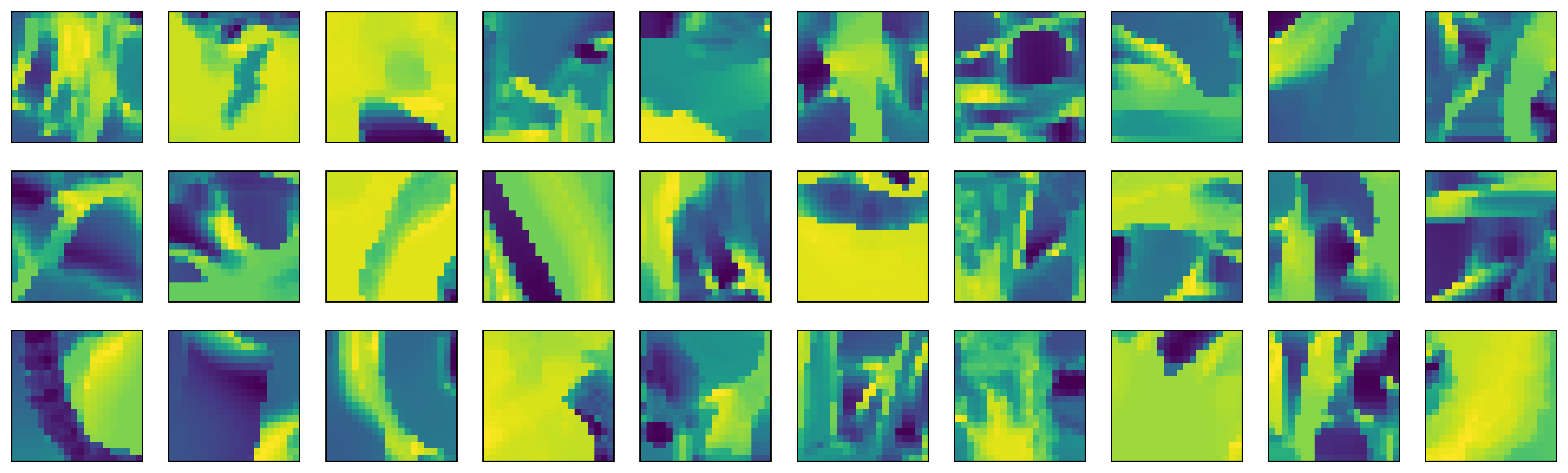}
    \caption{Example velocity slices from horizontal cross-sections of SedSimple generated 3D velocity models, also used as training samples for GNNs.}
    \label{fig:GNN_training_images_horizontal}
    
    \vspace{1em}
    
    \includegraphics[width=\columnwidth]{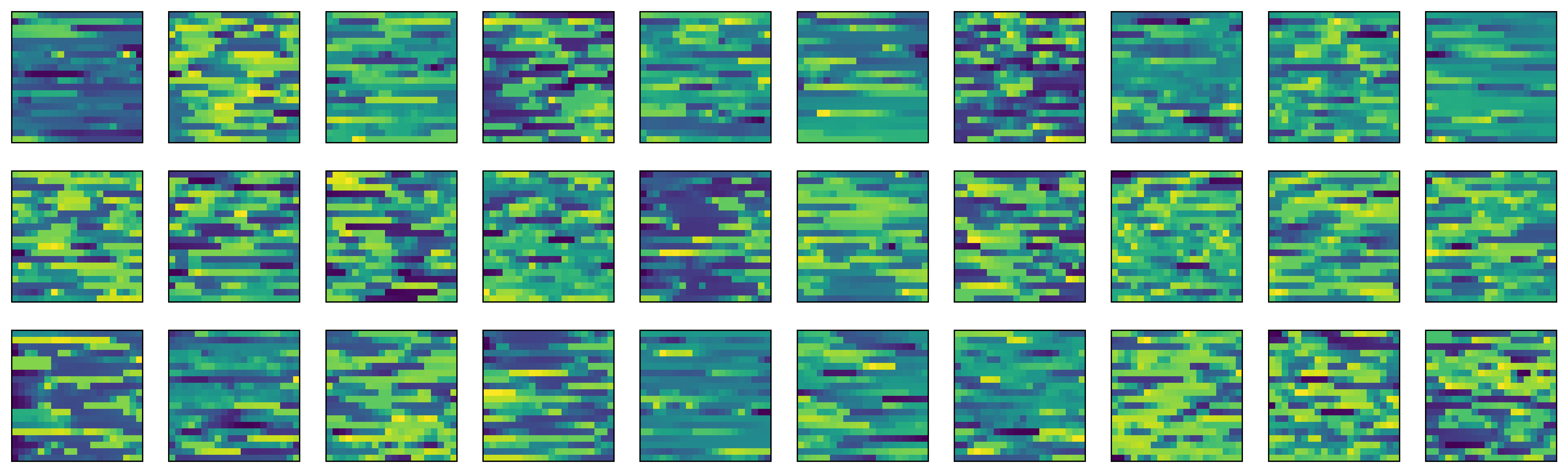}
    \caption{Vertical velocity slices from SedSimple, used as training samples for the GNNs.}
    \label{fig:GNN_training_images_vertical}
\end{figure}

In a third set of tests, we perform a higher-dimensional tomography experiment using true models with more realistic geological structures. We simulate geological processes such as deposition and erosion using SedSimple \citep{tetzlaff_stratigraphic_2023}, a geological process-modelling software, to generate 3D models of these processes. These models are then converted to 3D velocity models using the rock-physics relations (from \cite{strutz_variational_2024}). Horizontal and vertical cross-sections are extracted and further divided into $20\times20$ pixel patches to serve as training samples for the GNNs. By training the GNNs, we parameterise the geological information embedded in geological process models to impose geological realism in our solutions. Figures \ref{fig:GNN_training_images_horizontal} and \ref{fig:GNN_training_images_vertical} show some of the training samples used in our case. More examples on geological process modelling and parameterising prior information from them using GNNs can be found in \cite{bloem_introducing_2023}.

The extracted horizontal and vertical velocity cross-sections from the 3D velocity models mostly range from 1.1 km/s to 3.4 km/s; before treating them as training samples, they were normalised to a velocity range of [1, 3] km/s.  Normalising the training set prevents numerical instability and speeds up training - raw pixel/parameter values are often too large for the small, randomly initialised weights of a neural network, leading to gradients that explode or vanish during training \citep{ioffe_batch_2015}. By scaling model parameters to a uniform range, we create a more symmetrical loss landscape while training the GNNs.

In the following geological cases, the inversion is performed on 40 independent MCMC chains, each run in parallel and drawing nearly 2 million samples. Each chain was initialised from a different randomly sampled velocity model, with parameters drawn from a uniform distribution between 0.5 and 3.5 km/s. We compare the results from three sets of inversions: using only the uniform prior pdf (i.e., equation \ref{eq:total_prior_p(Mgen)Xuniform} with  $p(\mathbf{m}) = p_0(\mathbf{m})$), latent space inversion with three different GNNs (equation \ref{eq:latent_space_posterior_pdf_short}); and the model space inversion with the escaping prior using the hierarchical Bayesian algorithm (equation \ref{eq:hier_posterior}).

\subsection{Horizontal cross-sectional escaping priors}
\label{subsec:Horizontal cross-sectional escaping priors}

\begin{figure}
    \centering
    \includegraphics[width=1\columnwidth]{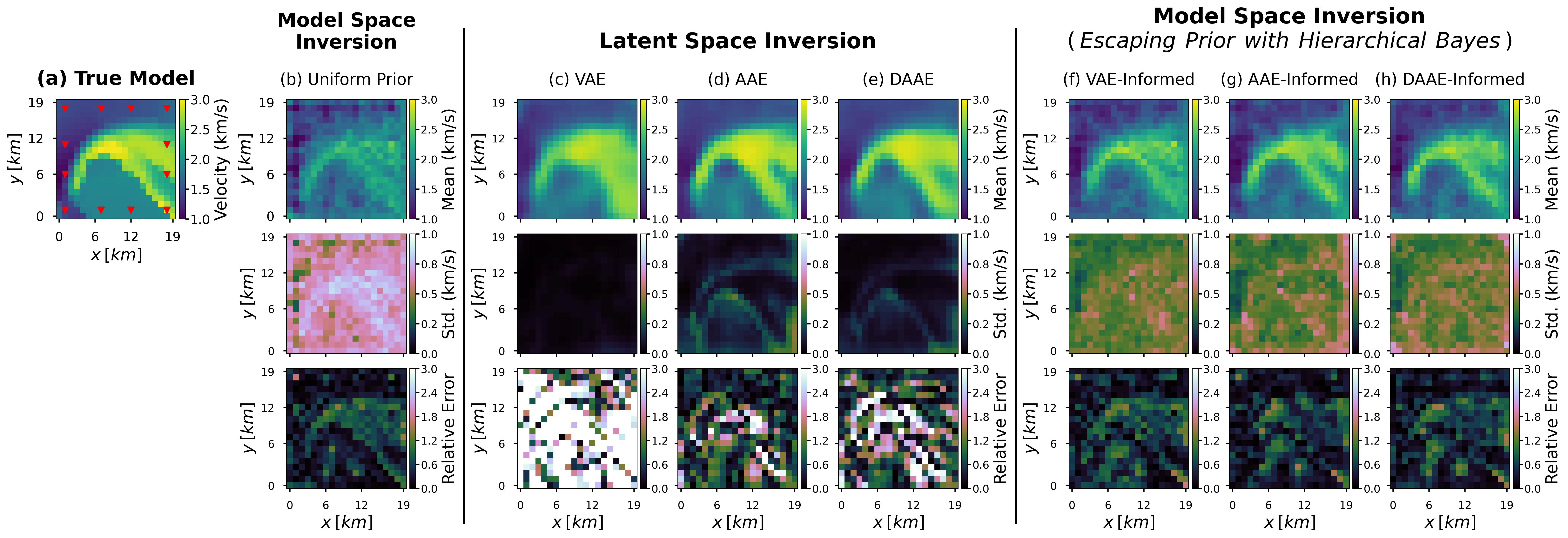}
    \caption{(a) A more geological true velocity model with 12 sources/receivers (red triangles). Each column represents different inversion methods with: (b) model space inversion with a uniform prior, (c-e) latent space inversion and (f-h) model space inversion with an escaping prior using a hierarchical Bayesian algorithm. Each row, from top to bottom, shows the posterior mean, standard deviation and relative error defined in the main text.}
    \label{fig:vel_geoKN_lat_vs_hier}

    \vspace{3em}
    
    \includegraphics[width=1\columnwidth]{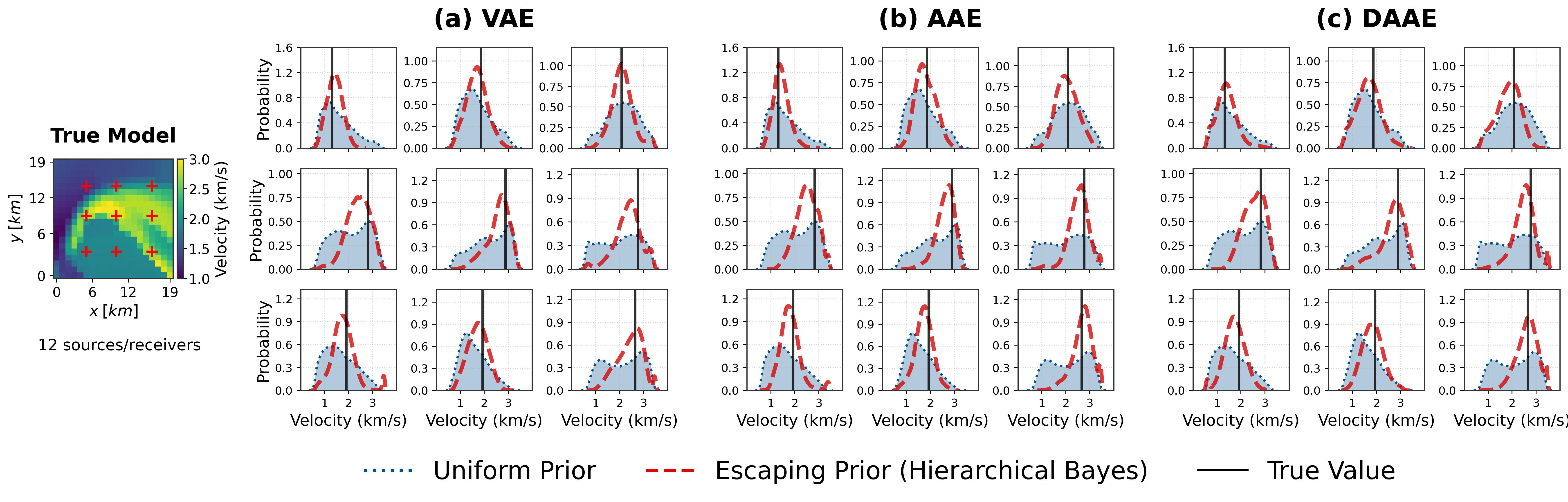}
    \caption{Marginal pdfs from posterior MCMC samples incorporating a non-informative uniform prior (grey dotted curve) comparing with (a) VAE, (b) AAE, and (c) DAAE-informed escaping prior (red dashed curve), the true value marked by black vertical lines.}
    \label{fig:pdf_geoKN_12x12}
\end{figure}

Figure \ref{fig:vel_geoKN_lat_vs_hier}a shows a true model with a reasonably geologically realistic velocity structure. We use a sparse geometry with 12 sources and receivers for travel-time observations. The true velocity model in this case is generated using SedSimple with another simulation process and is therefore not included in the GNN training set. Using the observed travel-time data of a more geologically reasonable velocity model, the model space inversion with a uniform prior (Figure \ref{fig:vel_geoKN_lat_vs_hier}a) recovers a faint and less clear posterior mean velocity model. The latent space inversion results produce overly smoothed approximations that capture the large-scale features of the true model's geological structure at an apparent lower resolution than the hierarchical inversion. The VAE-based latent space inversion (Figure \ref{fig:vel_geoKN_lat_vs_hier}c) results in a very confident posterior mean with misleadingly low uncertainty, leading to very high posterior relative error throughout. 

In Figures \ref{fig:vel_geoKN_lat_vs_hier}d,e, the AAE and DAAE-based latent space inversion produce solutions that are notably better than the VAE case. However, they still show high relative error in some areas, about 3 standard deviations from the true model, because the uncertainty estimate is low. Therefore, despite providing smooth mean velocity structures, the results of AAE and DAAE-based latent space inversion are overly confident due to the near-zero standard deviation of the posterior uncertainty. For the VAE-based latent space inversion, the reason for the overly smoothed and inaccurate posterior mean velocity model, with a high relative error, is that the VAE learned only a limited subset of the structural features of the training samples, resulting in a very restricted generative manifold $\mathcal{M}_{gen}$, with the true model beyond its reach. 

Despite the true solution lying outside of manifold $\mathcal{M}_{gen}$, the VAE-informed escaping prior (Figure \ref{fig:vel_geoKN_lat_vs_hier}f) effectively constrains the hierarchical solution, producing a structurally consistent posterior mean model with lower uncertainty than using a uniform prior and achieving very low relative error. The AAE- and DAAE-informed hierarchical inversions show lower posterior uncertainty at certain layer boundaries, while maintaining a similar level of relative error to results from the uniform prior. Additionally, the posterior mean model is more interpretable, less ambiguous and offers better resolution.

\begin{figure}
    \centering
    \includegraphics[width=1\linewidth]{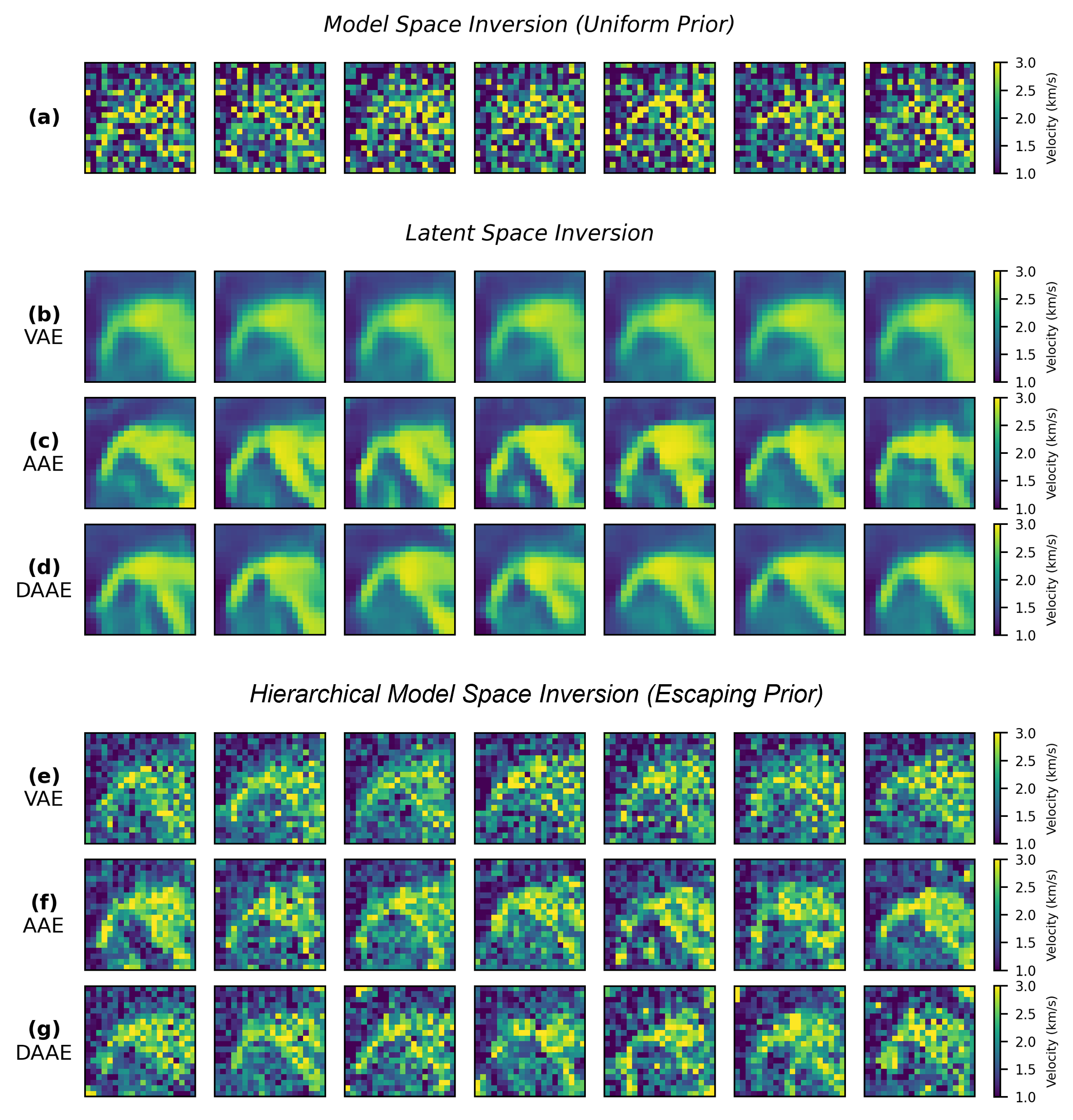}
    \caption{Posterior samples from independent MCMC chains of the velocity model shown in Figure \ref{fig:vel_geoKN_lat_vs_hier}, with (a) a uniform prior, (b-d) from latent space inversion, and (e-g) using the escaping prior.}
    \label{fig:12x12_mcmc_chain_samples_KN}
\end{figure}

We show the posterior marginal pdfs for 9 different locations (Figure \ref{fig:pdf_geoKN_12x12}) from the inversion results demonstrated in Figure \ref{fig:vel_geoKN_lat_vs_hier}. The marginal pdfs from the hierarchical Bayesian model space inversion with the escaping prior (red dashed curve) indicate a higher probability near the true value (black vertical line) and generally show lower variance compared to those using a uniform prior (blue dotted curve). The marginal distributions resulting from the latent space inversions are not displayed because their very low standard deviations make them overly narrow and sharply peaked, which makes them difficult to visualise meaningfully on similar axis scales. The true Earth model shown in Figure \ref{fig:vel_geoKN_lat_vs_hier}a and the training samples for the GNNs (Figure \ref{fig:GNN_training_images_horizontal}) are both generated with SedSimple and share similar structural features and horizontal correlations. During training, the AAE and DAAE captured these structural features more efficiently than the VAE, which explains why latent space inversion with the AAE and DAAE produces better results than with the VAE. Despite producing erroneous results in latent space inversion, using a VAE-informed escaping prior shows significant improvements in marginal plots. Nevertheless, if the true Earth model is on the generative manifold defined by the GNNs (that we have no information \textit{a priori}), it should be recoverable through both hierarchical inversion with the escaping prior and latent space inversion, due to the strong geological prior information from $\mathcal{M}_{gen}$.

Figure \ref{fig:12x12_mcmc_chain_samples_KN}a shows that samples produced by a non-informative uniform prior are noisy and lack a clearly visible structure. The latent space inversion velocity model samples (Figures \ref{fig:12x12_mcmc_chain_samples_KN}b-d) are overly smoothed and exhibit very little variation. The samples produced from VAE-based latent space inversion show no variation throughout, as shown in Figure \ref{fig:12x12_mcmc_chain_samples_KN}b. This lack of variation explains why the posterior standard deviation is close to zero in Figure \ref{fig:vel_geoKN_lat_vs_hier}c. Additionally, we observe that samples from latent space inversion have very low pixel-level random noise, because they are confined to the low-dimensional latent space; the generator produces structured and spatially coherent samples learned during training, thereby fundamentally precluding random pixel-level noise. The samples from the hierarchical inversion with escaping prior distributions (Figures \ref{fig:12x12_mcmc_chain_samples_KN}e-g) show well-resolved structural features with continuous structural boundaries, and clearly demonstrate that the escaping prior effectively constrains the inversion solutions more than the uniform prior. Layer boundaries are also visible in samples derived from the AAE and DAAE-informed escaping priors (Figures \ref{fig:12x12_mcmc_chain_samples_KN}f,g). The VAE, when used for latent space inversion, produces incorrect results. However, as a VAE-informed escaping prior, it offers better solution constraints than a uniform prior and latent space inversion, and produces consistent samples across different MCMC chains.

\newpage
\subsection{Vertical cross-sectional escaping priors}
\label{subsec:Vertical cross-sectional escaping priors}

\begin{figure}
    \centering
    \includegraphics[width=1\columnwidth]{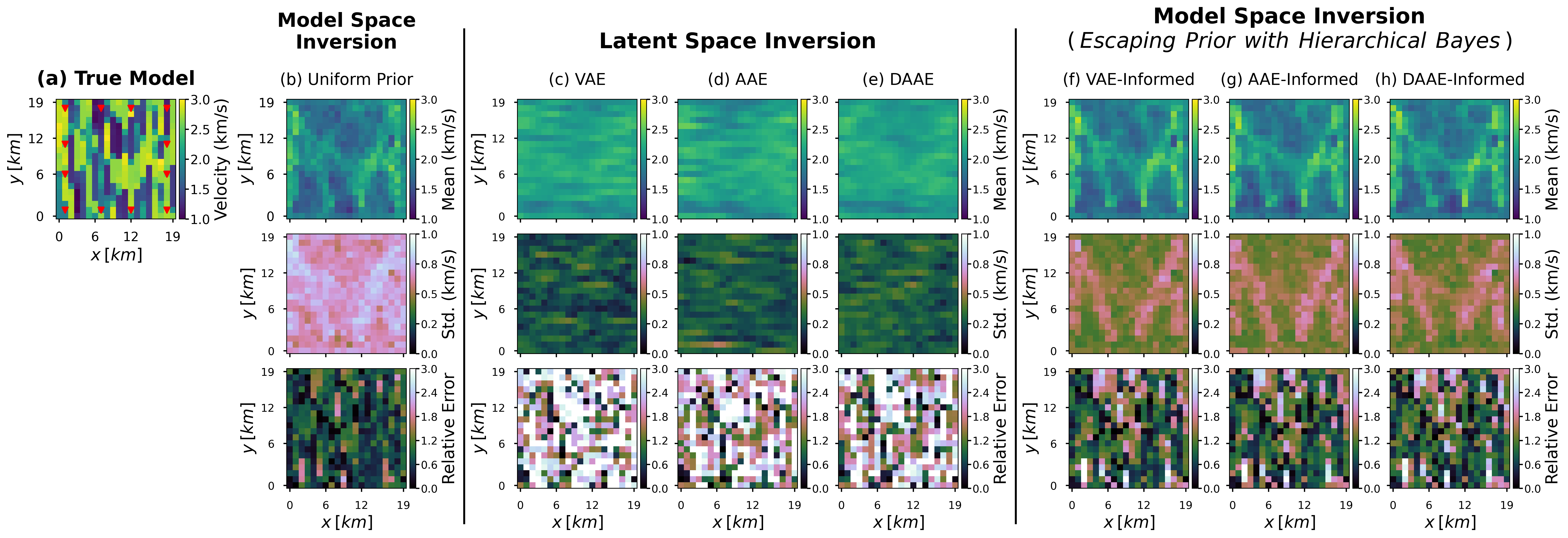}
    \caption{Inversion results obtained from travel-time data observations with (a) 12 sources/receivers (red triangles) of a true velocity model representing vertical strata. Rows display the posterior mean, standard deviation, and relative error defined in the main text. Columns denote the inversion method: (b) model space inversion with a uniform prior, (c–e) latent space inversion, and (f–h) hierarchical model space inversion with an escaping prior.}
    \label{fig:vel_geoUNKN_lat_vs_hier_12x12_Z8}

    \vspace{3em}
    
    \includegraphics[width=1\columnwidth]{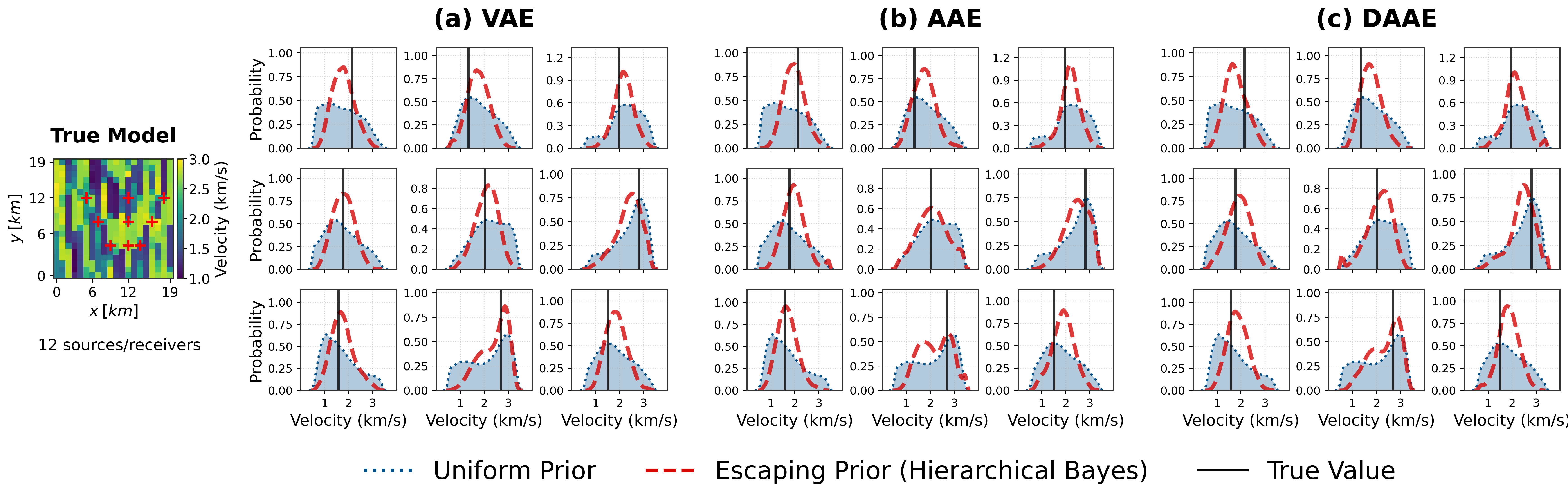}
    \caption{Marginal pdfs of the posterior MCMC samples using (a) VAE, (b) AAE, and (c) DAAE-informed escaping prior (red dashed curve), each compared with a uniform prior (grey dotted curve) across 9 locations of the true model, with true values shown as black vertical lines.}
    \label{fig:pdf_geoUNKN_12x12_Z8}
\end{figure}

Figure \ref{fig:vel_geoUNKN_lat_vs_hier_12x12_Z8} shows an example in which the true Earth model is far from the generative manifold $\mathcal{M}_{gen}$. Figure \ref{fig:vel_geoUNKN_lat_vs_hier_12x12_Z8}a shows a velocity model that includes structural features of a vertically stratified layer to serve as the true Earth model in this case. However, we now employ GNNs that can only generate examples resembling the structural features of horizontally stratified layers that were used as their training set (Figure \ref{fig:GNN_training_images_vertical}).

Figure \ref{fig:vel_geoUNKN_lat_vs_hier_12x12_Z8}b shows that model space inversion with a uniform prior was ineffective at constraining the inversion, resulting in a posterior mean with high uncertainty and only faint indications of vertical strata. In contrast, the latent space inversion produced blurry posterior means that only displayed horizontal strata as expected, showing high relative errors despite high confidence (low posterior standard deviations) (Figures \ref{fig:vel_geoUNKN_lat_vs_hier_12x12_Z8}c-e). By using the escaping prior in the hierarchical Bayesian approach (Figures \ref{fig:vel_geoUNKN_lat_vs_hier_12x12_Z8}f-h), the resulting posterior mean models appear more coherent and exhibit clearer, more distinct structures compared to those from the uniform prior. The associated uncertainty is lower and more realistic than in the latent space inversion. Overall, the inversion results with the escaping prior exhibit substantially reduced relative errors compared with latent space inversion - similar in magnitude to those obtained with the uniform prior.

The marginals from the inversion result of 9 different locations, using uniform prior and escaping priors in Figure \ref{fig:vel_geoUNKN_lat_vs_hier_12x12_Z8} are shown in Figure \ref{fig:pdf_geoUNKN_12x12_Z8}. Using the uniform prior results in high variance in posterior marginals at most locations. All the escaping priors consistently display low variance and high probability near the true solution across most locations.

\begin{figure}
    \centering
    \includegraphics[width=1\linewidth]{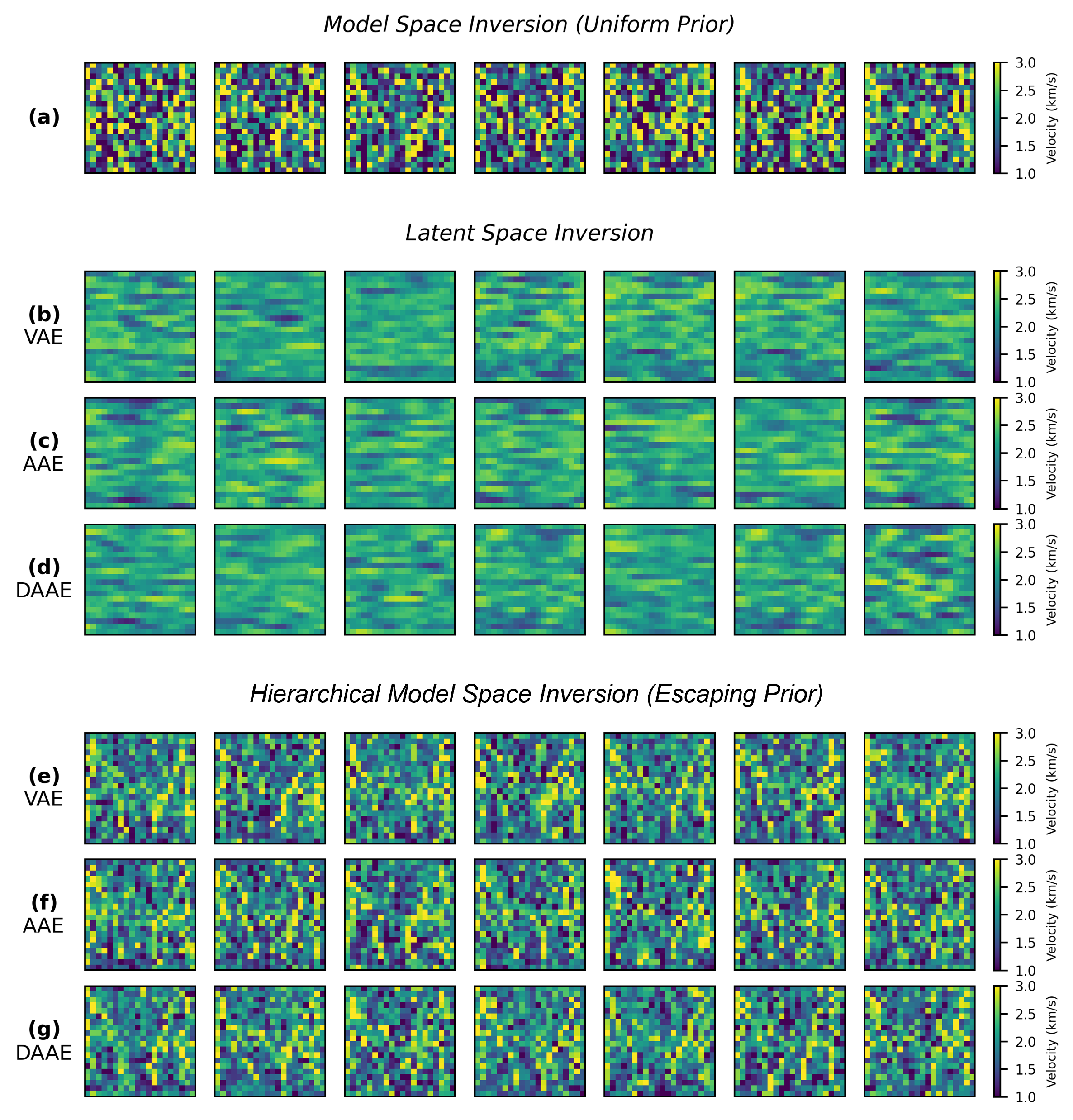}
    \caption{Posterior velocity samples drawn from independent MCMC chains based on Figure \ref{fig:vel_geoUNKN_lat_vs_hier_12x12_Z8}, with (a) a uniform prior, (b–d) latent space inversion, and (e–g) the escaping prior.}
    \label{fig:12x12_mcmc_chain_sediment_models_Z8}
\end{figure}

With a uniform prior, the inversion is primarily constrained by the likelihood, leading to high uncertainty when the data are sparse. However, the escaping prior imposes additional constraints, thereby reducing the uncertainty. Because the true model (with vertical strata) lies outside the generative manifold (family of GNN-generated structures of horizontal strata), latent space inversion fails to recover it. The samples from different MCMC chains show that the latent space inversion produced samples are horizontally stratified (Figures \ref{fig:12x12_mcmc_chain_sediment_models_Z8}b-d), whereas the true Earth model is vertically stratified. The Figures \ref{fig:12x12_mcmc_chain_sediment_models_Z8}a and \ref{fig:12x12_mcmc_chain_sediment_models_Z8}e-g show the difference between inversion samples using a uniform prior and escaping priors, respectively. If we examine the latter closely, we notice differences that result in a structurally consistent posterior mean model with lower uncertainty than a uniform prior, while achieving comparatively low relative error (Figure \ref{fig:vel_geoUNKN_lat_vs_hier_12x12_Z8}).

\begin{figure}
    \centering
    \includegraphics[width=0.9\linewidth]{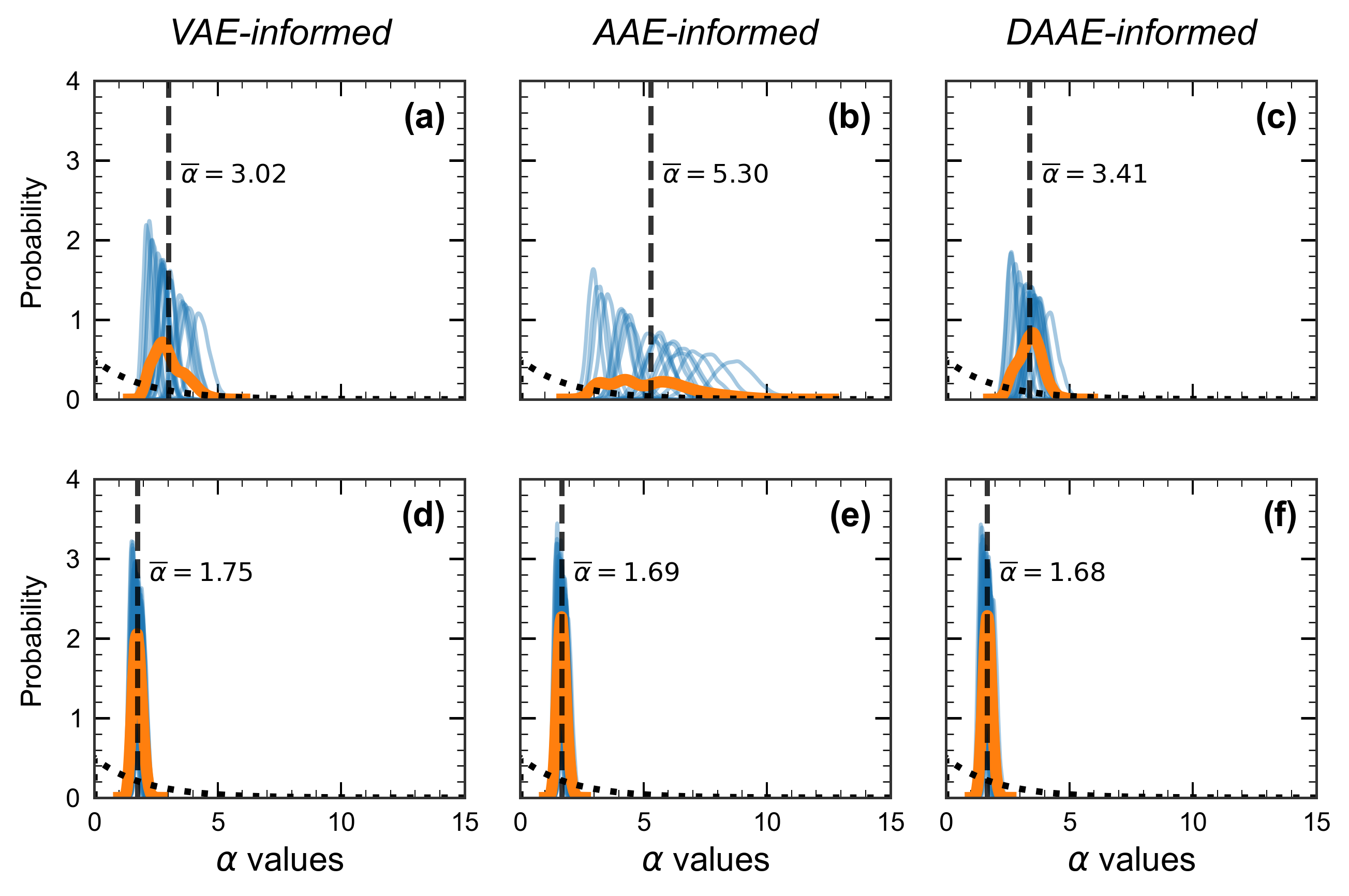}
    \caption{Marginal posterior pdfs of $\alpha$ derived from VAE, AAE, and DAAE-informed escaping priors. The top row (a-c) corresponds to Figure \ref{fig:vel_geoKN_lat_vs_hier}, and the bottom row (d-f) corresponds to Figure \ref{fig:vel_geoUNKN_lat_vs_hier_12x12_Z8}. Each panel displays individual MCMC chains (blue curve), the aggregated posterior (orange curve), and its mean $\overline{\alpha}$ (vertical black dashed line). In each case, we set the decay rate $\gamma = 0.5$ and the lower bound $\alpha_{lb} = 0.01$ for the hyperprior (black dotted curve) as defined in equation \ref{eq:hyperprior(alpha)}.}
    \label{fig:posterior_alpha_geos_K-UN}
\end{figure}

In Figures \ref{fig:posterior_alpha_geos_K-UN}a-c, which shows results corresponding to the experiments in Figure \ref{fig:vel_geoKN_lat_vs_hier}, the posterior distribution of $\alpha$ from each MCMC chain (blue curves) indicates that the $\alpha$ values tend to be higher than in Figures \ref{fig:posterior_alpha_geos_K-UN}d-f (which correspond to the experiment in Figure \ref{fig:vel_geoUNKN_lat_vs_hier_12x12_Z8}). This is again because, in the former case, the true Earth model lies on or near the generative manifold $\mathcal{M}_{gen}$ learned by the GNNs. In this case, the hierarchical inversion algorithm makes the escaping priors variance narrower (more informative) around $\mathcal{M}_{gen}$ by sampling high $\alpha$ values. When the true solution (vertically stratified layers) is distant from the $\mathcal{M}_{gen}$ (when it consists of horizontally stratified layers), the $\alpha$ posteriors (Figures \ref{fig:posterior_alpha_geos_K-UN}d-f) tend to shift towards the lower bound of the hyperprior, leading to the sampling of lower $\alpha$ values. Sampling lower $\alpha$ values broadens the escaping prior pdf, thereby introducing more freedom to sample solutions beyond the strict boundaries of $\mathcal{M}_{gen}$.

\subsection{Posterior correlations}
\label{subsec:Posterior Correlation}

In this section, we compare the posterior correlation coefficients among the model parameters estimated using a uniform prior, latent space inversion, and the escaping prior. The correlation matrices derived from a uniform prior in model space, including correlations from the training samples used for GNN training and from samples produced by the GNNs, are presented in Appendix \ref{app:prior_correlations}. A specific $7 \times 7$ grid within the model samples obtained from the inversion (resulting in 49 model parameters) is selected to compute the correlation coefficient matrices. Note that the colour scale is chosen to highlight subtler off-diagonal patterns and saturates for some correlation matrix elements. 

\begin{figure}
    \centering
    \includegraphics[width=1\linewidth]{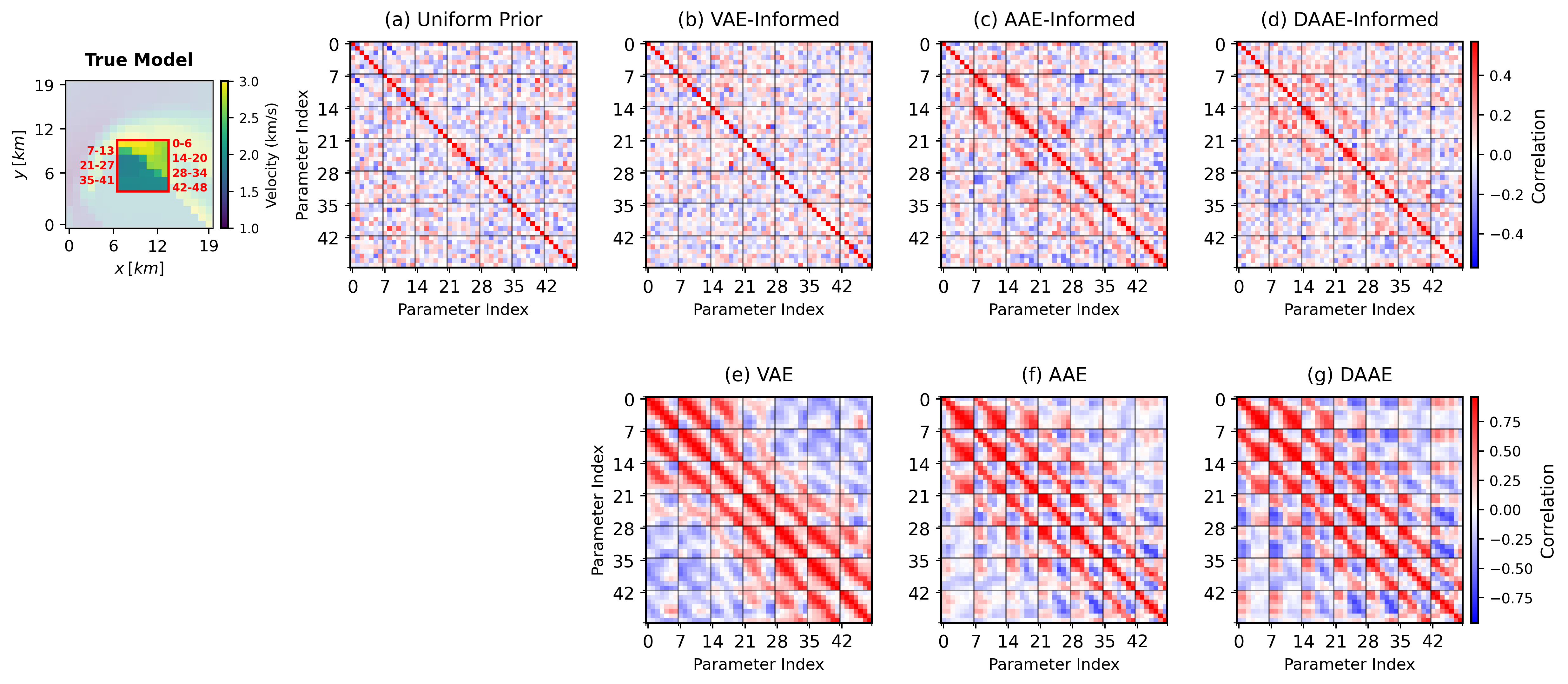}
    \caption{Posterior correlation matrices are shown from the results in Figure \ref{fig:vel_geoKN_lat_vs_hier}, using (a) a uniform prior, (b-d) escaping priors, and (e-g) latent space inversion. The true model features a red rectangle that highlights the $7\times7$ correlation grid, with the corresponding grid parameters written on both sides. The colour scale is adjusted to highlight the off-diagonal features.}
    \label{fig:12x12_posterior_cor_KN}
\end{figure}

In Figure \ref{fig:12x12_posterior_cor_KN}a, we observe very faint but visible negative correlations in the off-diagonals (both sides of the main diagonal) corresponding to cells offset vertically from one-another, for the case with a uniform prior. The overall correlation is mostly noisy, with a sharp red main diagonal. In VAE-informed hierarchical inversion samples (Figure \ref{fig:12x12_posterior_cor_KN}b), there is a faint but visible positive correlation between vertically adjacent cells (in the same off-diagonal elements that were negative for the uniform prior). The AAE-informed hierarchical inversion samples (Figure \ref{fig:12x12_posterior_cor_KN}c) demonstrate clear positive correlations between vertically neighbouring cells and horizontally neighbouring cells (the first off-diagonals, just below and above the main diagonal). The $7 \times 7$ grid contains both high and low-velocity zones; for this reason, we also observe negative correlations in some regions corresponding to trade-offs at high-contrast boundaries that produce uncertainty loops \citep{galetti_uncertainty_2015}. This pattern is also visible in the DAAE-informed hierarchical inversion samples (Figure \ref{fig:12x12_posterior_cor_KN}d), confirming that we successfully induced the correlations learned by the GNNs through the escaping priors in these experiments.

The posterior correlation matrices from the latent space inversion exhibit strong positive correlations between neighbouring cells both horizontally and vertically (Figures \ref{fig:12x12_posterior_cor_KN}e-g). Additionally, negative correlations are observed at the transition zones between high- and low-velocity regions in the inversion samples. These correlation matrices are more distinct and lack the ``speckle'' of non-zero background correlations, since latent space inversion samples are passed through the generator or generator to generate samples in model space. In contrast, the inversion results with escaping prior are sampled in model space using a MCMC sampler, which introduces some spurious correlations.

\begin{figure}
    \centering
    \includegraphics[width=1\linewidth]{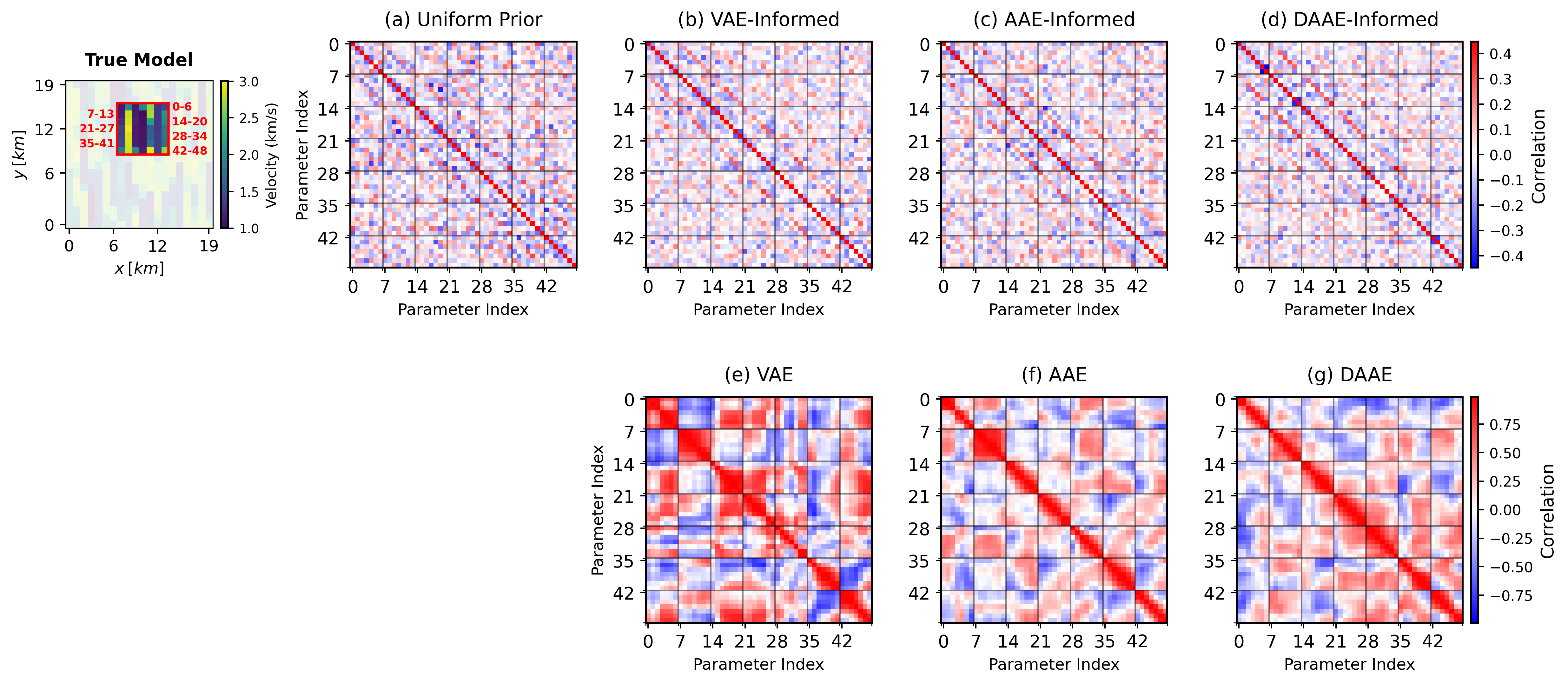}
    \caption{Posterior correlation matrices corresponding to the inversions in Figure \ref{fig:vel_geoUNKN_lat_vs_hier_12x12_Z8}, using (a) a uniform prior, (b–d) escaping priors, and (e–g) latent space inversion. A red rectangle on the true model highlights the $7\times7$ correlation grid, with its parameters indicated on the margins. The colour scale is set to show subtle off-diagonal features.}
    \label{fig:12x12_posterior_cor_sediment_Z8_2}
\end{figure}

The correlation matrices in Figure \ref{fig:12x12_posterior_cor_sediment_Z8_2} are obtained from the inversion results shown in Figure \ref{fig:12x12_mcmc_chain_sediment_models_Z8}. The uniform prior in Figure \ref{fig:12x12_posterior_cor_sediment_Z8_2}a indicates significant negative correlations among neighbouring cells (the first and seventh off-diagonal elements). In imaging inverse problems, the data typically impose negative correlations between neighbouring cells, since these parameters can trade off against each other while still fitting the data \citep{galetti_uncertainty_2015, zhang_bayesian_2021, zhao_physically_2024}. Figures \ref{fig:12x12_posterior_cor_sediment_Z8_2}b-d shows the escaping priors, as we expected, impose positive correlations. Interestingly, there are clear positive correlations between vertically adjacent cells (seventh off-diagonals) and noticeable negative correlations between horizontally adjacent cells (first off-diagonals, just below and above the main diagonal). This pattern arises because the true model consists of vertically stratified sediment layers with vertical correlations. However, the correlation patterns from the latent space inversion (Figures \ref{fig:12x12_posterior_cor_sediment_Z8_2}e-g) clearly indicate strong positive horizontal correlations, forming repeating square-like patterns along the main diagonal within a horizontal layer. Since the GNNs were trained on image samples of purely horizontal strata, the latent space inversion failed to produce samples that show vertical correlations.


\section{Discussion}
\label{sec:Discussion}

In practice, the true Earth is infinitely complex, with heterogeneity at every scale, and our prior knowledge is limited. The generative manifold $\mathcal{M}_{gen}$ (defined by the GNNs)  can only capture and parameterise a finite-dimensional subset of Earth's true heterogeneity due to training errors, limited latent parameters, or incorrect geological prior information \citep{bloem_introducing_2023}. For that reason, the true Earth almost always lies beyond $\mathcal{M}_{gen}$. Finding a solution strictly on $\mathcal{M}_{gen}$ using latent space inversion imposes prior information that is flawed and thus produces incorrect results. However, our approach shows that, unlike traditional latent space inversion, inversion solutions with the escaping prior are not confined to $\mathcal{M}_{gen}$, enabling sampling of model parameters beyond its strict limits.

We also demonstrate that when the true solution is far from the generative manifold $\mathcal{M}_{gen}$, decreasing $\alpha$ (the prior-strength hyperparameter), which implies weak prior information, is helpful. In contrast, if the true solution is located near $\mathcal{M}_{gen}$, increasing the prior-strength (implying strong prior information) successfully constrains the posterior distribution. The hierarchical Bayesian inversion algorithm simultaneously estimates the model parameters and an appropriate distribution for $\alpha$. 

\cite{mosegaard_inconsistency_2024} showed that hierarchical inversion does not provide an improved prior distribution, but rather provides a posterior distribution of hyperparameters that are consistent with the prior information, the observed data, and the forward model. We therefore do not contend that the hierarchical methods in this work are producing improved prior pdf's. Instead, the hierarchical flexibility is explicitly being used for exactly the purpose expressed by \citep{mosegaard_inconsistency_2024}: to estimate the level of consistency or otherwise of the prior information in $\mathcal{M}_{gen}$, with the data and forward relation. This level of consistency is expressed by $\alpha$, and the resulting distributions are all posterior pdf's, not priors.

We experimented with different distributions of hyperprior $p(\alpha)$, ranging from a non-informative uniform distribution to the weakly informative Half-Cauchy distribution. Using a uniform hyperprior causes the $\alpha$ posterior distribution to have a broad variance that includes excessively high $\alpha$ values, which in our examples adds unwanted strong prior information into the inversion \citep{gelman_prior_2006}. For the Half-Cauchy distribution, in the limit $\alpha \rightarrow \infty$, it behaves like a uniform prior density on $p(\alpha)$ \citep{gelman_prior_2006}, thus higher $\alpha$ values become equally probable, which causes the escaping prior pdf to be more confined to $\mathcal{M}_{gen}$.

In our examples, we used a global $\alpha$ parameter to represent the overall prior-strength or variance for all model parameters, assuming a uniform level of uncertainty across all dimensions of the model parameter space. Future work might usefully incorporate anisotropic uncertainty into the prior-strength of the escaping prior. This can be achieved by using a full, positive-semidefinite covariance matrix, which allows not only different variances for each dimension but also correlations among them.

\begin{figure}
    \centering
    \includegraphics[width=1\linewidth]{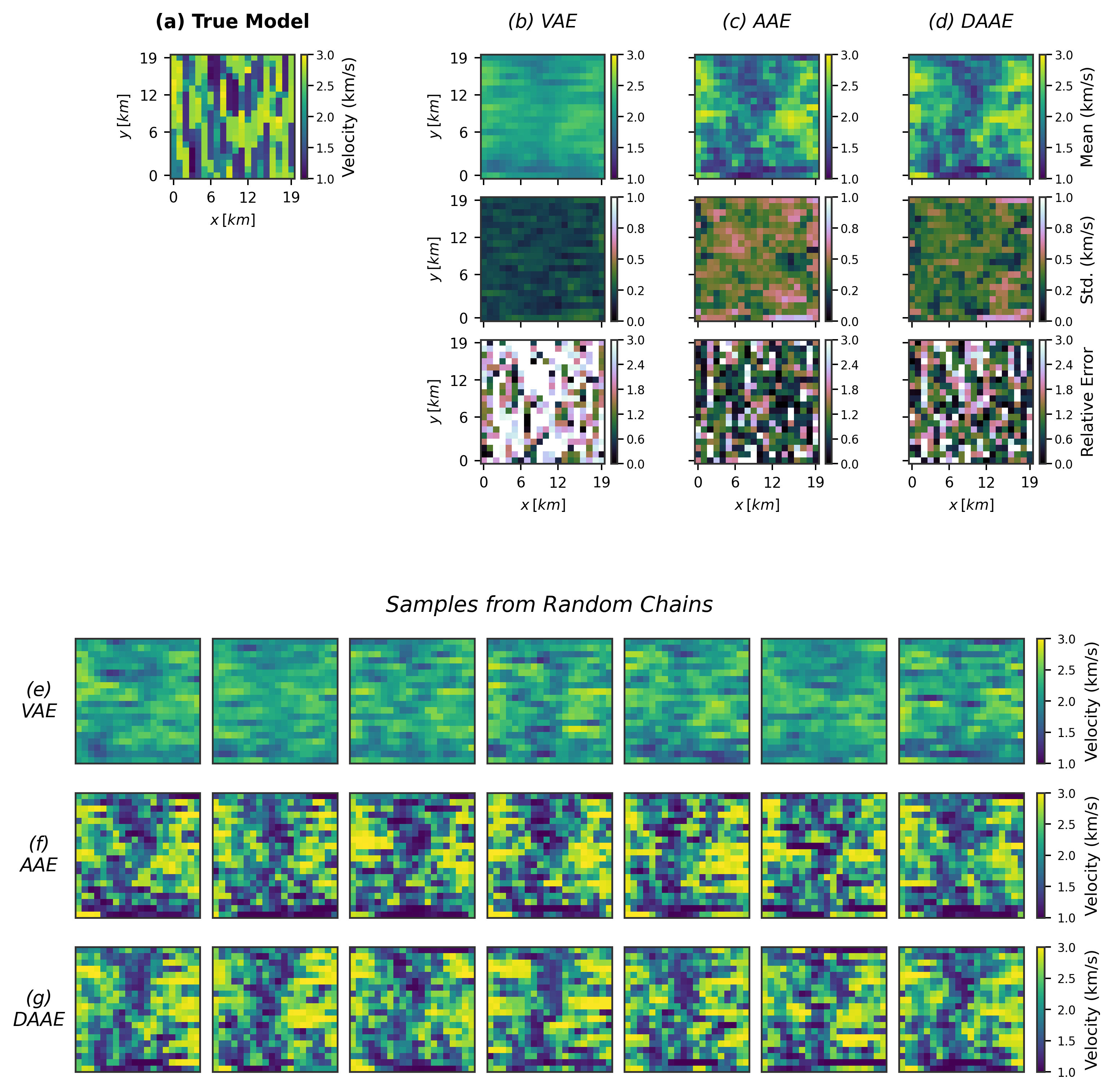}
    \caption{(a) True model representing vertically stratified layers, with corresponding (b) VAE, (c) AAE, and (d) DAAE-based latent space inversion results with latent space dimensionality of 128. Each row corresponds to the posterior mean, standard deviation and relative error (defined in the main text) in model space. Posterior samples are drawn from random MCMC chains for (e) VAE, (f) AAE, and (g) DAAE-based latent space inversion results.}
    \label{fig:vel_geoUNKN_lat_vs_hier_12x12_Z128}
\end{figure}

In tomography experiments with more geologically realistic velocity models, latent space inversion performs well when the true model has geological features and structural correlations similar to those present in the GNNs' training set. However, when the true solution (represented by vertically stratified sediment layers) lies outside the manifold of models that GNNs can generate (i.e., horizontal strata), latent space inversion fails. GNNs should never produce a vertical stratum when trained solely on horizontally stratified layers, and introducing additional data does not improve the results. However, in other experiments we noticed that increasing the latent space dimensionality improves GNN reconstruction quality and, consequently, the quality of latent space inversion results. This is likely because GNNs can sometimes produce spurious or meaningless samples \citep{bengio_generalized_2013, goodfellow_nips_2016, salimans_improved_2016, kegl_spurious_2018}, which are regarded as failures or hallucinations. For example, \cite{kazakci_digits_2016} ended up generating non-numeric symbols (non-digits) with GNNs using a training set limited to MNIST digits. In our case, hallucination becomes more pronounced as we increase the latent space dimensionality; as a result, the generator occasionally produces spurious samples that by chance fit the observed data, even though those features are absent in the training set. This should be regarded as a failing in the Bayesian methodology, because it indicates that the GNN training was ineffective.

\begin{figure}
    \centering
    \includegraphics[width=1\linewidth]{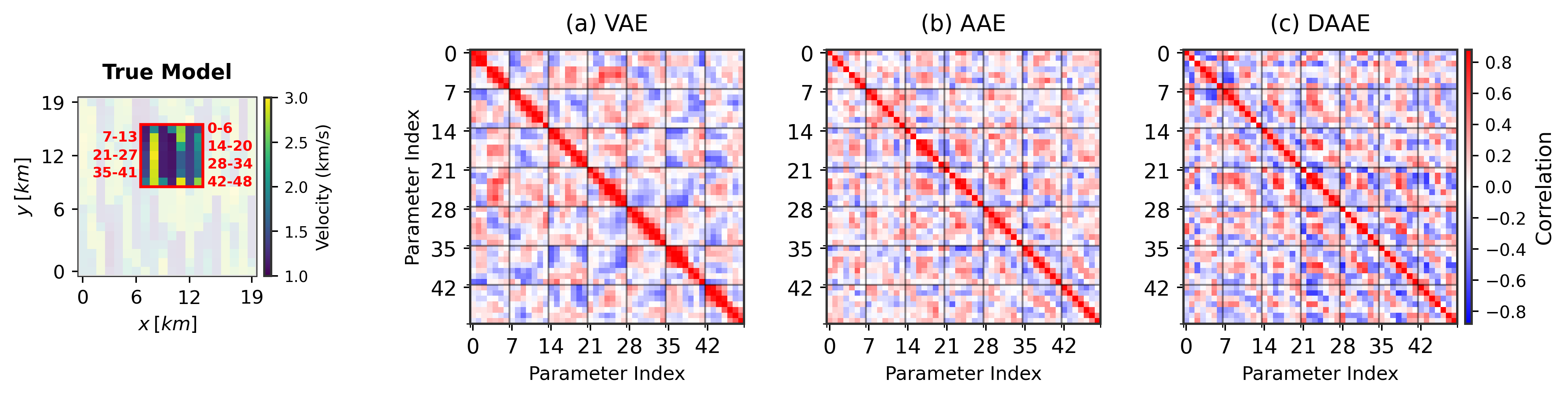}
    \caption{Posterior correlation matrices corresponding to the latent space inversion results in Figure \ref{fig:vel_geoUNKN_lat_vs_hier_12x12_Z128}, for (a) VAE, (b) AAE, and (c) DAAE with the colour scale adjusted to highlight subtle off-diagonal variations. A red rectangle outlines the $7\times7$ correlation grid on the true model, with its parameters indicated.}
    \label{fig:12x12_cor_lat_sediments_Z128_2}
\end{figure}

This is demonstrated in Figure \ref{fig:vel_geoUNKN_lat_vs_hier_12x12_Z128}, in which we show results using a latent space dimensionality of 128. While the VAE-based latent space inversion produces results similar to those in Figure \ref{fig:vel_geoUNKN_lat_vs_hier_12x12_Z8}c, the results from AAE and DAAE demonstrate significant improvements (Figures \ref{fig:vel_geoUNKN_lat_vs_hier_12x12_Z128}c,d) by producing hallucinated samples that display a form of vertically layered structures, as shown by individual samples from random MCMC chains (Figures \ref{fig:vel_geoUNKN_lat_vs_hier_12x12_Z128}f,g). Furthermore, examining the posterior correlations from the latent space inversion results in the model space (Figures \ref{fig:12x12_cor_lat_sediments_Z128_2}b,c), shows significant evidence of positive correlation between vertically adjacent cells (the higher-order off-diagonals parallel to the main diagonal). Appendix \ref{app:prior_correlations} highlights notable differences in correlation matrices for samples generated by the VAE, AAE and DAAE with latent space dimensions of 8 (see Figures \ref{fig:prior_cor_vertical_Z8}c-e) and 128 (see Figures \ref{fig:prior_cor_vertical_Z128}c-e). From a geological standpoint, this hallucination raises a serious concern regarding the reliability of the prior information incorporated by GNNs in latent space inversion. Specifically, whether the trained GNN produces hallucinated samples through the generator.

Conducting inversion through random sampling in MCMC methods can be computationally expensive. On the other hand, variational inference offers a potentially more efficient alternative by solving Bayesian inversion through optimisation. A family of known, tractable distributions, known as the variational family, is initially defined. From this, an optimal member is chosen to approximate the true posterior probability density function by minimising the difference between the variational and posterior distributions \citep{zhang_seismic_2020, zhang_variational_2020, zhang_chapter_2021, zhao_bayesian_2021, zhang_3d_2023, zhao_physically_2024, zhao_geopvi_2026}. Additionally, implementing more advanced and expressive GNNs, such as Generative Adversarial Networks (GANs, \cite{goodfellow_generative_2014}) or diffusion models \citep{sohl-dickstein_deep_2015, ho_denoising_2020}, might be a better alternative to autoencoder-based architectures. 

\section{Conclusion}
\label{sec:Conclusion}

We demonstrate an inversion method that incorporates Generative Neural Network (GNN) learned generative manifolds of geological velocity models as informative Bayesian priors in model space. GNNs trained on velocity models derived from geologically realistic sedimentary process simulations help preserve geological realism in the solutions. This study also investigates the extent to which prior information can be incorporated to obtain a geophysically plausible posterior by implementing a hierarchical Bayesian inversion algorithm. Experiments ranging from simple low-dimensional problems to complex geological travel-time tomography showed that the escaping prior produces better results than standard inversion techniques, such as latent space inversion and model space inversion with simple, non-informative priors. A comparative study of three GNNs (Variational Autoencoders, Adversarial Autoencoders, and Denoising Adversarial Autoencoders) provided valuable insights into the characteristics of the learned generative manifold and its role in the escaping prior. We also discuss the structural hallucinations caused by the GNNs. Finally, we conclude that the escaping prior allows us to introduce complex, geologically plausible prior information parametrised by GNNs into geophysical inversions and to find solutions that lie beyond the generative manifold.

\section{Acknowledgements}
We thank the sponsors of Edinburgh Imaging Project (EIP - https://blogs.ed.ac.uk/imaging/), BP and TotalEnergies, for supporting this research. To promote open access, we have applied a Creative Commons Attribution (CC BY) license to any Author Accepted Manuscript version resulting from this submission.

\section{Data Availability}
No real data is used in this study. Computer code and training dataset are available upon request by contacting the corresponding authors.

\bibliographystyle{gji}
\bibliography{references_final.bib}

@misc{bowman_generating_2016,
	title = {Generating {Sentences} from a {Continuous} {Space}},
	url = {http://arxiv.org/abs/1511.06349},
	doi = {10.48550/arXiv.1511.06349},
	urldate = {2025-02-18},
	publisher = {arXiv},
	author = {Bowman, Samuel R. and Vilnis, Luke and Vinyals, Oriol and Dai, Andrew M. and Jozefowicz, Rafal and Bengio, Samy},
	month = may,
	year = {2016},
}

@article{galetti_uncertainty_2015,
	title = {Uncertainty {Loops} in {Travel}-{Time} {Tomography} from {Nonlinear} {Wave} {Physics}},
	volume = {114},
	copyright = {http://link.aps.org/licenses/aps-default-license},
	issn = {0031-9007, 1079-7114},
	url = {https://link.aps.org/doi/10.1103/PhysRevLett.114.148501},
	doi = {10.1103/PhysRevLett.114.148501},
	language = {en},
	number = {14},
	urldate = {2026-08-14},
	journal = {Physical Review Letters},
	author = {Galetti, Erica and Curtis, Andrew and Meles, Giovanni Angelo and Baptie, Brian},
	month = apr,
	year = {2015},
	pages = {148501},
}

@article{bao_variational_2022,
	title = {Variational {Autoencoder} or {Generative} {Adversarial} {Networks}? {A} {Comparison} of {Two} {Deep} {Learning} {Methods} for {Flow} and {Transport} {Data} {Assimilation}},
	volume = {54},
	issn = {1874-8953},
	shorttitle = {Variational {Autoencoder} or {Generative} {Adversarial} {Networks}?},
	url = {https://doi.org/10.1007/s11004-022-10003-3},
	doi = {10.1007/s11004-022-10003-3},
	language = {en},
	number = {6},
	urldate = {2026-08-13},
	journal = {Mathematical Geosciences},
	author = {Bao, Jichao and Li, Liangping and Davis, Arden},
	month = aug,
	year = {2022},
	pages = {1017--1042},
}

@article{chen_fracture_2023,
	title = {Fracture network characterization with deep generative model based stochastic inversion},
	volume = {273},
	issn = {03605442},
	url = {https://linkinghub.elsevier.com/retrieve/pii/S0360544223006965},
	doi = {10.1016/j.energy.2023.127302},
	language = {en},
	urldate = {2026-08-13},
	journal = {Energy},
	author = {Chen, Guodong and Luo, Xin and Jiao, Jiu Jimmy and Jiang, Chuanyin},
	month = jun,
	year = {2023},
	pages = {127302},
}

@article{liu_uncertainty_2022,
	title = {Uncertainty quantification in stochastic inversion with dimensionality reduction using variational autoencoder},
	volume = {87},
	issn = {0016-8033, 1942-2156},
	url = {https://pubs.geoscienceworld.org/geophysics/article/87/2/M43/610488/Uncertainty-quantification-in-stochastic-inversion},
	doi = {10.1190/geo2021-0138.1},
	language = {en},
	number = {2},
	urldate = {2026-08-13},
	journal = {Geophysics},
	author = {Liu, Mingliang and Grana, Dario and De Figueiredo, Leandro Passos},
	month = mar,
	year = {2022},
	pages = {M43--M58},
}

@article{taufik_latent_2025,
	title = {Latent {Representation} {Learning} in {Physics}-{Informed} {Neural} {Networks} for {Full} {Waveform} {Inversion}},
	volume = {12},
	issn = {2333-5084},
	url = {https://onlinelibrary.wiley.com/doi/abs/10.1029/2024EA004107},
	doi = {10.1029/2024EA004107},
	language = {en},
	number = {9},
	urldate = {2026-08-13},
	journal = {Earth and Space Science},
	author = {Taufik, Mohammad H. and Huang, Xinquan and Alkhalifah, Tariq},
	year = {2025},
	pages = {e2024EA004107},
}

@article{liu_geostatistical_2024,
	title = {Geostatistical {Inversion} for {Subsurface} {Characterization} {Using} {Stein} {Variational} {Gradient} {Descent} {With} {Autoencoder} {Neural} {Network}: {An} {Application} to {Geologic} {Carbon} {Sequestration}},
	volume = {129},
	issn = {2169-9356},
	shorttitle = {Geostatistical {Inversion} for {Subsurface} {Characterization} {Using} {Stein} {Variational} {Gradient} {Descent} {With} {Autoencoder} {Neural} {Network}},
	url = {https://onlinelibrary.wiley.com/doi/abs/10.1029/2024JB029073},
	doi = {10.1029/2024JB029073},
	language = {en},
	number = {7},
	urldate = {2026-08-13},
	journal = {Journal of Geophysical Research: Solid Earth},
	author = {Liu, Mingliang and Grana, Dario and Mukerji, Tapan},
	year = {2024},
	pages = {e2024JB029073},
}

@misc{mosegaard_inconsistency_2024,
	title = {Inconsistency and {Acausality} in {Bayesian} {Inference} for {Physical} {Problems}},
	url = {http://arxiv.org/abs/2411.13570},
	doi = {10.48550/arXiv.2411.13570},
	urldate = {2025-10-04},
	publisher = {arXiv},
	author = {Mosegaard, Klaus and Curtis, Andrew},
	year = {2024},
	note = {arXiv:2411.13570 [stat]},
}

@article{burgess_sensitive_2004,
	title = {Sensitive dependence, divergence and unpredictable behaviour in a stratigraphic forward model of a carbonate system},
	volume = {239},
	copyright = {https://doi.org/10.15223/policy-002},
	issn = {0305-8719, 2041-4927},
	url = {https://www.lyellcollection.org/doi/10.1144/GSL.SP.2004.239.01.06},
	doi = {10.1144/GSL.SP.2004.239.01.06},
	language = {en},
	number = {1},
	urldate = {2026-08-12},
	journal = {Geological Society, London, Special Publications},
	author = {Burgess, Peter M. and Emery, David J.},
	month = jan,
	year = {2004},
	pages = {77--94},
}

@book{curtis_geological_2004,
	title = {Geological {Prior} {Information}: {Informing} {Science} and {Engineering}},
	isbn = {978-1-86239-171-0},
	shorttitle = {Geological {Prior} {Information}},
	language = {en},
	publisher = {Geological Society of London},
	author = {Curtis, Andrew and Wood, Rachel},
	year = {2004},
}

@misc{ioffe_batch_2015,
	title = {Batch {Normalization}: {Accelerating} {Deep} {Network} {Training} by {Reducing} {Internal} {Covariate} {Shift}},
	shorttitle = {Batch {Normalization}},
	url = {http://arxiv.org/abs/1502.03167},
	doi = {10.48550/arXiv.1502.03167},
	urldate = {2026-08-01},
	publisher = {arXiv},
	author = {Ioffe, Sergey and Szegedy, Christian},
	month = mar,
	year = {2015},
	note = {arXiv:1502.03167 [cs.LG]},
}

@article{hinton_reducing_2006,
	title = {Reducing the {Dimensionality} of {Data} with {Neural} {Networks}},
	volume = {313},
	issn = {0036-8075, 1095-9203},
	url = {https://www.science.org/doi/10.1126/science.1127647},
	doi = {10.1126/science.1127647},
	language = {en},
	number = {5786},
	urldate = {2026-08-01},
	journal = {Science},
	author = {Hinton, G. E. and Salakhutdinov, R. R.},
	month = jul,
	year = {2006},
	pages = {504--507},
}

@article{zhao_geopvi_2026,
	title = {{GeoPVI}: a {Python} package for geoscientific inversion using parametric variational inference},
	volume = {245},
	copyright = {https://creativecommons.org/licenses/by/4.0/},
	issn = {0956-540X, 1365-246X},
	shorttitle = {{GeoPVI}},
	url = {https://academic.oup.com/gji/article/doi/10.1093/gji/ggag138/8651301},
	doi = {10.1093/gji/ggag138},
	language = {en},
	number = {3},
	urldate = {2026-08-01},
	journal = {Geophysical Journal International},
	author = {Zhao, Xuebin and Curtis, Andrew},
	month = apr,
	year = {2026},
	pages = {ggag138},
}

@article{zhao_bayesian_2021,
	title = {Bayesian seismic tomography using normalizing flows},
	volume = {228},
	copyright = {https://academic.oup.com/journals/pages/open\_access/funder\_policies/chorus/standard\_publication\_model},
	issn = {0956-540X, 1365-246X},
	url = {https://academic.oup.com/gji/article/228/1/213/6332286},
	doi = {10.1093/gji/ggab298},
	language = {en},
	number = {1},
	urldate = {2026-07-16},
	journal = {Geophysical Journal International},
	author = {Zhao, Xuebin and Curtis, Andrew and Zhang, Xin},
	month = sep,
	year = {2021},
	pages = {213--239},
}

@article{zhang_variational_2020,
	title = {Variational full-waveform inversion},
	volume = {222},
	issn = {0956-540X},
	url = {https://doi.org/10.1093/gji/ggaa170},
	doi = {10.1093/gji/ggaa170},
	language = {en-US},
	number = {1},
	urldate = {2024-10-12},
	journal = {Geophysical Journal International},
	author = {Zhang, Xin and Curtis, Andrew},
	month = jul,
	year = {2020},
	pages = {406--411},
}

@article{zhang_seismic_2020,
	title = {Seismic {Tomography} {Using} {Variational} {Inference} {Methods}},
	volume = {125},
	copyright = {©2019. American Geophysical Union. All Rights Reserved.},
	issn = {2169-9356},
	url = {https://onlinelibrary.wiley.com/doi/abs/10.1029/2019JB018589},
	doi = {10.1029/2019JB018589},
	language = {en},
	number = {4},
	urldate = {2024-10-12},
	journal = {Journal of Geophysical Research: Solid Earth},
	author = {Zhang, Xin and Curtis, Andrew},
	year = {2020},
	pages = {e2019JB018589},
}

@misc{zeiler_visualizing_2013,
	title = {Visualizing and {Understanding} {Convolutional} {Networks}},
	url = {http://arxiv.org/abs/1311.2901},
	doi = {10.48550/arXiv.1311.2901},
	urldate = {2026-03-13},
	publisher = {arXiv},
	author = {Zeiler, Matthew D. and Fergus, Rob},
	month = nov,
	year = {2013},
	note = {(arXiv:1311.2901)},
}

@article{tenenbaum_global_2000,
	title = {A {Global} {Geometric} {Framework} for {Nonlinear} {Dimensionality} {Reduction}},
	volume = {290},
	issn = {0036-8075, 1095-9203},
	url = {https://www.science.org/doi/10.1126/science.290.5500.2319},
	doi = {10.1126/science.290.5500.2319},
	language = {en},
	number = {5500},
	urldate = {2025-08-04},
	journal = {Science},
	author = {Tenenbaum, Joshua B. and Silva, Vin De and Langford, John C.},
	month = dec,
	year = {2000},
	pages = {2319--2323},
}

@article{malinverno_expanded_2004,
	title = {Expanded uncertainty quantification in inverse problems: {Hierarchical} {Bayes} and empirical {Bayes}},
	volume = {69},
	issn = {0016-8033, 1942-2156},
	shorttitle = {Expanded uncertainty quantification in inverse problems},
	url = {https://library.seg.org/doi/10.1190/1.1778243},
	doi = {10.1190/1.1778243},
	language = {en},
	number = {4},
	urldate = {2025-12-06},
	journal = {GEOPHYSICS},
	author = {Malinverno, Alberto and Briggs, Victoria A.},
	month = jul,
	year = {2004},
	pages = {1005--1016},
}

@article{laloy_trainingimage_2018,
	title = {Training-{Image} {Based} {Geostatistical} {Inversion} {Using} a {Spatial} {Generative} {Adversarial} {Neural} {Network}},
	volume = {54},
	issn = {1944-7973},
	url = {https://onlinelibrary.wiley.com/doi/abs/10.1002/2017WR022148},
	doi = {10.1002/2017WR022148},
	language = {en},
	number = {1},
	urldate = {2026-03-07},
	journal = {Water Resources Research},
	author = {Laloy, Eric and Hérault, Romain and Jacques, Diederik and Linde, Niklas},
	year = {2018},
	pages = {381--406},
}

@article{laloy_highdimensional_2012,
	title = {High-dimensional posterior exploration of hydrologic models using multiple-try {DREAM}({ZS}) and high-performance computing},
	volume = {48},
	issn = {1944-7973},
	url = {https://onlinelibrary.wiley.com/doi/abs/10.1029/2011WR010608},
	doi = {10.1029/2011WR010608},
	language = {en},
	number = {1},
	urldate = {2026-03-06},
	journal = {Water Resources Research},
	author = {Laloy, Eric and Vrugt, Jasper A.},
	year = {2012},
}

@misc{kegl_spurious_2018,
	title = {Spurious samples in deep generative models: bug or feature?},
	shorttitle = {Spurious samples in deep generative models},
	url = {http://arxiv.org/abs/1810.01876},
	doi = {10.48550/arXiv.1810.01876},
	language = {en-US},
	urldate = {2026-04-14},
	publisher = {arXiv},
	author = {Kégl, Balázs and Cherti, Mehdi and Kazakçı, Akın},
	month = oct,
	year = {2018},
	note = {(arXiv:1810.01876)},
}

@misc{kingma_autoencoding_2014,
	title = {Auto-{Encoding} {Variational} {Bayes}},
	url = {http://arxiv.org/abs/1312.6114},
	doi = {10.48550/arXiv.1312.6114},
	language = {en-US},
	urldate = {2024-10-29},
	publisher = {arXiv},
	author = {Kingma, Diederik P. and Welling, Max},
	year = {2014},
	note = {(arXiv:1312.6114)},
}

@article{fefferman_testing_2016,
	title = {Testing the manifold hypothesis},
	volume = {29},
	issn = {0894-0347},
	url = {https://www.jstor.org/stable/jamermathsoci.29.4.983},
	language = {en-US},
	number = {4},
	urldate = {2025-08-04},
	journal = {Journal of the American Mathematical Society},
	author = {Fefferman, Charles and Mitter, Sanjoy and Narayanan, Hariharan},
	year = {2016},
	pages = {983--1049},
}

@misc{bengio_generalized_2013,
	title = {Generalized {Denoising} {Auto}-{Encoders} as {Generative} {Models}},
	url = {http://arxiv.org/abs/1305.6663},
	doi = {10.48550/arXiv.1305.6663},
	language = {en-US},
	urldate = {2025-07-30},
	publisher = {arXiv},
	author = {Bengio, Yoshua and Yao, Li and Alain, Guillaume and Vincent, Pascal},
	month = nov,
	year = {2013},
}

@article{zhang_3d_2023,
	title = {3-{D} {Bayesian} variational full waveform inversion},
	volume = {234},
	issn = {0956-540X},
	url = {https://doi.org/10.1093/gji/ggad057},
	doi = {10.1093/gji/ggad057},
	number = {1},
	urldate = {2026-04-16},
	journal = {Geophysical Journal International},
	author = {Zhang, Xin and Lomas, Angus and Zhou, Muhong and Zheng, York and Curtis, Andrew},
	month = jul,
	year = {2023},
	pages = {546--561},
}

@article{goodfellow_nips_2016,
	title = {{NIPS} 2016 {Tutorial}: {Generative} {Adversarial} {Networks}},
	language = {en},
	author = {Goodfellow, Ian},
	year = {2016},
}

@article{kazakci_digits_2016,
	title = {Digits that are not: {Generating} new types through deep neural nets},
	language = {en},
	author = {Kazakçi, Akin Osman and Mehdi, Cherti and Kégl, Balázs},
	year = {2016},
}

@inproceedings{salimans_improved_2016,
	title = {Improved {Techniques} for {Training} {GANs}},
	volume = {29},
	url = {https://proceedings.neurips.cc/paper_files/paper/2016/hash/8a3363abe792db2d8761d6403605aeb7-Abstract.html},
	urldate = {2026-04-15},
	booktitle = {Advances in {Neural} {Information} {Processing} {Systems}},
	publisher = {Curran Associates, Inc.},
	author = {Salimans, Tim and Goodfellow, Ian and Zaremba, Wojciech and Cheung, Vicki and Radford, Alec and Chen, Xi and Chen, Xi},
	year = {2016},
}

@article{zhang_bayesian_2021,
	title = {Bayesian {Geophysical} {Inversion} {Using} {Invertible} {Neural} {Networks}},
	volume = {126},
	issn = {2169-9356},
	url = {https://onlinelibrary.wiley.com/doi/abs/10.1029/2021JB022320},
	doi = {10.1029/2021JB022320},
	language = {en},
	number = {7},
	urldate = {2026-03-30},
	journal = {Journal of Geophysical Research: Solid Earth},
	author = {Zhang, Xin and Curtis, Andrew},
	year = {2021},
	note = {\_eprint: https://agupubs.onlinelibrary.wiley.com/doi/pdf/10.1029/2021JB022320},
	pages = {e2021JB022320},
}

@article{zhao_physically_2024,
	title = {Physically {Structured} {Variational} {Inference} for {Bayesian} {Full} {Waveform} {Inversion}},
	volume = {129},
	issn = {2169-9356},
	url = {https://onlinelibrary.wiley.com/doi/abs/10.1029/2024JB029557},
	doi = {10.1029/2024JB029557},
	language = {en},
	number = {11},
	urldate = {2026-03-28},
	journal = {Journal of Geophysical Research: Solid Earth},
	author = {Zhao, Xuebin and Curtis, Andrew},
	year = {2024},
	note = {\_eprint: https://agupubs.onlinelibrary.wiley.com/doi/pdf/10.1029/2024JB029557},
	pages = {e2024JB029557},
}

@article{gelman_prior_2006,
	title = {Prior distributions for variance parameters in hierarchical models},
	volume = {1},
	issn = {1936-0975, 1931-6690},
	url = {https://projecteuclid.org/journals/bayesian-analysis/volume-1/issue-3/Prior-distributions-for-variance-parameters-in-hierarchical-models-comment-on/10.1214/06-BA117A.full},
	doi = {10.1214/06-BA117A},
	language = {en},
	number = {3},
	urldate = {2026-03-19},
	journal = {Bayesian Analysis},
	publisher = {International Society for Bayesian Analysis},
	author = {Gelman, Andrew},
	month = sep,
	year = {2006},
	pages = {515--534},
}

@article{gelman_bayesian_2013,
	title = {Bayesian {Data} {Analysis}, {Third} {Edition}},
	language = {en},
	author = {Gelman, A and Carlin, J B and Stern, H S and Dunson, D B and Vehtari, A and Rubin, D B and Christensen, R and Johnson, W and Branscum, A and Hanson, T E and Carlin, B P and Louis, T A and Miller, R G and Lunn, D and Jackson, C and Best, N and Thomas, A and Spiegelhalter, D and Leonard, T and Ferguson, T S and Sprent, P and Smith, J Q and Lawson, J and Tuckwell, H C and Morgan, B J T and Woodward, M and Rees, D A G and Kupper, L L and Neelon, B H and O’Brien, S M},
	year = {2013},
}

@inproceedings{vincent_extracting_2008,
	address = {New York, NY, USA},
	series = {{ICML} '08},
	title = {Extracting and composing robust features with denoising autoencoders},
	isbn = {978-1-60558-205-4},
	url = {https://dl.acm.org/doi/10.1145/1390156.1390294},
	doi = {10.1145/1390156.1390294},
	language = {en-US},
	urldate = {2026-03-15},
	booktitle = {Proceedings of the 25th international conference on {Machine} learning},
	publisher = {Association for Computing Machinery},
	author = {Vincent, Pascal and Larochelle, Hugo and Bengio, Yoshua and Manzagol, Pierre-Antoine},
	month = jul,
	year = {2008},
	pages = {1096--1103},
}

@article{alder_quantifying_2017,
	title = {Quantifying seismic anisotropy induced by small-scale chemical heterogeneities},
	volume = {211},
	issn = {0956-540X, 1365-246X},
	url = {https://academic.oup.com/gji/article/211/3/1585/4191274},
	doi = {10.1093/gji/ggx389},
	language = {en},
	number = {3},
	urldate = {2026-03-07},
	journal = {Geophysical Journal International},
	author = {Alder, C and Bodin, T and Ricard, Y and Capdeville, Y and Debayle, E and Montagner, J P},
	month = dec,
	year = {2017},
	pages = {1585--1600},
}

@book{bishop_pattern_2006,
	address = {New York},
	series = {Information science and statistics},
	title = {Pattern recognition and machine learning},
	isbn = {978-0-387-31073-2},
	language = {eng},
	publisher = {Springer},
	author = {Bishop, Christopher M.},
	year = {2006},
}

@article{vrugt_accelerating_2009,
	title = {Accelerating {Markov} {Chain} {Monte} {Carlo} {Simulation} by {Differential} {Evolution} with {Self}-{Adaptive} {Randomized} {Subspace} {Sampling}},
	volume = {10},
	issn = {2191-0294, 1565-1339},
	url = {https://www.degruyter.com/document/doi/10.1515/IJNSNS.2009.10.3.273/html},
	doi = {10.1515/IJNSNS.2009.10.3.273},
	language = {en},
	number = {3},
	urldate = {2026-03-06},
	journal = {International Journal of Nonlinear Sciences and Numerical Simulation},
	author = {Vrugt, J. A. and Ter Braak, C.J.F. and Diks, C.G.H. and Robinson, B. A. and Hyman, J. M. and Higdon, D.},
	month = jan,
	year = {2009},
}

@article{laloy_inversion_2017,
	title = {Inversion using a new low-dimensional representation of complex binary geological media based on a deep neural network},
	volume = {110},
	issn = {03091708},
	url = {https://linkinghub.elsevier.com/retrieve/pii/S0309170817306243},
	doi = {10.1016/j.advwatres.2017.09.029},
	language = {en},
	urldate = {2026-03-06},
	journal = {Advances in Water Resources},
	author = {Laloy, Eric and Hérault, Romain and Lee, John and Jacques, Diederik and Linde, Niklas},
	month = dec,
	year = {2017},
	pages = {387--405},
}

@article{mosser_stochastic_2020,
	title = {Stochastic {Seismic} {Waveform} {Inversion} {Using} {Generative} {Adversarial} {Networks} as a {Geological} {Prior}},
	volume = {52},
	issn = {1874-8953},
	url = {https://doi.org/10.1007/s11004-019-09832-6},
	doi = {10.1007/s11004-019-09832-6},
	language = {en},
	number = {1},
	urldate = {2026-03-06},
	journal = {Mathematical Geosciences},
	author = {Mosser, Lukas and Dubrule, Olivier and Blunt, Martin J.},
	month = jan,
	year = {2020},
	pages = {53--79},
}

@misc{nguyen_plug_2017,
	title = {Plug \& {Play} {Generative} {Networks}: {Conditional} {Iterative} {Generation} of {Images} in {Latent} {Space}},
	shorttitle = {Plug \& {Play} {Generative} {Networks}},
	url = {http://arxiv.org/abs/1612.00005},
	doi = {10.48550/arXiv.1612.00005},
	urldate = {2026-03-06},
	publisher = {arXiv},
	author = {Nguyen, Anh and Clune, Jeff and Bengio, Yoshua and Dosovitskiy, Alexey and Yosinski, Jason},
	month = apr,
	year = {2017},
	note = {arXiv:1612.00005 [cs]},
}

@article{roberts_optimal_1998,
	title = {Optimal {Scaling} of {Discrete} {Approximations} to {Langevin} {Diffusions}},
	volume = {60},
	issn = {1369-7412},
	url = {https://www.jstor.org/stable/2985986},
	number = {1},
	urldate = {2026-03-06},
	journal = {Journal of the Royal Statistical Society. Series B (Statistical Methodology)},
	publisher = {[Royal Statistical Society, Oxford University Press]},
	author = {Roberts, Gareth O. and Rosenthal, Jeffrey S.},
	year = {1998},
	pages = {255--268},
}

@article{roberts_exponential_1996,
	title = {Exponential {Convergence} of {Langevin} {Distributions} and {Their} {Discrete} {Approximations}},
	volume = {2},
	issn = {1350-7265},
	url = {https://www.jstor.org/stable/3318418},
	doi = {10.2307/3318418},
	number = {4},
	urldate = {2026-03-06},
	journal = {Bernoulli},
	publisher = {International Statistical Institute (ISI) and Bernoulli Society for Mathematical Statistics and Probability},
	author = {Roberts, Gareth O. and Tweedie, Richard L.},
	year = {1996},
	pages = {341--363},
}

@article{ruderman_statistics_1994,
	title = {The statistics of natural images},
	volume = {5},
	issn = {0954-898X},
	url = {https://doi.org/10.1088/0954-898X/5/4/006},
	doi = {10.1088/0954-898X/5/4/006},
	language = {en},
	number = {4},
	urldate = {2026-03-05},
	journal = {Network: Computation in Neural Systems},
	author = {Ruderman, D L},
	month = nov,
	year = {1994},
	pages = {517},
}

@article{scholkopf_nonlinear_1998,
	title = {Nonlinear {Component} {Analysis} as a {Kernel} {Eigenvalue} {Problem}},
	volume = {10},
	issn = {0899-7667, 1530-888X},
	url = {https://direct.mit.edu/neco/article/10/5/1299-1319/6193},
	doi = {10.1162/089976698300017467},
	language = {en},
	number = {5},
	urldate = {2026-03-05},
	journal = {Neural Computation},
	author = {Schölkopf, Bernhard and Smola, Alexander and Müller, Klaus-Robert},
	month = jul,
	year = {1998},
	pages = {1299--1319},
}

@article{roweis_nonlinear_2000,
	title = {Nonlinear {Dimensionality} {Reduction} by {Locally} {Linear} {Embedding}},
	volume = {290},
	issn = {0036-8075, 1095-9203},
	url = {https://www.science.org/doi/10.1126/science.290.5500.2323},
	doi = {10.1126/science.290.5500.2323},
	language = {en},
	number = {5500},
	urldate = {2026-03-05},
	journal = {Science},
	author = {Roweis, Sam T. and Saul, Lawrence K.},
	month = dec,
	year = {2000},
	pages = {2323--2326},
}

@article{bloem_introducing_2023,
	title = {Introducing conceptual geological information into {Bayesian} tomographic imaging},
	volume = {36},
	issn = {1365-2117},
	url = {https://onlinelibrary.wiley.com/doi/abs/10.1111/bre.12811},
	doi = {10.1111/bre.12811},
	language = {en},
	number = {1},
	urldate = {2024-10-12},
	journal = {Basin Research},
	author = {Bloem, Hugo and Curtis, Andrew and Tetzlaff, Daniel},
	year = {2023},
}

@article{galetti_transdimensional_2017,
	title = {Transdimensional {Love}-wave tomography of the {British} {Isles} and shear-velocity structure of the {East} {Irish} {Sea} {Basin} from ambient-noise interferometry},
	volume = {208},
	issn = {0956-540X},
	url = {https://doi.org/10.1093/gji/ggw286},
	doi = {10.1093/gji/ggw286},
	number = {1},
	urldate = {2024-10-12},
	journal = {Geophysical Journal International},
	author = {Galetti, Erica and Curtis, Andrew and Baptie, Brian and Jenkins, David and Nicolson, Heather},
	month = jan,
	year = {2017},
	pages = {36--58},
}

@incollection{zhang_chapter_2021,
	series = {Inversion of {Geophysical} {Data}},
	title = {Chapter {Two} - {An} introduction to variational inference in geophysical inverse problems},
	volume = {62},
	url = {https://www.sciencedirect.com/science/article/pii/S0065268721000030},
	doi = {10.1016/bs.agph.2021.06.003},
	urldate = {2024-10-12},
	booktitle = {Advances in {Geophysics}},
	publisher = {Elsevier},
	author = {Zhang, Xin and Nawaz, Muhammad Atif and Zhao, Xuebin and Curtis, Andrew},
	editor = {Schmelzbach, Cedric},
	month = jan,
	year = {2021},
	pages = {73--140},
}

@misc{goodfellow_generative_2014,
	title = {Generative {Adversarial} {Networks}},
	url = {http://arxiv.org/abs/1406.2661},
	doi = {10.48550/arXiv.1406.2661},
	urldate = {2024-10-29},
	publisher = {arXiv},
	author = {Goodfellow, Ian J. and Pouget-Abadie, Jean and Mirza, Mehdi and Xu, Bing and Warde-Farley, David and Ozair, Sherjil and Courville, Aaron and Bengio, Yoshua},
	month = jun,
	year = {2014},
}

@misc{meles_bayesian_2023,
	title = {Bayesian tomography using polynomial chaos expansion and deep generative networks},
	url = {http://arxiv.org/abs/2307.04228},
	doi = {10.48550/arXiv.2307.04228},
	language = {en-US},
	urldate = {2024-10-29},
	publisher = {arXiv},
	author = {Meles, Giovanni Angelo and Amaya, Macarena and Levy, Shiran and Marelli, Stefano and Linde, Niklas},
	month = oct,
	year = {2023},
}

@article{santos_refining_2024,
	title = {Refining tomography with generative neural networks trained from geodynamics},
	volume = {238},
	issn = {1365-246X},
	url = {https://doi.org/10.1093/gji/ggae240},
	doi = {10.1093/gji/ggae240},
	number = {3},
	urldate = {2024-10-29},
	journal = {Geophysical Journal International},
	author = {Santos, T and Bodin, T and Soulez, F and Ricard, Y and Capdeville, Y},
	month = sep,
	year = {2024},
	pages = {1676--1695},
}

@article{levy_variational_2023,
	title = {Variational {Bayesian} inference with complex geostatistical priors using inverse autoregressive flows},
	volume = {171},
	issn = {0098-3004},
	url = {https://www.sciencedirect.com/science/article/pii/S0098300422002126},
	doi = {10.1016/j.cageo.2022.105263},
	language = {en-US},
	urldate = {2024-10-29},
	journal = {Computers \& Geosciences},
	author = {Levy, Shiran and Laloy, Eric and Linde, Niklas},
	month = feb,
	year = {2023},
	pages = {105263},
}

@book{tarantola_inverse_2005,
	series = {Other {Titles} in {Applied} {Mathematics}},
	title = {Inverse {Problem} {Theory} and {Methods} for {Model} {Parameter} {Estimation}},
	isbn = {978-0-89871-572-9},
	url = {https://epubs.siam.org/doi/book/10.1137/1.9780898717921},
	urldate = {2024-12-26},
	publisher = {Society for Industrial and Applied Mathematics},
	author = {Tarantola, Albert},
	month = jan,
	year = {2005},
}

@misc{creswell_denoising_2018,
	title = {Denoising {Adversarial} {Autoencoders}},
	url = {http://arxiv.org/abs/1703.01220},
	doi = {10.48550/arXiv.1703.01220},
	language = {en-US},
	urldate = {2025-04-22},
	publisher = {arXiv},
	author = {Creswell, Antonia and Bharath, Anil Anthony},
	month = jan,
	year = {2018},
}

@article{mosegaard_monte_1995,
	title = {Monte {Carlo} sampling of solutions to inverse problems},
	volume = {100},
	copyright = {Copyright 1995 by the American Geophysical Union.},
	issn = {2156-2202},
	url = {https://onlinelibrary.wiley.com/doi/abs/10.1029/94JB03097},
	doi = {10.1029/94JB03097},
	language = {en},
	number = {B7},
	urldate = {2025-04-15},
	journal = {Journal of Geophysical Research: Solid Earth},
	author = {Mosegaard, Klaus and Tarantola, Albert},
	year = {1995},
	pages = {12431--12447},
}

@article{curtis_prior_2001,
	title = {Prior information, sampling distributions, and the curse of dimensionality},
	volume = {66},
	issn = {0016-8033},
	url = {https://doi.org/10.1190/1.1444928},
	doi = {10.1190/1.1444928},
	language = {en-US},
	number = {2},
	urldate = {2025-04-15},
	journal = {Geophysics},
	author = {Curtis, Andrew and Lomax, Anthony},
	month = jan,
	year = {2001},
	pages = {372--378},
}

@misc{obmann_sparse_2020,
	title = {Sparse {aNETT} for {Solving} {Inverse} {Problems} with {Deep} {Learning}},
	url = {http://arxiv.org/abs/2004.09565},
	doi = {10.48550/arXiv.2004.09565},
	language = {en-US},
	urldate = {2025-07-04},
	publisher = {arXiv},
	author = {Obmann, Daniel and Nguyen, Linh and Schwab, Johannes and Haltmeier, Markus},
	month = apr,
	year = {2020},
}

@misc{makhzani_adversarial_2016,
	title = {Adversarial {Autoencoders}},
	url = {http://arxiv.org/abs/1511.05644},
	doi = {10.48550/arXiv.1511.05644},
	language = {en-US},
	urldate = {2025-07-21},
	publisher = {arXiv},
	author = {Makhzani, Alireza and Shlens, Jonathon and Jaitly, Navdeep and Goodfellow, Ian and Frey, Brendan},
	month = may,
	year = {2016},
}

@misc{sohl-dickstein_deep_2015,
	title = {Deep {Unsupervised} {Learning} using {Nonequilibrium} {Thermodynamics}},
	url = {http://arxiv.org/abs/1503.03585},
	doi = {10.48550/arXiv.1503.03585},
	urldate = {2025-10-30},
	publisher = {arXiv},
	author = {Sohl-Dickstein, Jascha and Weiss, Eric A. and Maheswaranathan, Niru and Ganguli, Surya},
	month = nov,
	year = {2015},
}

@article{bayes_essay_1763,
	title = {An {Essay} towards {Solving} a {Problem} in the {Doctrine} of {Chances}. {By} the {Late} {Rev}. {Mr}. {Bayes}, {F}. {R}. {S}. {Communicated} by {Mr}. {Price}, in a {Letter} to {John} {Canton}, {A}. {M}. {F}. {R}. {S}.},
	volume = {53},
	issn = {0260-7085},
	url = {https://www.jstor.org/stable/105741},
	urldate = {2025-10-26},
	journal = {Philosophical Transactions (1683-1775)},
	author = {Bayes, Mr. and Price, Mr.},
	year = {1763},
	pages = {370--418},
}

@article{jackson_use_1979,
	title = {The use of a priori data to resolve non-uniqueness in linear inversion},
	volume = {57},
	issn = {0956-540X},
	url = {https://doi.org/10.1111/j.1365-246X.1979.tb03777.x},
	doi = {10.1111/j.1365-246X.1979.tb03777.x},
	number = {1},
	urldate = {2025-10-24},
	journal = {Geophysical Journal International},
	author = {Jackson, David D.},
	month = apr,
	year = {1979},
	pages = {137--157},
}

@article{tarantola_generalized_1982,
	title = {Generalized nonlinear inverse problems solved using the least squares criterion},
	volume = {20},
	issn = {1944-9208},
	url = {https://onlinelibrary.wiley.com/doi/abs/10.1029/RG020i002p00219},
	doi = {10.1029/RG020i002p00219},
	language = {en},
	number = {2},
	urldate = {2025-10-21},
	journal = {Reviews of Geophysics},
	author = {Tarantola, Albert and Valette, Bernard},
	year = {1982},
	pages = {219--232},
}

@article{strutz_variational_2024,
	title = {Variational {Bayesian} experimental design for geophysical applications: seismic source location, amplitude versus offset inversion, and estimating {CO2} saturations in a subsurface reservoir},
	volume = {236},
	issn = {1365-246X},
	shorttitle = {Variational {Bayesian} experimental design for geophysical applications},
	url = {https://doi.org/10.1093/gji/ggad492},
	doi = {10.1093/gji/ggad492},
	language = {en-US},
	number = {3},
	urldate = {2025-11-10},
	journal = {Geophysical Journal International},
	author = {Strutz, Dominik and Curtis, Andrew},
	month = mar,
	year = {2024},
	pages = {1309--1331},
}

@article{metropolis_equation_1953,
	title = {Equation of {State} {Calculations} by {Fast} {Computing} {Machines}},
	volume = {21},
	issn = {0021-9606},
	url = {https://doi.org/10.1063/1.1699114},
	doi = {10.1063/1.1699114},
	number = {6},
	urldate = {2025-11-07},
	journal = {The Journal of Chemical Physics},
	author = {Metropolis, Nicholas and Rosenbluth, Arianna W. and Rosenbluth, Marshall N. and Teller, Augusta H. and Teller, Edward},
	month = jun,
	year = {1953},
	pages = {1087--1092},
}

@misc{ho_denoising_2020,
	title = {Denoising {Diffusion} {Probabilistic} {Models}},
	url = {http://arxiv.org/abs/2006.11239},
	doi = {10.48550/arXiv.2006.11239},
	urldate = {2025-10-30},
	publisher = {arXiv},
	author = {Ho, Jonathan and Jain, Ajay and Abbeel, Pieter},
	month = dec,
	year = {2020},
}

@misc{tetzlaff_stratigraphic_2023,
	title = {Stratigraphic forward modeling software package for research and education},
	url = {http://arxiv.org/abs/2302.05272},
	doi = {10.48550/arXiv.2302.05272},
	language = {en-US},
	urldate = {2025-12-19},
	publisher = {arXiv},
	author = {Tetzlaff, Daniel},
	month = feb,
	year = {2023},
	note = {https://wsoftc.com/},
}

@article{casella_explaining_1992,
	title = {Explaining the {Gibbs} {Sampler}},
	volume = {46},
	issn = {0003-1305},
	url = {https://www.jstor.org/stable/2685208},
	doi = {10.2307/2685208},
	number = {3},
	urldate = {2026-02-13},
	journal = {The American Statistician},
	publisher = {[American Statistical Association, Taylor \& Francis, Ltd.]},
	author = {Casella, George and George, Edward I.},
	year = {1992},
	pages = {167--174},
}

@article{hastings_monte_1970,
	title = {Monte {Carlo} sampling methods using {Markov} chains and their applications},
	volume = {57},
	issn = {0006-3444},
	url = {https://doi.org/10.1093/biomet/57.1.97},
	doi = {10.1093/biomet/57.1.97},
	number = {1},
	urldate = {2026-01-30},
	journal = {Biometrika},
	author = {Hastings, W. K.},
	month = apr,
	year = {1970},
	pages = {97--109},
}

@article{malinverno_two_2006,
	title = {Two ways to quantify uncertainty in geophysical inverse problems},
	volume = {71},
	issn = {0016-8033},
	url = {https://doi.org/10.1190/1.2194516},
	doi = {10.1190/1.2194516},
	number = {3},
	urldate = {2026-01-26},
	journal = {Geophysics},
	author = {Malinverno, Alberto and Parker, Robert L.},
	month = may,
	year = {2006},
	pages = {W15--W27},
}

@article{holden_bayesian_2022,
	title = {Bayesian {Imaging} with {Data}-{Driven} {Priors} {Encoded} by {Neural} {Networks}},
	volume = {15},
	url = {https://epubs.siam.org/doi/abs/10.1137/21M1406313},
	doi = {10.1137/21M1406313},
	language = {en-US},
	number = {2},
	urldate = {2026-01-19},
	journal = {SIAM Journal on Imaging Sciences},
	publisher = {Society for Industrial and Applied Mathematics},
	author = {Holden, Matthew and Pereyra, Marcelo and Zygalakis, Konstantinos C.},
	month = jun,
	year = {2022},
	pages = {892--924},
}

@article{kullback_information_1951,
	title = {On {Information} and {Sufficiency}},
	volume = {22},
	issn = {0003-4851, 2168-8990},
	url = {https://projecteuclid.org/journals/annals-of-mathematical-statistics/volume-22/issue-1/On-Information-and-Sufficiency/10.1214/aoms/1177729694.full},
	doi = {10.1214/aoms/1177729694},
	language = {en-US},
	number = {1},
	urldate = {2026-01-12},
	journal = {The Annals of Mathematical Statistics},
	author = {Kullback, S. and Leibler, R. A.},
	month = mar,
	year = {1951},
	pages = {79--86},
}

@misc{hu_bayesian_2025,
	title = {Bayesian full waveform inversion with learned prior using deep convolutional autoencoder},
	url = {http://arxiv.org/abs/2511.02737},
	doi = {10.48550/arXiv.2511.02737},
	language = {en-US},
	urldate = {2025-12-27},
	publisher = {arXiv},
	author = {Hu, Shuhua and Sen, Mrinal K. and Zhao, Zeyu and Elmeliegy, Abdelrahman and Zhang, Shuo},
	month = nov,
	year = {2025},
	note = {arXiv:2511.02737 [physics]},
}

@article{bodin_transdimensional_2012,
	title = {Transdimensional tomography with unknown data noise},
	volume = {189},
	issn = {1365-246X},
	url = {https://onlinelibrary.wiley.com/doi/abs/10.1111/j.1365-246X.2012.05414.x},
	doi = {10.1111/j.1365-246X.2012.05414.x},
	language = {en},
	number = {3},
	urldate = {2025-12-05},
	journal = {Geophysical Journal International},
	author = {Bodin, Thomas and Sambridge, Malcolm and Rawlinson, Nick and Arroucau, Pierre},
	year = {2012},
	pages = {1536--1556},
}

@misc{lecun_mnist_1998,
	title = {The {MNIST} database of handwritten digits},
	url = {http://yann.lecun.com/exdb/mnist/},
	author = {LeCun, Yann and Cortes, Corinna and Burges, Christopher JC},
	year = {1998},
}

@article{chib_understanding_1995,
	title = {Understanding the {Metropolis}-{Hastings} {Algorithm}},
	volume = {49},
	issn = {0003-1305},
	url = {https://www.jstor.org/stable/2684568},
	doi = {10.2307/2684568},
	number = {4},
	urldate = {2026-02-13},
	journal = {The American Statistician},
	publisher = {[American Statistical Association, Taylor \& Francis, Ltd.]},
	author = {Chib, Siddhartha and Greenberg, Edward},
	year = {1995},
	pages = {327--335},
}

\appendix

\section{Training of Generative Neural Networks}
\label{app:Training of Generative Neural Networks}
\subsection{Variational Autoencoder (VAE)}

For a training set consisting of $N$ independent and identically distributed (i.i.d.) model samples, the training of a VAE \citep{kingma_autoencoding_2014} involves maximising an evidence lower bound:
\begin{equation}
\label{eq:vae_ELBO}
   \mathcal{L}(\mathbf{m}) = \mathbb{E}_{q(\mathbf{z}| \mathbf{m})}\big[\log p(\mathbf{m}| \mathbf{z})\big] - \mathrm{KL}\big(q(\mathbf{z}| \mathbf{m}) || p(\mathbf{z})\big) 
\end{equation}
The first RHS (Right-Hand Side) term in equation \ref{eq:vae_ELBO} is the reconstruction loss (see Figure \ref{fig:GNN_architectures}a) that ensures the models generated from latent samples $\mathbf{z}$ resemble the input models (minimising the difference between $\mathbf{m}$ and its reconstruction $\mathbf{m}'$). In our case, we calculated the reconstruction loss using the mean squared error. The second RHS term is the Kullback-Leibler divergence (KL, \cite{kullback_information_1951}) (illustrated in Figure \ref{fig:GNN_architectures}a). This KL-divergence term regularises the latent space by encouraging the encoder output distribution $q(\mathbf{z}| \mathbf{m})$ (the distribution of all latent samples produced by the encoder mapping from the entire training set) to match a simple target distribution $p(\mathbf{z})$, which is typically set to a standard Gaussian $\mathcal{N}(\mathbf{0}, \mathbf{I}_r)$, with the aim of ensuring the latent space is continuous and well-structured.

\subsection{Adversarial Autoencoder (AAE)}

Similarly to a VAE, AAE \citep{makhzani_adversarial_2016} has the same encoder-generator architecture; however, it regularises its latent space through adversarial training of the encoder $\mathbf{E}$ and the discriminator $\mathcal{D}_\mathbf{\mathbf{z}}$ and consists of two stages. In this adversarial training, first, we freeze the encoder and train the discriminator $\mathcal{D}_\mathbf{\mathbf{z}}$ to distinguish between ``real" samples drawn from a target distribution (a simple, predefined distribution like $\mathcal{N}(\mathbf{0}, \mathbf{I}_r)$ that we want the encoder output distribution in latent space to follow) and ``fake" samples from the encoder output distribution $q(\mathbf{z}| \mathbf{m})$ (see Figure \ref{fig:GNN_architectures}b). The loss function to train the discriminator is,
\begin{equation}
\mathcal{L}_{\mathcal{D}_\mathbf{z}} = -  \mathbb{E}_{\mathbf{z} \sim \mathcal{N}(\mathbf{0}, \mathbf{I})}[\log \mathcal{D}_\mathbf{z}(\mathbf{z})] - \mathbb{E}_{\mathbf{m} \sim p_{model}(\mathbf{m})}[\log(1 - \mathcal{D}_\mathbf{z}(q(\mathbf{z}|\mathbf{m})))] 
\label{eq:AAE_dis_loss}
\end{equation}
where $p_{model}(\mathbf{m})$ is the model parameter distribution from a training set. Symbol $\sim$ means ``is distributed as" the following probability distribution. 

Then, in the second stage, the encoder is trained to ``fool" the trained discriminator from the previous stage, thereby implicitly forcing the encoder's output distribution $q(\mathbf{z} | \mathbf{m})$ over the latent parameters to match the target distribution $\mathcal{N}(\mathbf{0}, \mathbf{I}_r)$ in the latent space. The encoder loss function can be written as:
\begin{equation}
    \mathcal{L}_{\mathbf{E}} = - \mathbb{E}_{\mathbf{m} \sim p_{model}(\mathbf{m})}[\log(\mathcal{D}_\mathbf{\mathbf{z}}(q(\mathbf{z} | \mathbf{m})))]
    \label{eq:AAE_adv_loss}
\end{equation}
By minimising these two loss functions, the encoder is trained to produce latent parameters that the discriminator $\mathcal{D}_\mathbf{\mathbf{z}}$ mistakes for ``real" samples. This adversarial training allows the encoder to learn more complex mappings and can prevent the formation of ``holes" in the latent space (sub-volumes that contain no latent space samples, potentially creating a discontinuous latent distribution) that sometimes occur in VAEs \citep{makhzani_adversarial_2016}.

\subsection{Denoising Adversarial Autoencoder (DAAE)}

During training, the DAAE \citep{creswell_denoising_2018} applies a corruption process $c(\mathbf{\widetilde{m}}|\mathbf{m})$ (e.g., adding white Gaussian noise) and learns to reconstruct clean models $\mathbf{m}$ from its corrupted version $\widetilde{\mathbf{m}}$ (see Figure \ref{fig:GNN_architectures}c). The encoder learns a mapping $\mathbf{E: \mathbf{\widetilde{m}} \rightarrow \mathbf{z}}$ that maps all corrupted models $\widetilde{\mathbf{m}}$ to the encoder output distribution $q(\mathbf{z}|\widetilde{\mathbf{m}})$. The encoded latent samples are then mapped back to a denoised version of the corrupted models through a mapping $\mathbf{\mathcal{G}: \mathbf{z}\rightarrow \mathbf{m'}}$. This mapping is learned by the generator network $\mathcal{G}$ by minimising a reconstruction loss (Figure \ref{fig:GNN_architectures}c) that reduces the difference between the input model $\mathbf{m}$ and its reconstruction $\mathbf{m}'$. DAAE regularises its latent space through an adversarial training process, similar to AAE, by matching the distribution of latent samples conditioned on corrupted model samples $q(\mathbf{z}|\widetilde{\mathbf{m}})$, to a target distribution in latent space, such as $\mathcal{N}(\mathbf{0}, \mathbf{I}_r)$ using the discriminator network $\mathcal{D}_\mathbf{\mathbf{z}}$. 

In the first stage, the discriminator is trained to distinguish between ``real" samples from the target distribution $\mathcal{N}(\mathbf{0}, \mathbf{I}_r)$ and ``fake" samples from the encoder output distribution $q(\mathbf{z} | \widetilde{\mathbf{m}})$, through a loss function:
\begin{equation}
    \mathcal{L}_{\mathcal{D}_\mathbf{z}} = - \mathbb{E}_{\mathbf{z} \sim \mathcal{N}(\mathbf{0}, \mathbf{I})}[\log \mathcal{D}_{\mathbf{z}}(\mathbf{z})] - \mathbb{E}_{\mathbf{\widetilde{m}} \sim c(\mathbf{\widetilde{m}}|\mathbf{m})}[\log(1 - \mathcal{D}_{\mathbf{z}}(q(\mathbf{z} | \widetilde{\mathbf{m}})))]
    \label{eq:DAAE_cost_fn}
\end{equation}
In the second stage, the encoder's objective to ``fool" the discriminator is achieved by minimising its own loss function:
\begin{equation}
    \mathcal{L}_{\mathbf{E}} = - \mathbb{E}_{\mathbf{\widetilde{m}} \sim c(\mathbf{\widetilde{m}}|\mathbf{m})}[\log(\mathcal{D}_\mathbf{\mathbf{z}}(q(\mathbf{z} | \widetilde{\mathbf{m}})))]
\end{equation}

\section{Escaping prior samples}
\label{app:Escaping prior samples}

\begin{figure}
    \centering
    \includegraphics[width=1\linewidth]{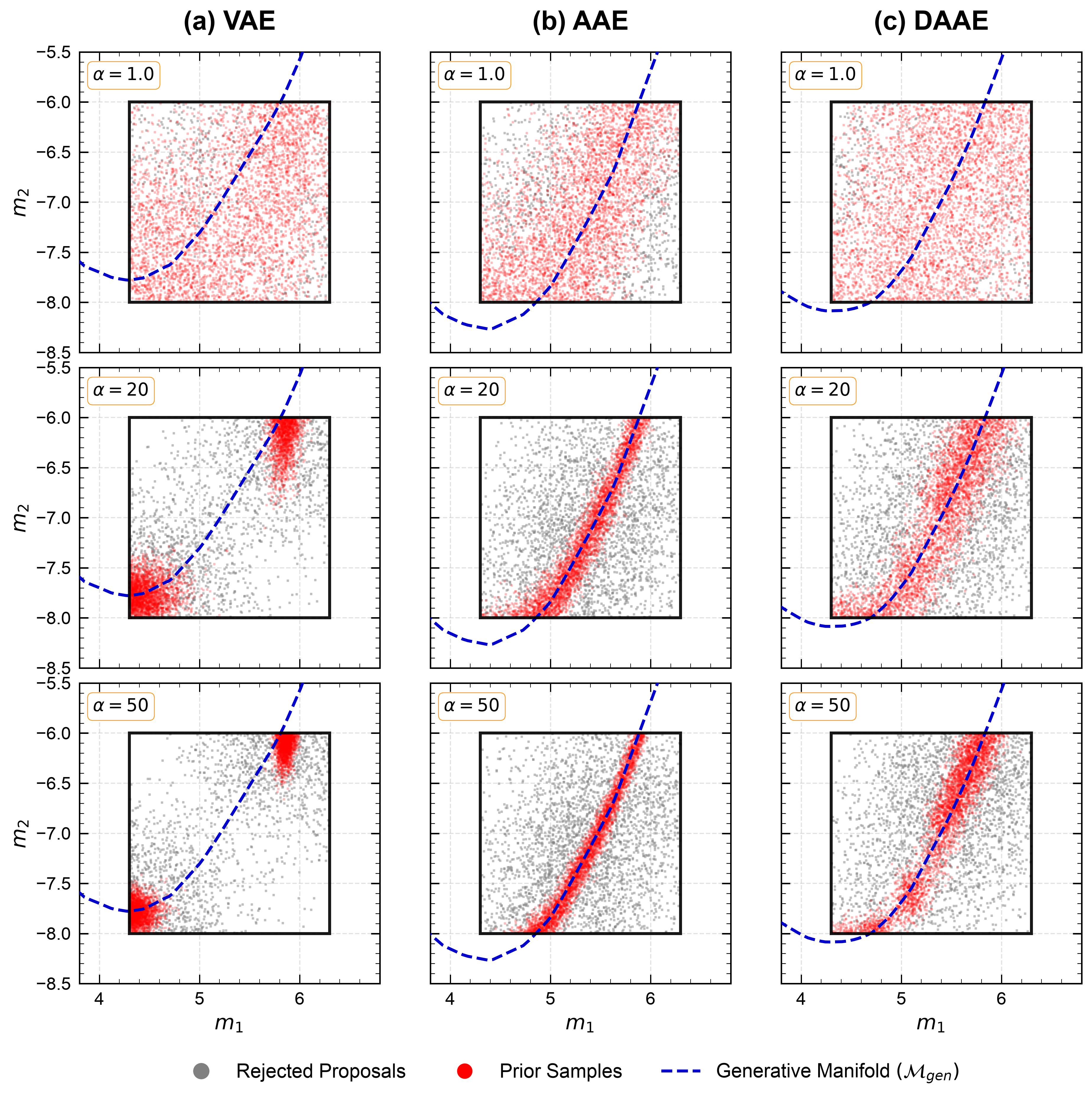}
    \caption{MCMC results from the black square outlined region in Figure \ref{fig:toy_problem_2}, comparing the escaping priors (a) VAE, (b) AAE and (c) DAAE (columns) at three prior-strengths $\alpha = \{1, 10, 50\}$ (rows). Every 250th sample from the prior (red) is plotted along with the rejected proposals (grey). The generative manifold $\mathcal{M}_{gen}$ is shown as a blue dashed curve.}
    \label{fig:toy_problem2_black_box}
\end{figure}

Here we present additional tests to compare the escaping prior samples with the region marked in Figure \ref{fig:toy_problem_2}. We run another 10 MCMC chains (each with $10^5$ samples) with three prior-strengths, $\alpha = 1, 20$ and $50$ to sample from the escaping prior in Figure \ref{fig:toy_problem2_black_box}, particularly within the black square outlined region in Figure \ref{fig:toy_problem_2}. The results show that at higher prior-strengths ($\alpha = 20$ and $50$), the prior sample clusters are again observable for the VAE (Figure \ref{fig:toy_problem2_black_box}a) and a faintly discernible cluster of prior samples appears for the DAAE with $\alpha = 50$, located near the coordinate $(5, -7.5)$. 

\begin{figure}
    \centering
    \includegraphics[width=1\linewidth]{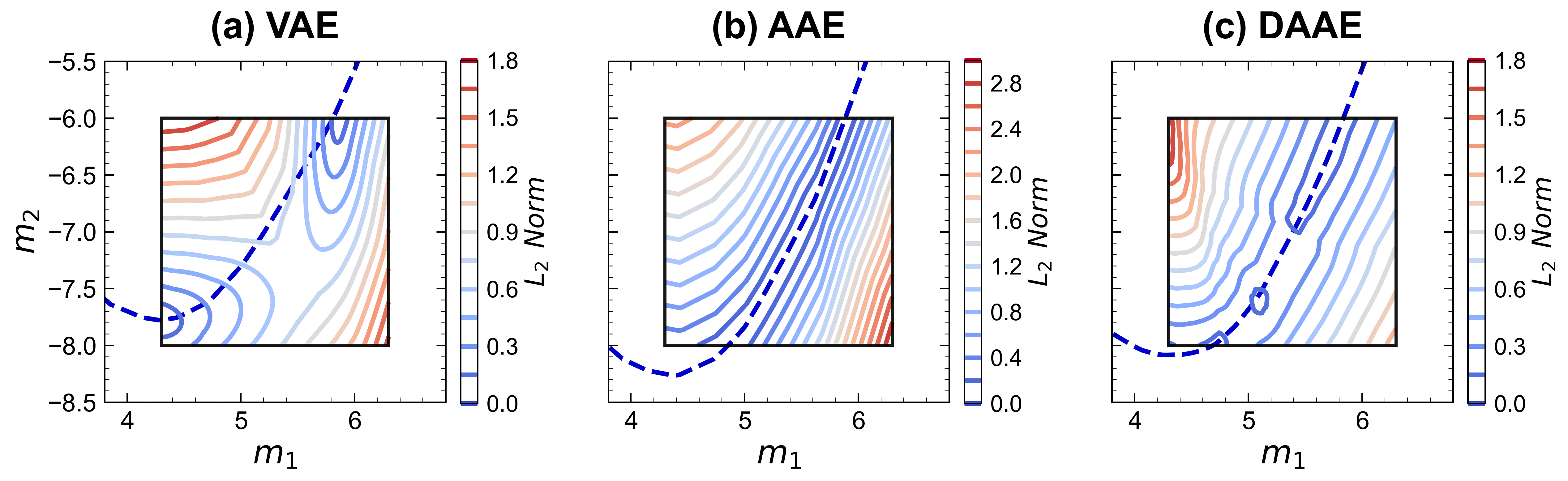}
    \caption{Contour lines represent the $L_2$ norm (Euclidean distance) between the input models $\mathbf{m}$ and their resulting reconstructions $\mathcal{G}(\mathbf{E}(\mathbf{m}))$ on a dense 2D grid, using (a) VAE, (b) AAE and (c) DAAE within the black square outlined region in Figure \ref{fig:toy_problem2_black_box}. The dashed blue curve represents the generative manifold $\mathcal{M}_{gen}$.}
    \label{fig:black_box_L2_norm}
\end{figure}

\begin{figure}
    \centering
    \includegraphics[width=0.85\linewidth]{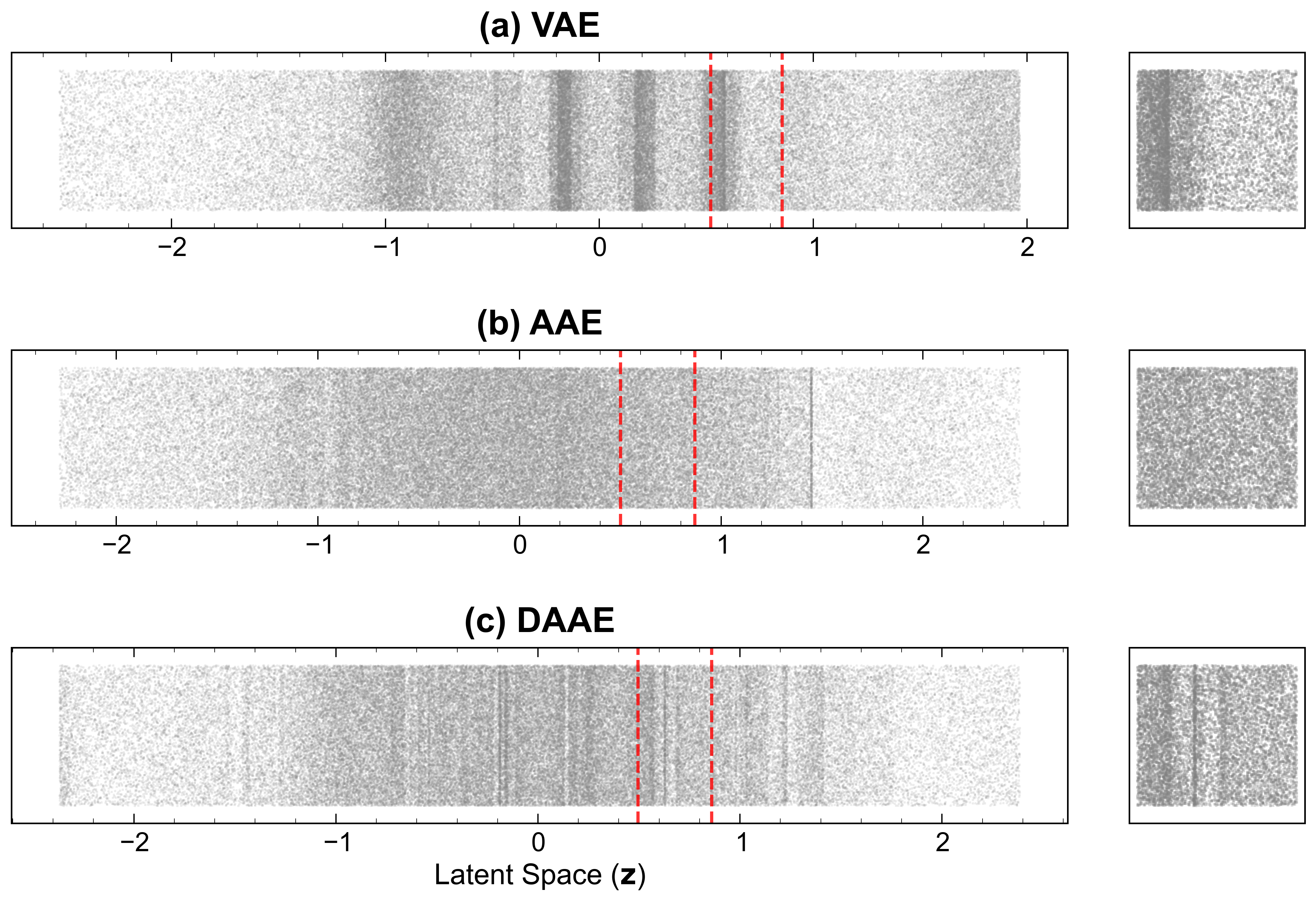}
    \caption{Latent space samples (grey), obtained by passing training data (2D) through the encoder, are displayed in 1D latent parameter space (left panels) and are randomly shifted vertically for better visibility of the under-explored regions or gaps for (a) VAE, (b) AAE, and (c) DAAE. Vertical red dashed lines indicate the upper and lower bounds on latent samples from $\mathcal{M}_{gen}$ marked by the black square in Figure \ref{fig:black_box_L2_norm}. Right panels show a close-up view of the samples between the red dashed lines.}
    \label{fig:latent_space_gaps}
\end{figure}

\begin{figure}
    \centering
    \includegraphics[width=0.9\linewidth]{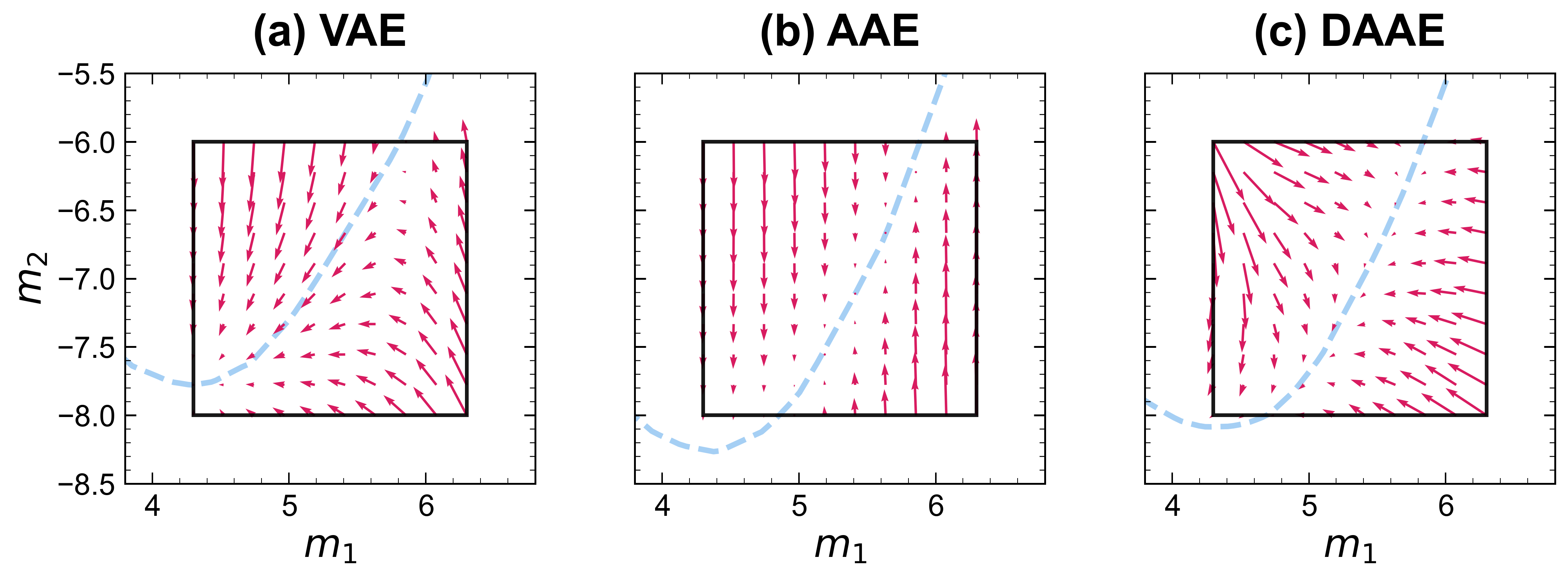}
    \caption{Vector field representing the direction of mapping of a model to its reconstruction on $\mathcal{M}_{gen}$ (dashed blue curve) using equation \ref{eq:E_D_mapping}, for (a) VAE, (b) AAE and (c) DAAE within the black square outlined region shown in Figure \ref{fig:toy_problem2_black_box} over a $10 \times 10$ uniform 2D grid.}
    \label{fig:black_box_vectors}
\end{figure}

Figure \ref{fig:black_box_L2_norm} shows the $L_2$ norm between the input models $\mathbf{m}$ and their resulting reconstructions $\mathcal{G}(\mathbf{E}(\mathbf{m}))$. Using a VAE, we observe higher $L_2$ norm in the central region near $\mathcal{M}_{gen}$, with two separate lower-value zones at the corners in Figure \ref{fig:black_box_L2_norm}a and with a DAAE we observe a small lower-value zone near coordinate $(5, -7.5)$ in Figure \ref{fig:black_box_L2_norm}c. Comparing Figure \ref{fig:toy_problem2_black_box} and Figure \ref{fig:black_box_L2_norm}, we notice that zones with higher $L_2$ norms have lower concentration of prior samples, because a higher $L_2$ norm implies a lower probability (see equation \ref{eq:modelspace_prior_pdf}), causing the MCMC sampler to reject those samples. For this reason, the DAAE-informed escaping prior samples in Figure \ref{fig:toy_problem2_black_box}c at $\alpha = 50$ also show clustering near the coordinate $(5, -7.5)$. 

The KL-divergence term (equation \ref{eq:vae_ELBO}) encourages the encoded latent parameters to resemble a simple predefined distribution (in our case, a standard Gaussian), but it does not strictly enforce the encoder output distribution to be a Gaussian distribution and can create under-explored regions in the latent parameter space \citep{makhzani_adversarial_2016}. Figure \ref{fig:latent_space_gaps} shows the encoder output distribution in the latent space for the VAE, AAE and DAAE. Figure \ref{fig:latent_space_gaps}a shows that the VAE's encoder output distribution does not follow a continuous Gaussian shape and contains five distinct regions with higher sample density or modes. The encoder produced latent samples from the black square outlined region in Figure \ref{fig:toy_problem2_black_box} reside near a mode in the VAE's latent space, the region between the red vertical lines in Figure \ref{fig:latent_space_gaps}a. There are some faint but observable modes in the encoder output distribution of DAAE as well (Figure \ref{fig:latent_space_gaps}c). However, the latent space of AAE seems to better follow the shape of a continuous Gaussian distribution (Figure \ref{fig:latent_space_gaps}b).

\begin{figure}
    \centering
    \includegraphics[width=0.9\columnwidth]{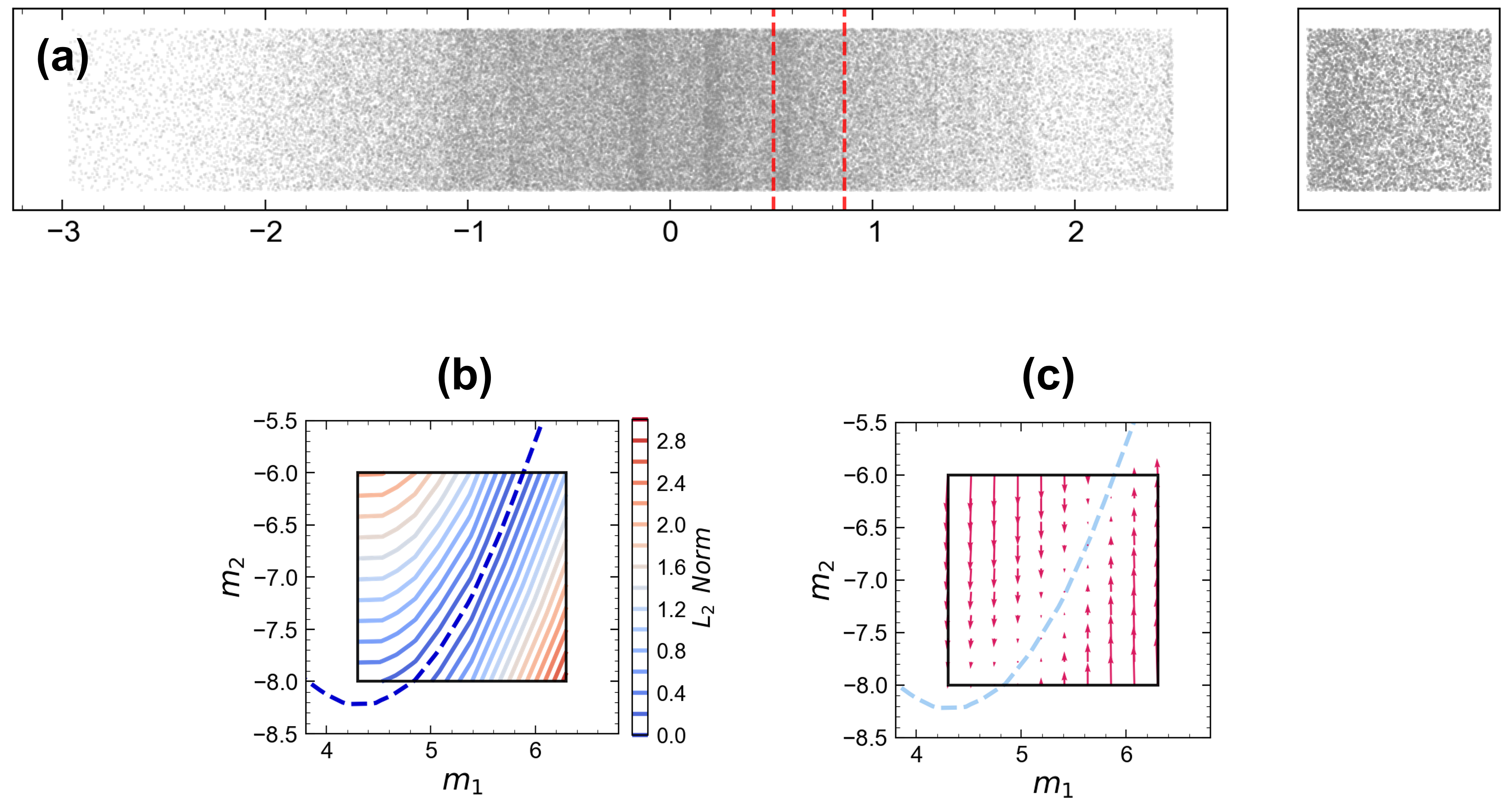}
    \caption{(a) Latent space samples (grey) from a VAE trained with a decreased value of KL divergence weight, and vertical red dashed lines represent latent samples (with a close-up view at the right panel) from $\mathcal{M}_{gen}$ marked by the black square outlined region in Figure \ref{fig:VAE-informed_prior_beta0.05}. (b) Contour lines showing $L_2$ norm between a model $\mathbf{m}$ and its reconstruction $\mathcal{G}(\mathbf{E}(\mathbf{m}))$ over a 2D grid using the same VAE. (c) Vector field representing the direction of mapping of a model to its VAE reconstruction on $\mathcal{M}_{gen}$ (dashed blue curve) and is similar to Figure \ref{fig:black_box_vectors}b.}
    \label{fig:VAE_latspace_beta0.05}

    \vspace{3em}

    \centering
    \includegraphics[width=0.75\columnwidth]{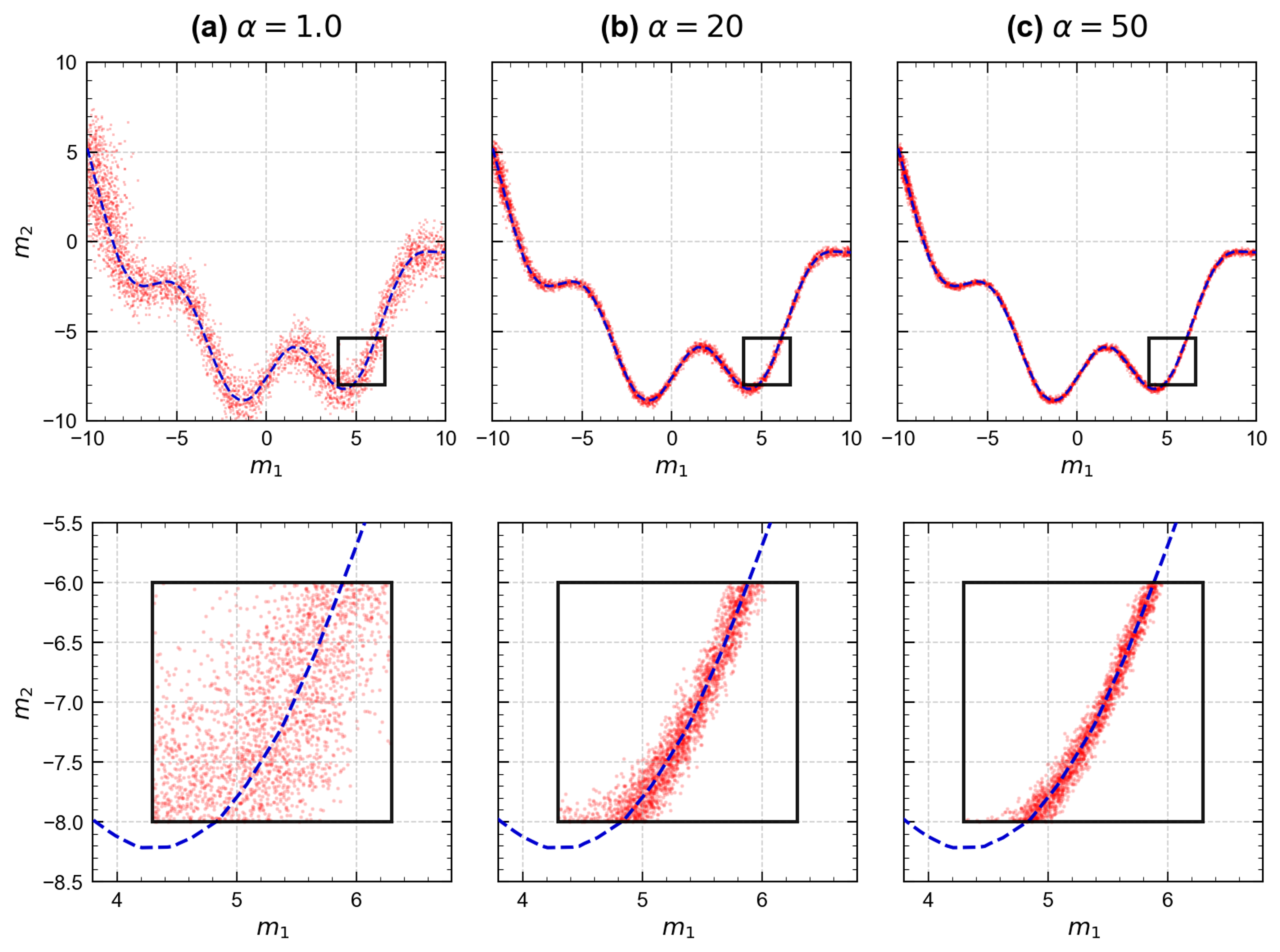}
    \caption{MCMC samples from the escaping prior informed by the VAE mentioned in Figure \ref{fig:VAE_latspace_beta0.05} at three different prior-strengths $\alpha = \{1, 20, 50\}$. Every 250th sample from the prior is shown here, with a close-up view of the black square outlined region plotted in the bottom row. The dashed blue curve represents the generative manifold $\mathcal{M}_{gen}$.}
    \label{fig:VAE-informed_prior_beta0.05}
\end{figure}

Ideally, the encoder maps training samples with similar features to nearby regions in the latent space that reflect their structural or spatial relationships in the model space \citep{hinton_reducing_2006}. Since the generator relies solely on those encoded latent parameters, it maps latent samples that are very close in the latent space to structurally similar or spatially nearby reconstructed model samples. For this reason, the VAE's generator maps latent samples from those denser regions shown in Figure \ref{fig:latent_space_gaps}a (right panel) to reconstructed model samples that are spatially close to each other on $\mathcal{M}_{gen}$ but far from their actual locations on $\mathcal{M}$. Figure \ref{fig:black_box_vectors} clearly demonstrates this phenomenon by displaying the reconstruction vector directions from model samples towards their reconstructions in a uniform 2D grid. The reconstruction vectors indicate that the VAE maps the model samples from the central region of the 2D grid towards a highly localised region on $\mathcal{M}_{gen}$ near the bottom-left corner in Figure \ref{fig:black_box_vectors}a. Using an AAE or DAAE (Figures \ref{fig:black_box_vectors}b and \ref{fig:black_box_vectors}c) causes the reconstruction vectors to converge uniformly towards $\mathcal{M}_{gen}$, rather than to localised regions. This means the reconstructed samples are evenly distributed across $\mathcal{M}_{gen}$. Therefore, maintaining a continuous latent space is crucial to prevent localisations or clustering of the escaping prior samples in the model space. A possible way to make the VAEs' latent space continuous is to adjust the weight of the KL divergence term (see Figures \ref{fig:VAE_latspace_beta0.05} and \ref{fig:VAE-informed_prior_beta0.05}), whether by trial and error or through an annealing strategy \citep{bowman_generating_2016}.
\clearpage
\section{Hyperprior distributions}
\label{app:Hyperprior distributions}

\begin{figure}
    \centering
    \includegraphics[width=0.85\linewidth]{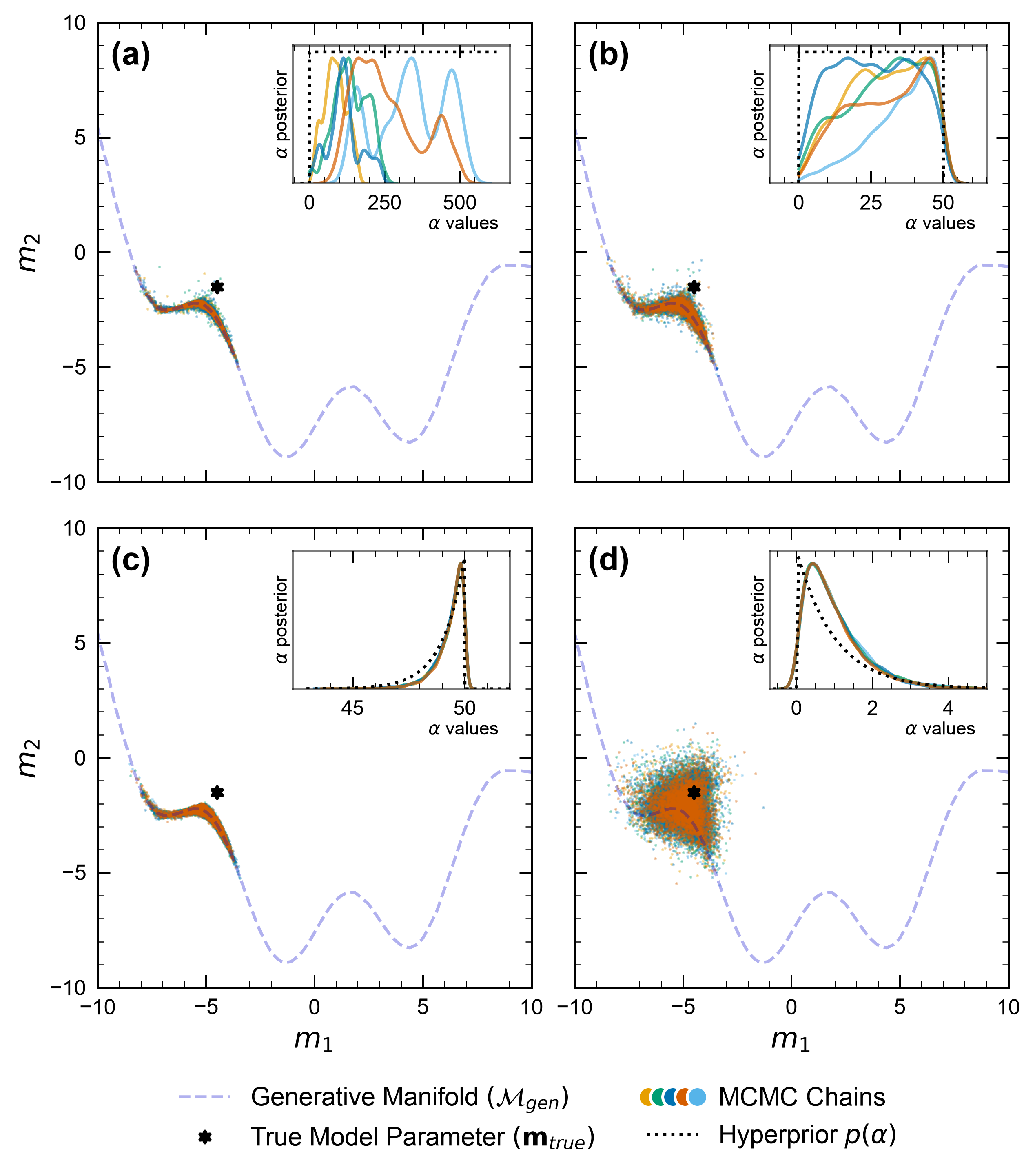}
    \caption{Posterior samples from 5 MCMC chains evaluated under 4 different hyperprior distributions: (a) uniform on $\alpha \in (0, \infty)$, (b) uniform on $\alpha \in [0.01, 50]$, (c) exponential $\exp(\alpha)$ for $\alpha \in (0.01, 50]$, and (d) equation \ref{eq:hyperprior(alpha)} with $(\gamma, \alpha_{lb}) = (1, 0.01)$. The main panels display the posterior samples (5 colours represent 5 MCMC chains), the generative manifold $\mathcal{M}_{gen}$ (blue dashed curve) and the true model parameter $\mathbf{m}_{true}$ (black star). Inset plots compare the sampled $\alpha$ posteriors with their corresponding hyperpriors (black dotted curves).}
    \label{fig:hyperprior_test_near}
\end{figure}

The hyperprior $p(\alpha)$ in equation \ref{eq:hier_posterior} represents the prior information on the $\alpha$ values. Here we compare 4 different hyperpriors using a low-dimensional (2D) problem in which $\mathbf{m}_{true}$ is close to $\mathcal{M}_{gen}$. If we use a non-informative uniform distribution for $p(\alpha)$ within a range [$\alpha_{min}, \alpha_{max}$], all we need to know is the minimum ($\alpha_{min}$) and the maximum value ($\alpha_{max}$) of the hyperparameter, which is also prior information. In Figure \ref{fig:hyperprior_test_near}a, we use a uniform distribution for $p(\alpha)$ within the range $\alpha \in (0, \infty)$, and there is no initial preference for any specific value of the bounds. In this case, we set the upper limit and starting point of the chains to very large alpha values, which can be any positive value; however, the chains immediately discard those for lower values during burn-in. Alternatively, in Figure \ref{fig:hyperprior_test_near}b, we use another uniform distribution for $p(\alpha)$ bounded within a sufficiently broad range, such as $[0.01, 50]$, so that all potential $\alpha$ values are included within those bounds. In both cases, the $\alpha$ posterior distribution shows unreasonably wide variance (see inset plots of Figures \ref{fig:hyperprior_test_near}a,b), which includes excessively high values, thus introducing unnecessary and strong prior information from $\mathcal{M}_{gen}$. This constrains the posterior samples to be on or very close to $\mathcal{M}_{gen}$ and $\mathbf{m}_{true}$ is not captured in the model parameter posterior distributions from any MCMC chains with uniform hyperpriors. \cite{gelman_prior_2006} showed that when observational data are sparse and provide limited information about the posterior variance of model parameters, a non-informative uniform hyperprior leads to an unrealistically wide posterior variance for $\alpha$. Using an exponential hyperprior (Figure \ref{fig:hyperprior_test_near}c), which assigns higher probability to larger $\alpha$ values and decreases monotonically towards zero as $\alpha$ decreases, the posterior samples are again restricted to $\mathcal{M}_{gen}$. Figure \ref{fig:hyperprior_test_near}d shows that $\mathbf{m}_{true}$ is effectively captured within the posterior distribution of model parameters using our hyperprior distribution described in equation \ref{eq:hyperprior(alpha)}.

\clearpage
\section{Prior correlations}
\label{app:prior_correlations}

\begin{figure}
    \centering
    \includegraphics[width=1\linewidth]{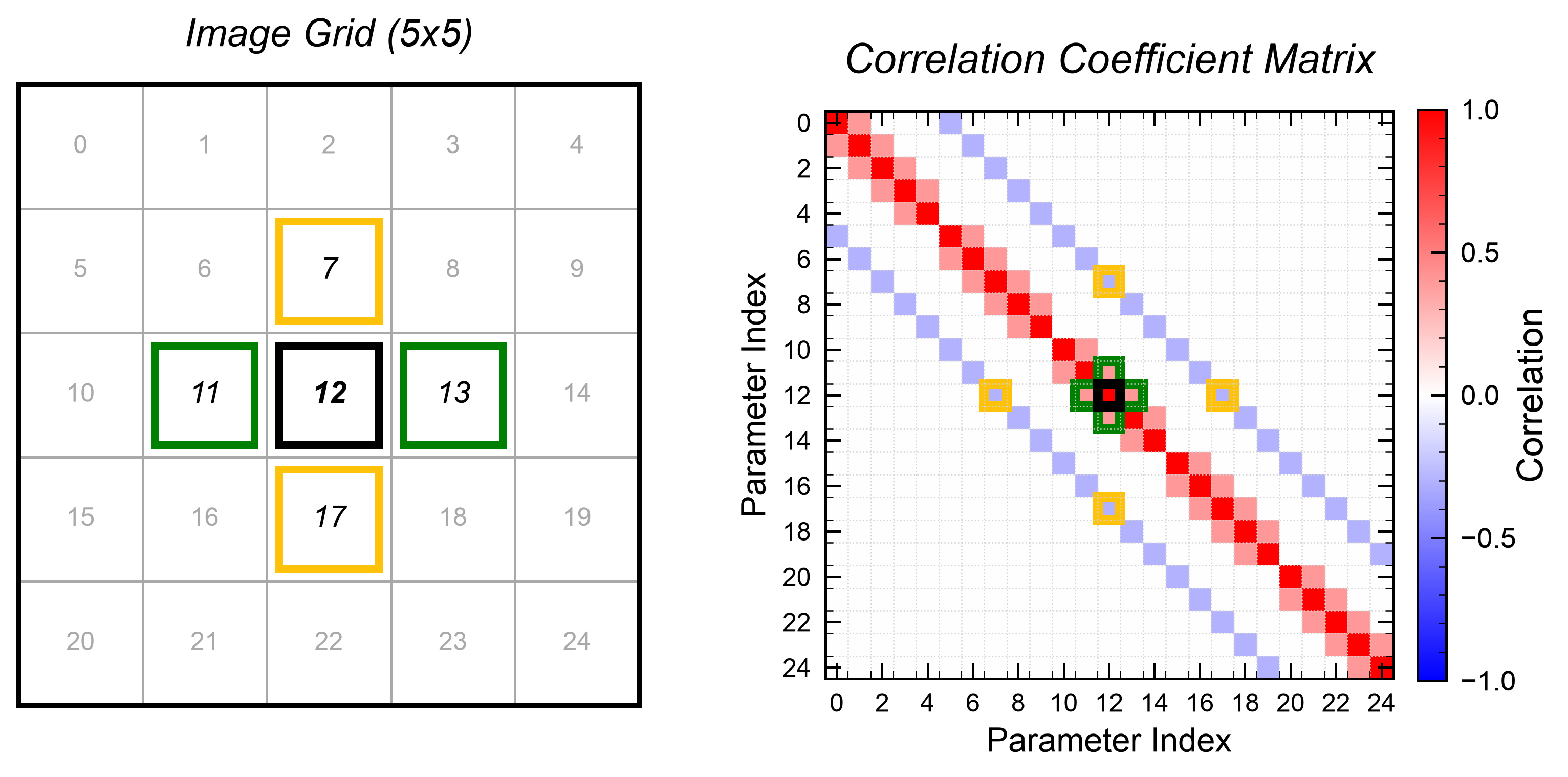}
    \caption{Schematic illustration showing the spatial correlation within a $5\times5$ image grid (left) to their corresponding representation in a $25\times25$ correlation coefficient matrix (right). The central parameter (index 12, black box) is highlighted alongside its immediate horizontal (indices 11 and 13, green boxes) and vertical (indices 7 and 17, golden-yellow boxes) neighbours. In the resulting correlation matrix, the variance of the central parameter (index 12, black box) is located on the main diagonal (black outline). Correlations with horizontal neighbours appear as immediately adjacent off-diagonal elements (green outlines). In contrast, correlations with vertical neighbours are displayed as secondary off-diagonal bands offset by the width of the spatial grid (golden-yellow outlines). The colour scale denotes the strength of the correlation, ranging from -1.0 (blue) to 1.0 (red).}
    \label{fig:corr_tutorial}
\end{figure}

In this Appendix, we present and compare correlation matrices among a uniform prior, the training samples used for the GNNs, and the samples they generate. Generally, the first off-diagonals directly below and above the main diagonal show correlation between horizontally adjacent cells (green, Figure \ref{fig:corr_tutorial}), while the other off-diagonals  (golden-yellow, Figure \ref{fig:corr_tutorial}) on both sides of the main diagonal indicate correlations between vertically adjacent cells.

\begin{figure}
    \centering
    \includegraphics[width=1\columnwidth]{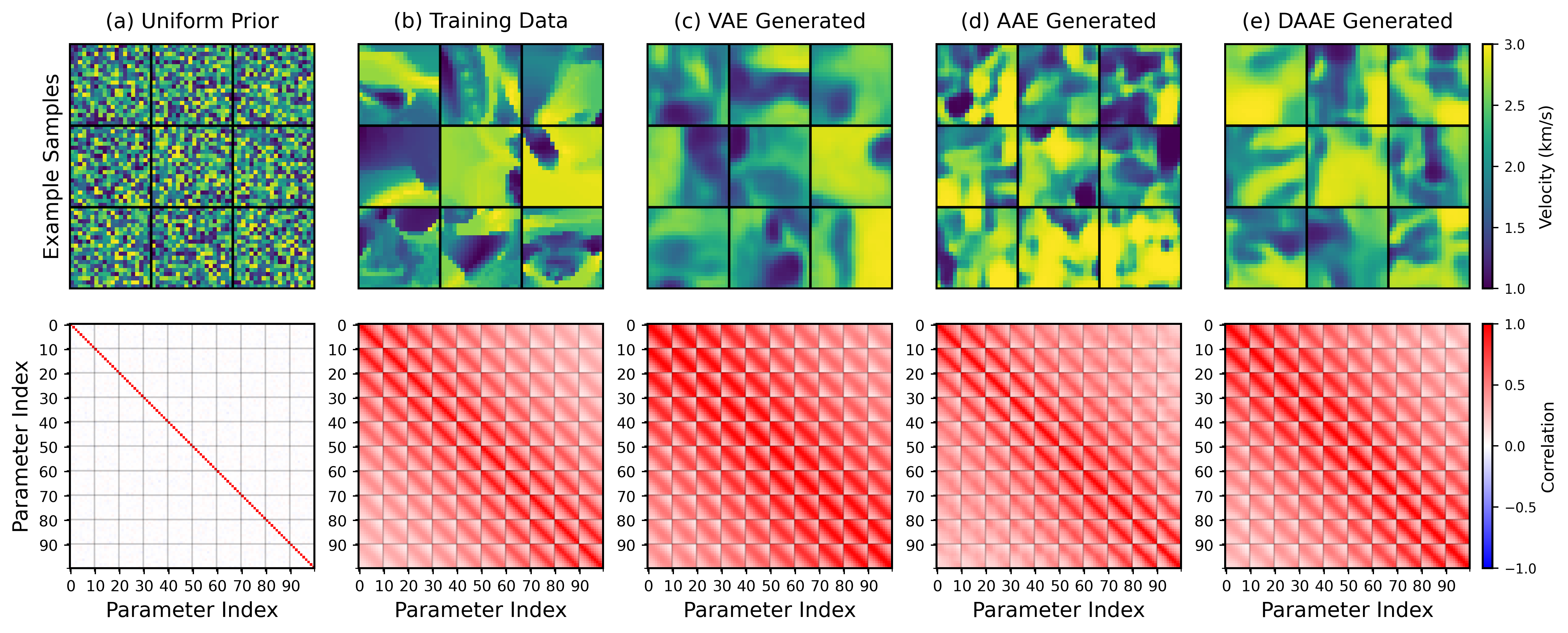}
    \caption{Random samples (top row) and correlation matrices (bottom row) obtained from (a) a non-informative uniform prior, (b) the training set shown in Figure \ref{fig:GNN_training_images_horizontal}, and samples generated using trained (c) VAE, (d) AAE, and (e) DAAE with 64 latent parameters. In the top row, 9 samples, divided by black lines, are shown in each panel.}
    \label{fig:prior_cor_horizontal}
\end{figure}

A $10\times10$ correlation kernel is used to compute the correlation coefficient across 100 model parameters. Figure \ref{fig:prior_cor_horizontal}b indicates correlations between laterally and vertically offset cells in the training images for the scenario in Figure \ref{fig:GNN_training_images_horizontal}. The GNNs learn these correlations, as demonstrated by the correlation coefficients found in the samples that they generate (Figures \ref{fig:prior_cor_horizontal}c-e). To produce GNN-generated image samples in model space, latent space samples are drawn from $\mathcal{N}(0, \mathbf{I})$ and passed through the generator. In the case of the uniform prior (Figure \ref{fig:prior_cor_horizontal}a), examples are generated by randomly sampling each parameter from a uniform distribution between [1, 3] km/s. Its correlation coefficient matrix shows only the main diagonal, indicating that each parameter correlates only with itself - there is no spatial correlation with neighbouring parameters, which is why the rest of the matrix appears white or grey.

\begin{figure}
    \centering
    \includegraphics[width=1\columnwidth]{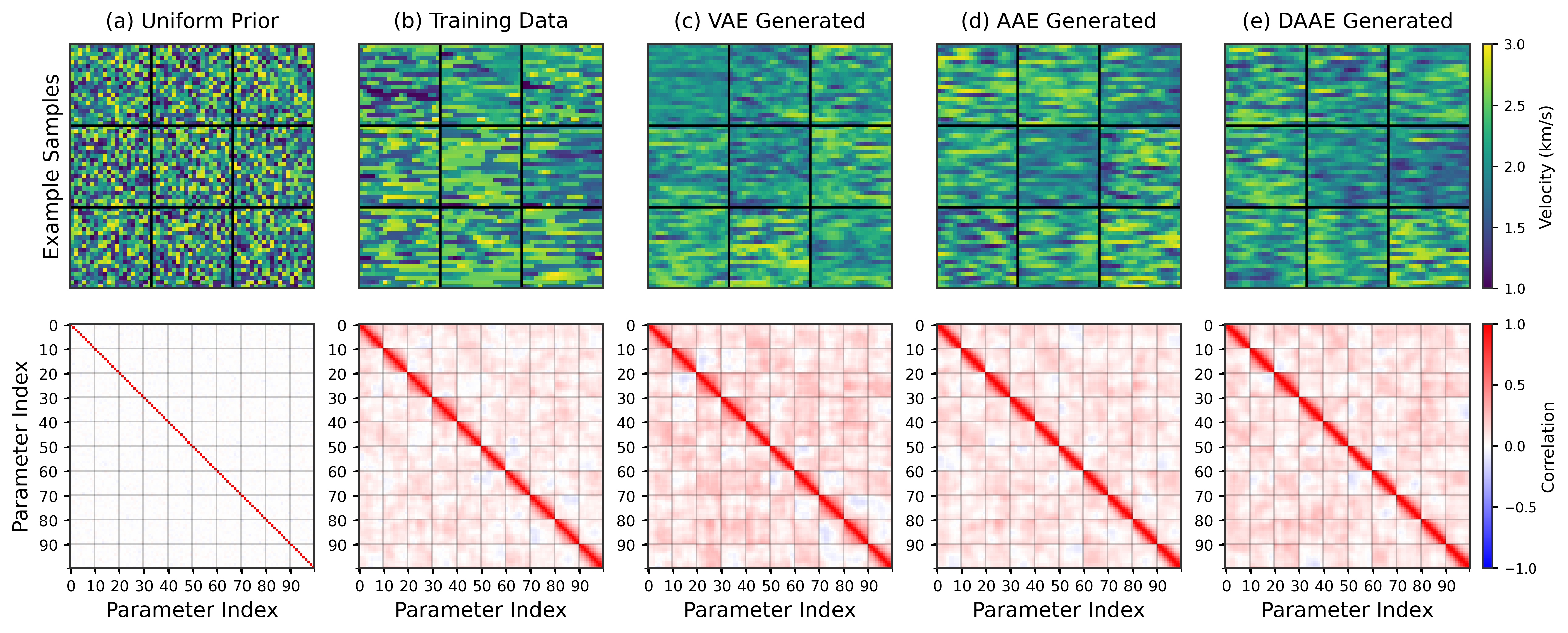}
    \caption{Random samples (top row) and correlation matrices (bottom row) obtained from (a) a non-informative uniform prior, (b) the training set shown in Figure \ref{fig:GNN_training_images_vertical}, and samples produced by trained (c) VAE, (d) AAE, and (e) DAAE with 8 latent parameters. In the top row, each panel shows 9 samples, separated by black lines.}
    \label{fig:prior_cor_vertical_Z8}

    \vspace{3em}

    \centering
    \includegraphics[width=1\columnwidth]{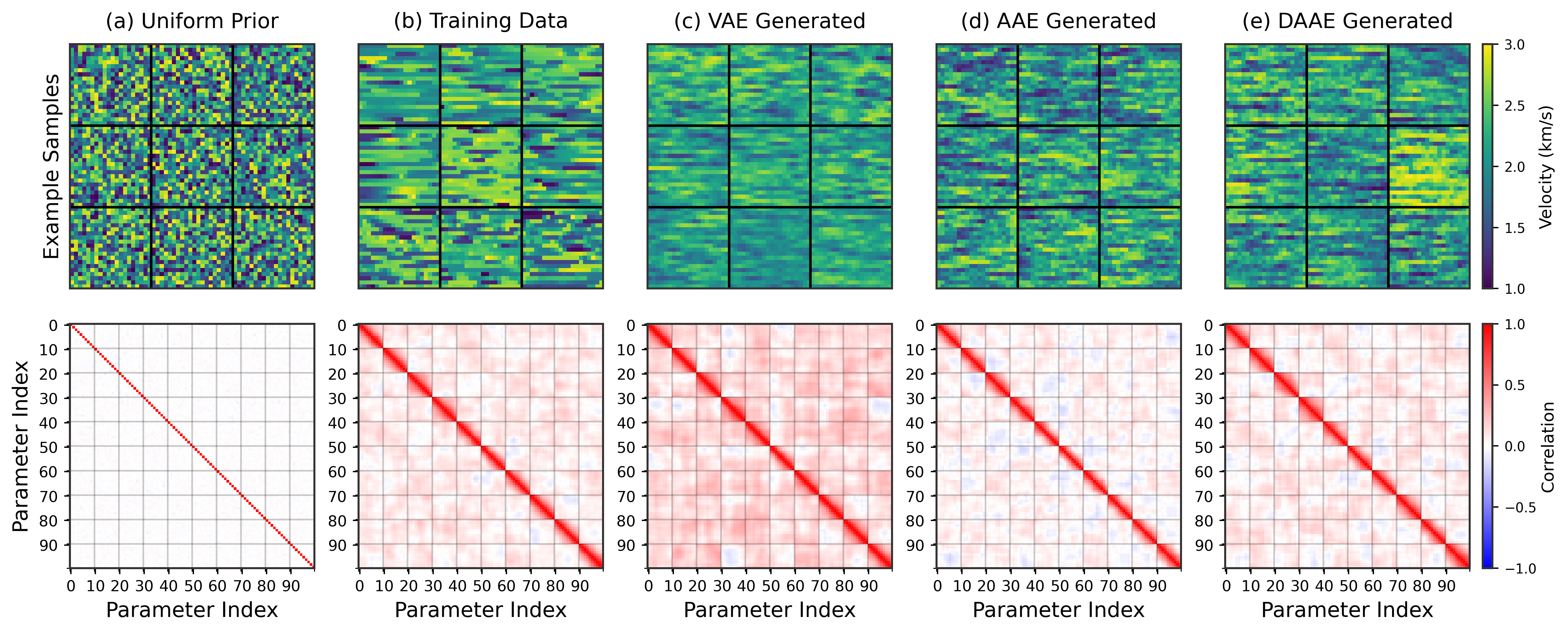}
    \caption{Random samples (top row) and correlation matrices (bottom row) obtained from (a) a non-informative uniform prior, (b) the training samples shown in Figure \ref{fig:GNN_training_images_vertical}, and samples generated using trained (c) VAE, (d) AAE, and (e) DAAE with 128 latent parameters. In the top row, 9 samples, separated by black lines, are shown in each panel.}
    \label{fig:prior_cor_vertical_Z128}
\end{figure}

Figure \ref{fig:prior_cor_vertical_Z8}b presents the training image set described in Figure \ref{fig:GNN_training_images_vertical} with sedimentary deposition, in which rock properties are relatively continuous horizontally but vary vertically across the layers. The broad band of high values on and around the diagonal, and repeating pattern within each grid square in the figure (representing one row in the training image) indicate strong horizontal correlations within a layer and weak vertical correlations across layers. Correlations from the GNN-generated samples are shown in Figures \ref{fig:prior_cor_vertical_Z8}c-e. All three GNNs learn the underlying spatial distribution of the training set, and their generated samples show nearly identical correlation matrices.

Similarly, the training set in Figure \ref{fig:prior_cor_vertical_Z128}b presents horizontally layered sedimentary rock strata alongside their correlation matrix. However, the GNNs were trained with a latent parameter dimension of 128 rather than 8 (Figure \ref{fig:prior_cor_vertical_Z8}). This gives the GNNs a significantly higher capacity to learn and reconstruct complex details. Although the correlation between the VAE-generated samples in Figure \ref{fig:prior_cor_vertical_Z8}c and Figure \ref{fig:prior_cor_vertical_Z128}c remains similar, the correlations for the case of AAE and DAAE-generated samples change slightly. With a higher latent space dimensionality, GNNs might capture slightly more high-frequency variation within the sediment layers, as shown in Figures \ref{fig:prior_cor_vertical_Z128}d,e.

\end{document}